\documentclass[aps,twocolumn,showlabels,showrefs,amsmath,amssymb,pre,superscriptaddress,floatfix,colors]{revtex4-1}

\usepackage{graphicx}
\usepackage{dcolumn}
\usepackage{bm}
\usepackage{amssymb}
\usepackage{hyperref}
\usepackage{multirow}
\usepackage{color}
\usepackage[normalem]{ulem}

\begin{document}

\title{Unified analysis of  Topological Defects in 2D systems of Active and Passive disks}

\date{\today}
 
\author{Pasquale~Digregorio}
\affiliation{Dipartimento  di  Fisica,  Universit\`a  degli  Studi  di  Bari  and  INFN,
Sezione  di  Bari,  via  Amendola  173,  Bari,  I-70126,  Italy}
\affiliation{Centre  Europ\'een  de  Calcul  Atomique  et  Mol\'eculaire (CECAM), Ecole  Polytechnique  F\'ed\'erale  de  Lausanne (EPFL),  Batochimie,  Avenue  Forel  2,  1015  Lausanne,  Switzerland}

\author{Demian~Levis} 
\affiliation{Departament de Fisica de la Materia Condensada, Universitat de Barcelona, Marti i Franques 1, 08028 Barcelona, Spain}
\affiliation{UBICS  University  of  Barcelona  Institute  of  Complex  Systems,  Mart\'{\i}  i  Franqu\`es  1,  E08028  Barcelona,  Spain}


\author{Leticia~F.~Cugliandolo}
\affiliation{Sorbonne Universit\'e, Laboratoire de Physique Th\'eorique et Hautes Energies, CNRS UMR 7589, 
4 Place Jussieu, 75252 Paris Cedex 05, France}
\affiliation{Institut Universitaire de France, 1 rue Descartes, 75231 Paris Cedex 05 France}

\author{Giuseppe~Gonnella}
\affiliation{Dipartimento  di  Fisica,  Universit\`a  degli  Studi  di  Bari  and  INFN,
Sezione  di  Bari,  via  Amendola  173,  Bari,  I-70126,  Italy}

\author{Ignacio~Pagonabarraga}
\affiliation{Centre  Europ\'een  de  Calcul  Atomique  et  Mol\'eculaire (CECAM), Ecole  Polytechnique  F\'ed\'erale  de  Lausanne (EPFL),  Batochimie,  Avenue  Forel  2,  1015  Lausanne,  Switzerland}
\affiliation{UBICS  University  of  Barcelona  Institute  of  Complex  Systems,  Mart\'{\i}  i  Franqu\`es  1,  E08028  Barcelona,  Spain}

\email[email: ]{name@}


\begin{abstract}
We provide a comprehensive  quantitative analysis of  localized and extended topological defects in the steady state of 
2D passive and active repulsive Brownian disk systems. 
We show that,  both in and out-of-equilibrium, the passage from the solid to the hexatic is driven by the unbinding of dislocations, in
 quantitative agreement with the KTHNY singularity. Instead, although 
disclinations dissociate as soon as the liquid phase appears, extended clusters of defects largely dominate below the solid-hexatic critical line. The latter percolate in the liquid phase very close to  the hexatic-liquid transition, both for continuous and discontinuous transitions, 
in the homogeneous liquid regime. At critical percolation the clusters of defects  are fractal with statistical and geometric 
properties that, within our numerical accuracy, are independent of the activity and compatible with the universality class of 
uncorrelated critical percolation. We also characterize the spatial organization of 
different kinds of point-like defects and  we show that the disclinations are not free, but rather always very near more complex defect structures.
At high activity, the bulk of the dense phase generated by Motility-Induced Phase Separation is characterized by a density of point-like defects, 
and statistics and morphology of defect clusters, set by the amount of activity and not the packing fraction.
Hexatic domains within the dense phase are separated by grain-boundaries along which a finite network of
topological defects resides, interrupted by gas bubbles in cavitation. The fractal dimension of this 
network diminishes for increasing activity. This structure is dynamic in the sense that the defect network allows for an
unzipping mechanism that leaves free space for gas bubbles to appear, close,  and even be released into the  dilute phase.
\end{abstract}

\maketitle


\section{Introduction}

%

In two-dimensions (2D) thermal fluctuations  often prevent the emergence of  Long-Range Order (LRO), as illustrated by the absence of spontaneous magnetization in 2D Heisenberg magnets~\cite{MerminWagner} and positional order in 2D particle systems~\cite{Mermin1968}. However, unconventional phase transitions driven by topological defects can still occur: 
for example, in 2D planar magnets, the binding-unbinding of vortices drive the so-called Berezenskii-Kosterlitz-Thouless (BKT)  transition between a paramagnet and a low temperature critical phase with Quasi-Long-Range Order (QLRO) \cite{Berezinskii1971, KosterlitzThouless}. 

The nature of the melting transition in 2D  
 is far more involved and controversial than the BKT magnetic one~\cite{StrandburgRev, GlaserRev,vonGrunberg07,Gasser09,Gasser10,Wang16}, 
 partly due to the fact that particle systems might have two types of order, 
 translational and orientational, and thus two kinds of topological defects: dislocations and disclinations. 
 The most standard picture of 2D melting  in  spherically symmetric particle systems 
 follows the work of Kosterlitz-Thouless-Halperin-Nelson-Young (KTHNY) \cite{Halperin1978, Nelson1979, Young1979}, according to which the transition from the solid (with positional QLRO and orientational LRO) to the isotropic liquid, occurs in two-steps, separated by an intermediate hexatic phase characterized by orientational QLRO.  In this picture, these solid-hexatic and hexatic-liquid transitions are of BKT type, driven by the unbinding of dislocations and disclinations, respectively.  While evidence for the KT\-HNY scenario was given by some experiments~\cite{Maret1999colloids} and simulations~\cite{ChenKaplan1995, Prestipino2005GC, QiYukawa, Wierschem2011}, alternative mechanisms were proposed~\cite{StrandburgRev, GlaserRev}. In particular, one in which 
 the continuous two-step scenario is preempted by a single solid-liquid first-order transition driven by the aggregation of defects into grain-boundary-like structures~\cite{Chui1982, Chui1983, Kleinert1983}. 
 Recent simulations~\cite{BernardKrauth, KapferKrauth,PRLino,QiDijkstra,PicaCiamarra20,Khali21}  (see also~\cite{Delhommelle04})  
 and experiments~\cite{Dullens} showed that the melting of equilibrium passive 
 repulsive disks shares aspects of both scenarios:   a BKT solid-hexatic transition but a first-order hexatic-liquid one, if the interaction potential is stiff enough. It was thus suggested that the dis\-cli\-na\-tion-unbinding mechanism should be preempted by a first-order transition involving the proliferation of clusters of defects forming a percolating network in the liquid regime~\cite{QiYukawa,KapferKrauth, QiDijkstra}. The mechanism \ for the hexatic-liquid transition  should then be similar to the one controlling melting in 3D colloidal crystals, i.e.,  the growth of the liquid along grain-boundaries~\cite{Alsayed2005,Wang16} (in this case delimiting regions with different hexatic order). Yet,  neither a quantitative analysis of such clusters nor the derivation of a theory for the stability of the hexatic phase against grain-boundaries have been conducted. Moreover, and surprisingly enough, no clear experimental evidence and little numerical one \cite{QiDijkstra,Mazars15,Mazars19}
for dislocation unbinding at the solid-hexatic transition exists. 

Besides the issues that still remain unclear for passive particle systems, the classical problem of 2D phase transitions is experiencing a resurge of interest 
in the context of active matter systems. These are collections of self-propelled particles which pump energy from their environment and convert it into motion in the presence of dissipation, in a way that breaks detailed balance.  Different aspects and approaches to active matter have been extensively reviewed in the litterature~\cite{Fletcher09,Menon10,Vicsek12,Romanczuk2012,CatesDB, Ramaswamy13,MarchettiRev,Marenduzzo15,CatesRev,Bechinger16,Bernheim-Groswasser18,Cugliandolo18,Szamel19,Allen19,Bar20,WinklerRev,MarchettiTopo,Martin21}. 
Recently, it has been shown that self-propelled hard disks follow the two-step melting scenario of their passive counterparts at weak activities,  
up to a threshold above which hexatic-liq\-uid coexistence, characteristic of the first-order nature of the transition, disappears \cite{PRLino} (see the phase diagram in Fig.~\ref{fig:phasediagram}). 
 Both the hexatic-liquid and solid-hexatic transitions are shifted to higher densities as the degree of activity is increased and, at sufficiently high {energy injection},   these transitions overlap with a coexistence region purely triggered by self-propulsion, the 
 so-called 
 Motility-Induced Phase-Separation (MIPS)~\cite{CatesRev, TailleurCates2008,Fily2012,Redner2013f,Bialke2013,Stenhammar2014,wysocki2014cooperative, KKK}. Although topological defects are known to be crucial to grasp 2D equilibrium phase transitions,  little attention has been paid to the study of how they influence phase transitions in active systems \cite{Frey2014,PaliwalDijkstra,Bartolo2021,Ylann}.
 
 In this paper we shed new light on the 2D melting mechanism, both in- and out-of-equilibrium. 
 With a huge computational effort we collected enough numerical data to perform a thorough quantitative analysis of 
 the full spectrum of topological defects in systems of passive  and active repulsive Brownian disks. 
 Besides  a careful  analysis of point-like defects, we also characterize more complex structures,  
 in the form of networks of strings and extended clusters of defects. Worthy of note is that we work in regimes in which the 
 topological defects are numerous and of different kind, and it would be very 
hard,  if not impossible, to follow their individual temporal evolution, contrary to what is sometimes possible to do
in studies of active 
nematics~\cite{Thampi13b,Thampi14,Pismen13,Giomi13a,Giomi13b,Giomi14, Shankar18,Shankar19}.

Let us summarize here our results.
 We first show that the solid-hexatic transition is driven by the unbinding of dislocations
 and agrees, even \emph{quantitatively}, with the KTHNY scenario at all activities.   Although we see dislocation unbinding and hence 
 proliferation of disclinations as soon as the liquid phase appears, either in coexistence or through a continuous phase transition, 
 we find that 2D melting is generically accompanied by the {\it percolation} of clusters of topological defects
 close to the hexatic-liquid transition, both for active and passive systems. The presence of a spanning cluster, with similar properties to the ones 
 of critical percolation,  seems to be independent of the order of the transition, and thus 
 constitutes a fundamental feature of the melting of the hexatic. At critical percolation, the nature and geometry of the clusters are independent of 
 activity within our numerical accuracy.  We complement the study of the point-like defect number densities with the one of the defect-defect spatial correlations. 
 This allows us to grasp the spatial organization of different kinds of point-like defects and, in particular, the preferred distance 
 between them.  We prove  in this way that disclinations are not free but lie 
 close to more complex defect structures at the interesting high densities where the thermodynamic transition takes place.
 Finally, we focus on the MIPS phase and we demonstrate that the defects are mostly located along 
 grain boundaries between regions with different local orientational order. These  give rise to a sort of unzipping dynamic mechanism that leaves
 free space for  gas bubbles to cavitate. The size of this network does not scale with the size of the 
 dense phase since  the bubbles interrupt its connectivity. Consistently with other measurements, the properties of the 
 defects in MIPS do not depend on  the packing fraction at fixed activity.
 
 The layout of the paper is the following. In Sec.~\ref{sec:model-methods} we introduce the model and 
 the numerical methods. Section~\ref{sec:topological-defects} is devoted to the definition  and computational identification of all kinds of 
 topological defects.  We also recall the definition of several observables and critical percolation properties.
We have chosen to order the following Sections in a way that goes from the analysis of the simpler point-like defects to the one of the 
larger structures. In Sec.~\ref{sec:point-like} we focus on the study dislocations and disclinations, and their influence on the 
 solid-hexatic and hexatic-liquid transitions, respectively. The MIPS phase and its network of strings of defects are studied in Sec.~\ref{sec:MIPS}. 
 The properties of the defect clusters, their percolation, and its interplay with the 
 melting mechanism are analyzed in Sec.~\ref{sec:clusters}. 
 Finally, we close the paper in Sec.~\ref{sec:conclusions}, where we put our results in the 
 context of generic melting studies on the one hand, and we explain how the topological defect organization in 
 MIPS relates to the micro hexatic phase separation on the other. We  also make a short summary of our findings in this Section. 
 Several Appendices complement our analysis. 

\section{Model and methods}
\label{sec:model-methods}

In this Section we present the model of Active Brownian Particles \cite{LowenABP,romanczuk2012active,Fily2012}, we discuss the numerical methods employed, and 
we recall the  P\'eclet number -- packing fraction phase diagram at fixed temperature~\cite{PRLino}, focusing on the 
regime of variation of parameters that will be relevant to our study.

\subsection{Active Brownian Particles}
\label{subsec:model}

We consider  $N$  
Active Brownian Particles (ABP) located at the positions ${\bold{r}}_i$ 
in an $L_x\times L_y$ rectangular box with periodic boundary conditions (PBC) and obeying 
\begin{equation}
\label{eq:langevin}
\begin{array}{rcl}
	m\ddot{\bold{r}}_i+ \gamma\dot{\bold{r}}_i &=& F_{\rm act} \bold{n}_i- \sum_{j(\neq i)}{\boldsymbol{\nabla}}_iU(r_{ij}) 
	\vspace{0.2cm}
	\\
	&& + \sqrt{2 \gamma k_B T}\bm{\xi}_i \; ,
		\vspace{0.2cm}
	\\
	\dot{\theta}_i &=& \sqrt{2 D_{\theta}}\eta_i 
	 \;  ,
	 \end{array} 
\end{equation}
where $F_{\rm act}$ is the self-propulsion force acting along the time-dependent direction 
$\bold{n}_i=( \cos{\theta_i},\sin{\theta_i})$, $r_{ij}=|{\bold{r}}_i-{\bold{r}}_j|$ is the inter-particle distance, 
and $U(r)$ is a repulsive potential. We consider two cases: 
(i) a hardcore form
\begin{equation}
U^H(r)=4\varepsilon [({\sigma}/{r})^{64}-({\sigma}/{r})^{32}]+\varepsilon
\label{eq:hard-disks}
\end{equation}
if $r< \sigma_d=2^{1/32}\sigma$ and $0$ otherwise
 or (ii) a soft  one~\cite{KapferKrauth}
\begin{equation}
U^S(r)= \varepsilon ({\sigma_d}/{r})^{6}
\label{eq:soft-disks}
\end{equation} 
if  $r< 2.6 \, \sigma_d$ and $0$ otherwise.
The  components of $\bm{\xi}$ and $\eta$  are zero-mean  and unit variance independent white Gaussian noises.
We fixed $D_\theta = 3 k_BT/(\sigma^2_d \gamma)$. A constant persistence time $\tau_P=1/D_{\theta}$ 
gives a persistence length $l_P=F_{\rm act} \tau_P/\gamma$.
We vary the packing fraction 
$\phi =\pi{\sigma^2_d}N/(4L_xL_y)$, where $L_x/L_y=2/\sqrt{3}$ (to allow for perfect hexagonal sphere packing),  
and the P\'eclet number Pe = $F_{\rm act} {\sigma_d}/(k_BT)$. {The persistence length is $l_P$ = Pe$\,\sigma_d$/3 and 
grows proportionally to the P\'eclet number.}
{Lengths and energies are later measured in} units of $ \sigma_d$ and $\varepsilon$, respectively. 
The friction coefficient and the mass are set to  $\gamma=10$ and $m=1$, consistently with 
an over-damped description.
The temperature $T$ controls the translational and orientational noise-noise temporal delta-correlations
and was held fixed to $k_BT=0.05$ for the hard potential case and $k_BT=1$ for the soft one. 
In most of the simulations $N=512^2$. 

\subsection{Numerical simulations}
\label{sec:numerical}

We used a velocity Verlet algorithm in the $NVT$ ensemble that solves Newton's equations of motion
with the addition of two force terms, a friction and a noise, which mimic a Langevin-type  thermostat. The number of particles $N$ and the volume $V$ are fixed  externally. 
 The algorithm integrates numerically   the stochastic evolution. 

We  used  perfectly ordered initial conditions --- with the disks arranged on a triangular lattice of a given spacing yielding the desired global surface fraction --- because they relax faster towards stationarity in dense cases, the ones on which we are mainly interested in, rather than disordered packings. We ensured that the analysis is performed in the steady state limit.

Concerning the numerical methods, we parallelized  the  numerical  
computation  with the help of the  open  source  software  Large-scale Atomic/Molecular Massively Parallel Simulator (LAMMPS), 
available at \url{www.github.com/lammps}.

We adapted the time-step of integration, $\Delta t$, to enforce numerical stability. In this study, we gradually reduced the value of 
$\Delta t$ in Molecular Dynamics simulation time units (MDs) for increasing Pe. For systems at {$ 0 \leqslant$ Pe $\leqslant 10$ we used $\Delta t=0.005$, 
for 10 $<$ Pe $\leqslant 50$} we used $\Delta t=0.002$, while for Pe $>50$ we used the smallest time-step $\Delta t=0.001$.
Typical simulations took $50 \times 10^4$ MDs to ensure stationarity. As a rule, we let the system further evolve in the
stationary regime for $50 \times 10^4$ MDs before gathering data. On average each simulation lasting $100
\times 10^4$ MDs was run on 48 processors  for a total of 96 hours for each CPU. 
We run  3-5 independent simulations per parameter value.

\subsection{The phase diagram}
\label{subsec:phase-diagram}

\begin{figure}[t!]
\vspace{0.25cm}
\centering
\includegraphics[width=\linewidth]{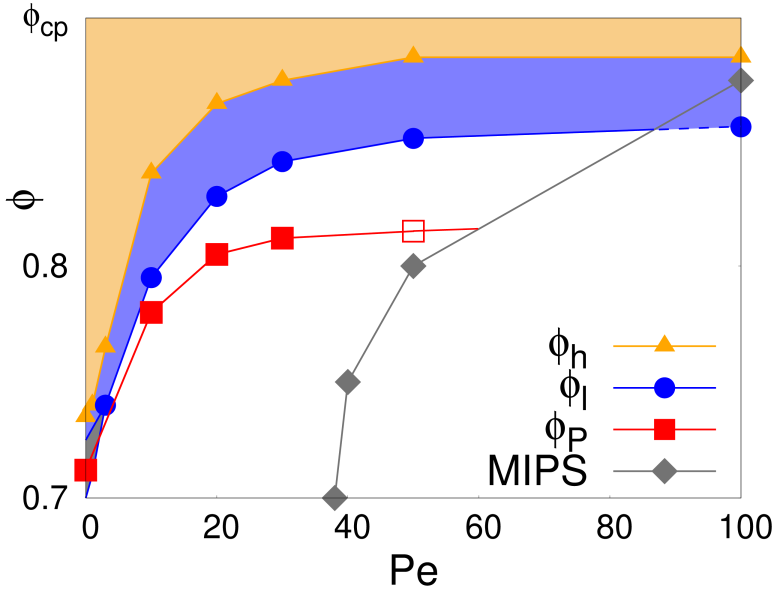}
\caption{Phase diagram of Active Brownian hard disks at high densities ($\phi>0.7$).
The blue line locates the hexatic-liquid transition, $\phi_l$, which can either be discontinuous at low Pe ($<3$, opening up a coexistence region,  
shown in shaded gray), 
continuous at higher values of Pe, and even penetrate within the MIPS region (dashed line at high Pe).
The orange curve is the solid-hexatic transition, $\phi_h$, 
which is continuous for all values of Pe.
The values of $\phi_h$ and $\phi_l$ at Pe = 0, 10, 20, 30, 40,  50 are 
given in Tables~\ref{table:fits-dislocations} and \ref{table:fits-disclinations}, respectively.
{(For Pe = 0, $\phi_l$ is the upper limit of the coexistence sector.)}
The black curve delimits the coexistence region opened up by Motility-Induced Phase Separation (MIPS).
The red curve indicates the percolation transition of (coarse-grained) clusters of defects
occurring at $\phi_P$. Filled red symbols are the $\phi_P$ extracted from the  analysis of  the 
probability of occurrence of a wrapping cluster,
$P^{\infty}$, at different system sizes, while the empty one at Pe = 50 is determined from the inspection of the cluster size distributions, $P(n)$.
The properties of this transition are largely discussed in Sec.~\ref{sec:clusters}. 
The  close packing of hard disks sets the upper limit of the plot, at
$\phi_{\rm cp} \simeq 0.91$.
}
\label{fig:phasediagram}
\end{figure}

From extensive numerical simulations as the ones described above, 
and a systematic analysis of the pressure, correlation functions, local density and hexatic order distributions, 
we obtained the phase diagram of the hard ABP,  with the potential defined in 
Eq.~(\ref{eq:hard-disks})~\cite{PRLino}. A zoom over the high packing fraction part of 
it is presented in Fig.~\ref{fig:phasediagram}. The diagram 
includes a solid, a hexatic and a liquid phase, with the addition of the MIPS sector at sufficiently high Pe
and a small co-existence region close to Pe = 0. The phase transitions are indicated with line-points 
(with different colors). We call $\phi_h$ the solid-hexatic transition density and $\phi_l$ 
the hexatic-liquid one.
Note that for a first-order hexatic-liquid transition we denote by $\phi_l$ the high-density branch of the coexistence region.
Clearly, both vary with Pe. The phase transition between liquid and hexatic, close to the passive 
limit, is of first order and it is accompanied by a narrow co-existence region which is shaded in blue in the figure. 
$\phi_{\rm cp}\approx 0.91$ is the close packing fraction for hard disks.
We refer to \cite{PRLino} for further details on how the MIPS, solid  and hexatic regions  were identified. 
On top of the latter ``thermodynamic'' transitions,  in the present work we identify and characterize a novel geometric  transition, 
associated with the percolation of clusters of defects, occurring at $\phi_P$ and represented in the phase diagram with red symbols.

Instrumental to measure orientational order, 
is the local hexatic order parameter
\begin{equation}
\psi_{6i} = \psi_6(\bold{r}_i)=\frac{1}{N_i}\sum_{k=1}^{N_i}e^{i6\theta_{ik}}
\; , 
\label{eq:local-hexatic}
\end{equation}
with $\bold{r}_i$ is the position of the $i$th disk and $\theta_{ik}$ the angle formed by the segment that connects 
its center and the one of its $k$th  out of its $N_i$ nearest neighbors, identified with a Voronoi construction.
Pictorially, the local orientational order is represented with color maps in which 
the colors designate different projections of the local hexatic order on the orientation of the global hexatic order parameter of the system. 
This representation was used in several recent publications, see e.g.~\cite{BernardKrauth,KapferKrauth,PRLino}.
 
 In this paper we also consider a soft disk system, see Eq.~(\ref{eq:soft-disks}). 
The phase diagram is similar, although with the important difference that the transition 
between hexatic and liquid in the passive limit, and near it, is continuous (and not first order). 
In App.~\ref{app:soft-disks} we present the orientational correlation function in 
equilibrium at different packing fractions, and the equation of state 
(pressure against packing fraction) for soft disks. None of them shows evidence for a first order
phase transition, in agreement with the  previous report in~\cite{KapferKrauth} for a slightly
different model.

\section{Topological defects}
\label{sec:topological-defects}

The topological defects are mis-coordinated particles, or cells in the Voronoi construction, with respect to  the perfect hexagonal lattice, that is to say, the perfectly ordered crystal. Free disclinations correspond to individual cells with 5 or 7 
neighbors (5-fold and 7-fold defects), while free dislocations are pairs of neighboring cells with 5 and 7 neighbors (5-7 pairs). 
Vacancies are point defects resulting from the removal of a particle from the hexagonal packing, 
and can be identified with different configurations involving, either groups of pairs of bounded 
dislocations (two or more), or higher-order mis-coordinated cells (with less than 5 or more than 7 neighbors). 
All these  localized topological defects might aggregate into clusters, giving rise to extended structures that we also identify and analyze.

\begin{figure}[t!]
\centering 
\includegraphics[width=\linewidth]{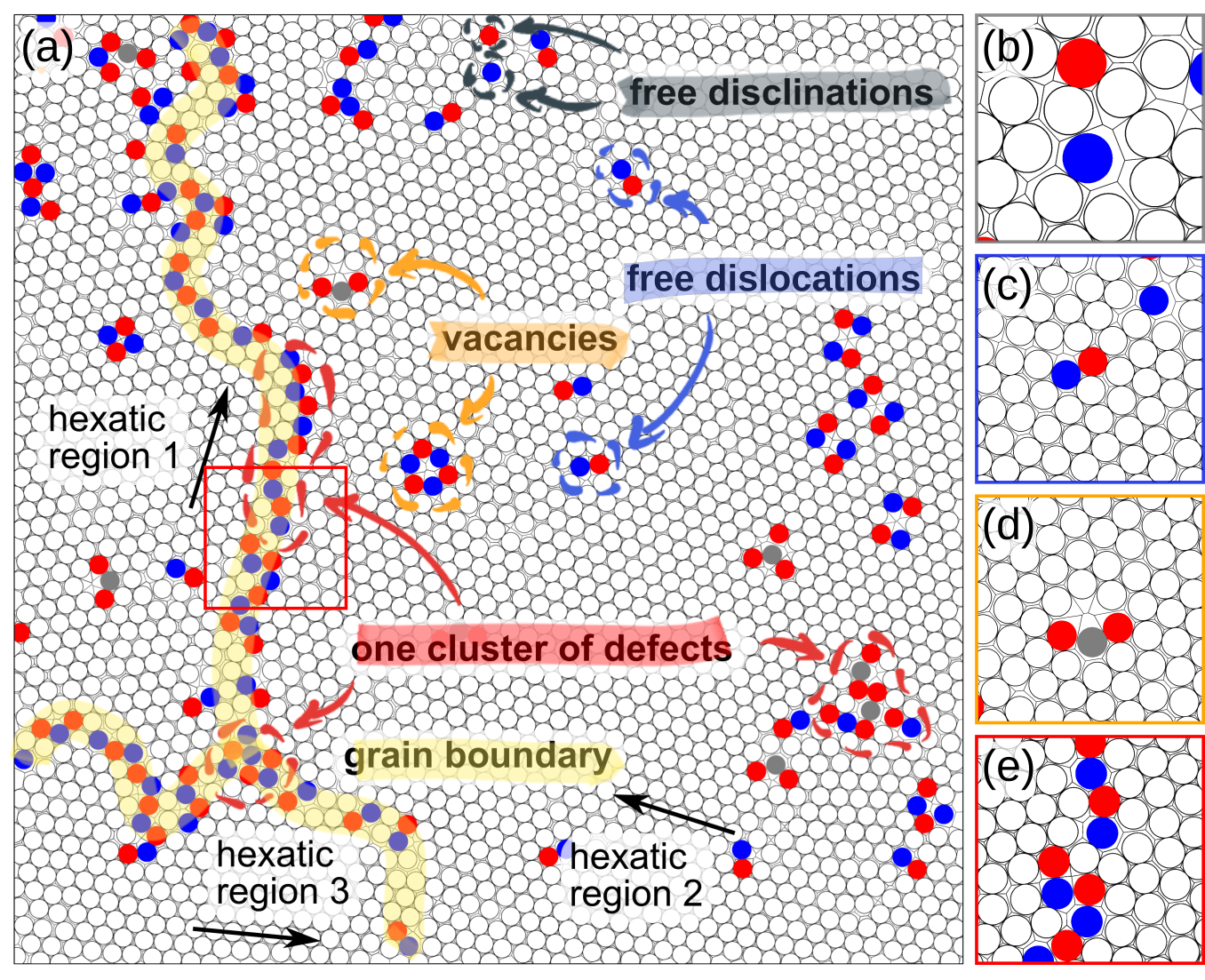} 
\caption{(a) Detailed view of a typical snapshot ($\phi=0.820$ and  Pe = 20) with different kinds of defects. 
Particles in 5-fold  cells are colored in red, 7-fold ones in blue, and all other mis-coordinated disks in gray.  
{The large black arrows have the direction of the orientational order parameter in three regions of the system.}
Two disclinations are shown in (b),  a dislocation  and a disclination
in (c),  a vacancy in (d), and a detailed view of two clusters of defects belonging to the same grain boundary 
(shown in yellow in (a) and delimiting regions with different hexatic order) in (e). 
 }
\label{fig:string}
\end{figure}

The various types of defects can be seen in Fig.~\ref{fig:string}, where we plot a selection of a configuration of the active system at a rather high packing fraction. 
The four kinds of defects mentioned in the previous paragraph are zoomed over in the boxes next to 
the configuration in panel (a). We see two disclinations in (b),  a dislocation in (c), a vacancy 
in (d), and a piece of two clusters of defects belonging to the same grain boundary in (e). 

 In this Section we explain  the way in which we classified the defects (Subsec.~\ref{subsec:classification}), 
 we define the number density of each defect kind  and their radial correlation function (Subsec.~\ref{subsec:densities-correlations}), 
we describe  the coarse-graining method used to build the clusters of defects (Subsec.~\ref{subsec:coarse-graining}), 
 and we recall some elements of percolation theory that we will  use later on (Subsec.~\ref{subsec:percolation0}).

\subsection{Classification}
\label{subsec:classification}

The analysis of the defects was done according to the classification proposed by Pertsinidis \& Ling in Ref.~\cite{Pertsinidis}. 
After having defined a defect core as a set of nearest-neighbor mis-coordinated particles, these authors
point out that one configuration of defects can be said to come from a local rearrangement of particles around a vacancy if it satisfies the following properties:

\begin{widetext}

 \begin{figure}[h!]
\vspace{0.25cm}
\centering
\includegraphics[width=\textwidth]{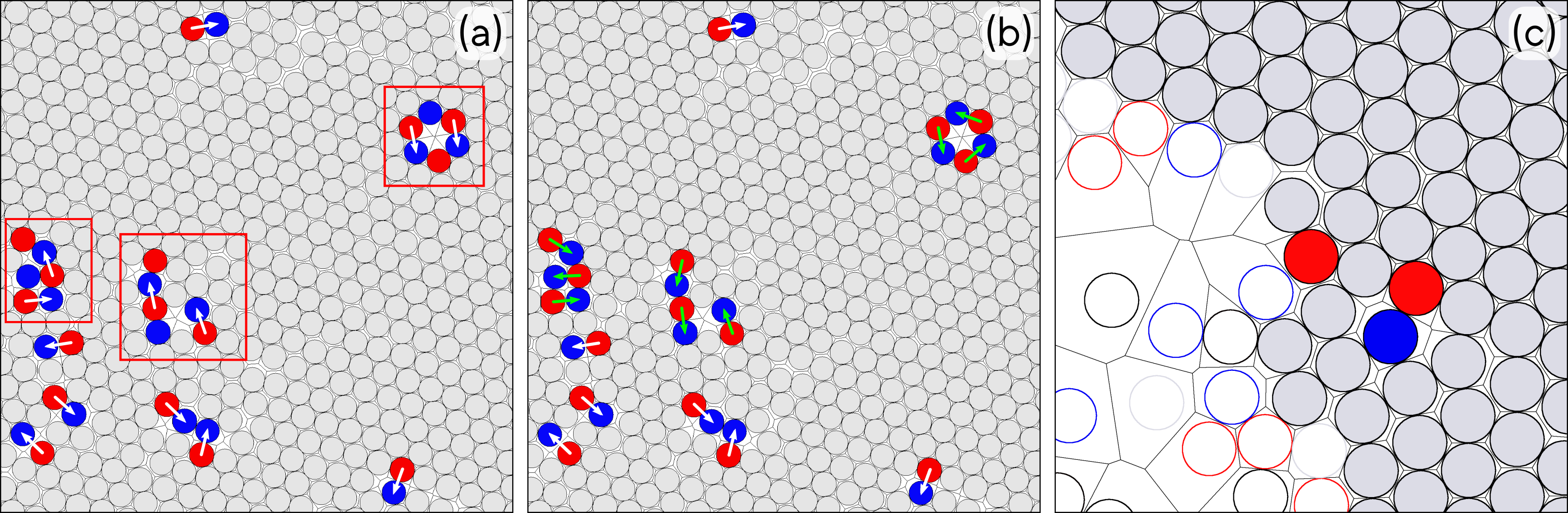}
\caption{
Voronoi tessellation, with defected particles painted in blue (7 neighbors) and red (5 neighbors).
The arrow assignment is explained in the text. We show here two cases:
(a) Wrong assignment in the three red boxes.
(b) Correct assignment for which an arrow is attributed to each pair of defected particles. Green arrows are the ones that changed from panel (a) and made the assignment correct.
(c) Voronoi construction close to the boundary between the dense drop and the gas in the MIPS regime (Pe = 200 and $\phi=0.890$).
The filled gray disks belong to the dense phase and the empty ones to the gas.
The empty disks which sit on defected Voronoi cells are not included in the counting of defects since they belong to the interface
or to the gaseous phase.
}
\label{fig:snap-construction}
\end{figure} 

\end{widetext}

\noindent
\vspace{-0.2cm}
\begin{itemize}
\item[--] 
The mean coordination number is equal to six.
\vspace{-0.2cm}
\item[--]
The modulus $S$ of the sum of vectors assigned as follows is equal to zero. 
Consider a set of unit vectors $\boldsymbol e_{ij}$ within the defect core, each one starting from a particle with coordination number $m_i<6$ 
and ending in a nearest neighbor with coordination number $m_j>6$. The vector ${\mathbf S}$ is the sum of these vectors.
This protocol is justified by noting that, for a single dislocation, the Burgers vector 
$\boldsymbol b$ is in  one-to-one correspondence to the vector defined above, being in particular $\boldsymbol b$ equal to $e_{ij}$ 
rotated
counterclockwise by $2\pi/6$. Therefore, an arrangement of cells with vanishing $S$ corresponds to a defect with zero
Burgers vector.
\end{itemize}

\noindent
Given the definition of a vacancy as a defect carrying zero Burgers vectors, all the other defects are classified in the following way.
\begin{itemize}
\item[--]
Free disclinations: one-particle defects.
\vspace{-0.2cm}
\item[--]
Free dislocations: a pair of nearest-neghbor 5-fold and 7-fold defected particles.
\vspace{-0.2cm}
\item[--]
Vacancies of size $n$: any configuration of $n$ nearest-neighbor mis-coordinated particles with mean coordination number equal to six 
and $S<S_{\rm th}$, where $S_{\rm th}$ is a threshold value (replacing zero) taken to be a half of the mean separation between particles.
\vspace{-0.2cm}
\item[--]
A cluster of size $n$ is any set of $n$ nearest-neighbor mis-coordinated particles with $S>S_{\rm th}$.
\end{itemize}
    
For large and complex groups of nearest-neighbor mis-coordinated Voronoi cells, there are many ways to arrange the vectors $e_{ij}$ 
among all possible 5-7 pairs. One could attempt a complete enumeration that will eventually lead to the optimal assignment of 
vectors. Another option is to generate a large but not complete number of vector assignments and choose the best one among them, which is the one that accommodates the 
largest number of arrows, resulting in a maximally-covered configuration.
We checked that this strategy is good enough to pinpoint the ``maximally-covered'' configuration for most of the defect groups.
As two examples, Fig.~\ref{fig:snap-construction} shows a ``wrong'' vector assignment in panel (a) and a ``correct'' one, which we will choose, in panel (b).

Finally, in panel~(c), we zoom over the interface between the dense drop and the gas in a system with Pe = 200 and $\phi=0.880$, within the MIPS 
coexistence region of the phase diagram (see App.~\ref{app:particle-clustering} for the algorithm used to single out the dense particle cluster). 
We want to identify defects within the dense phase only, and avoid counting the ones 
placed at the boundary between the two phases. For this reason, we spotted the particles 
or Voronoi cells at the interface and we excluded them from the statistics, see Fig. \ref{fig:snap-construction} (c). 
For example, in the snapshot shown, this led us to disregard from the counting the blue and red empty particles that are placed on defected cells but lie 
right on the 
boundary. 

 \subsection{Number densities and correlations}
\label{subsec:densities-correlations}

We define the number density of defects, $\rho_d$, as the number of defected 
particles involved in all the defects of a given type (e.g., 2 for each dislocation) divided by $N$, 
the total number of particles, 
\begin{equation}
\rho_d = \frac{1}{N} \; \# \, \mbox{particles in the selected kind of defect}
\; . 
\label{eq:def-rhod}
\end{equation}

In two dimensions, the radial pair distribution function 
of any kind of point-like object is defined as
\begin{equation}
g(r) \equiv \frac{dn_r}{2\pi r dr \rho} \,.
\label{eq:radial}
\end{equation}
Given the center of mass of  such an object at the origin, $g(r)$  represents the probability of finding another 
one at distance $r$ from it, relative to the one of an ideal gas. It therefore helps describing how the 
density of the objects under study varies as a function of distance from a reference 
one: $dn_r$ is the number of items that are a distance between $r$ and $r + dr$ away from the reference, 
and the denominator ensures normalization 
with respect to the angular average, with $2\pi r dr$ the 
area of a bidimensional spherical shell, and  $\rho$ the averaged number density. 
This ensures that $g(r)=1$ when the system is structureless (for instance in the dilute or $r\to \infty$ limits).

\subsection{Coarse-graining}
\label{subsec:coarse-graining}

\begin{figure}[h!]
\vspace{0.25cm}
\centering
\includegraphics[width=8cm]{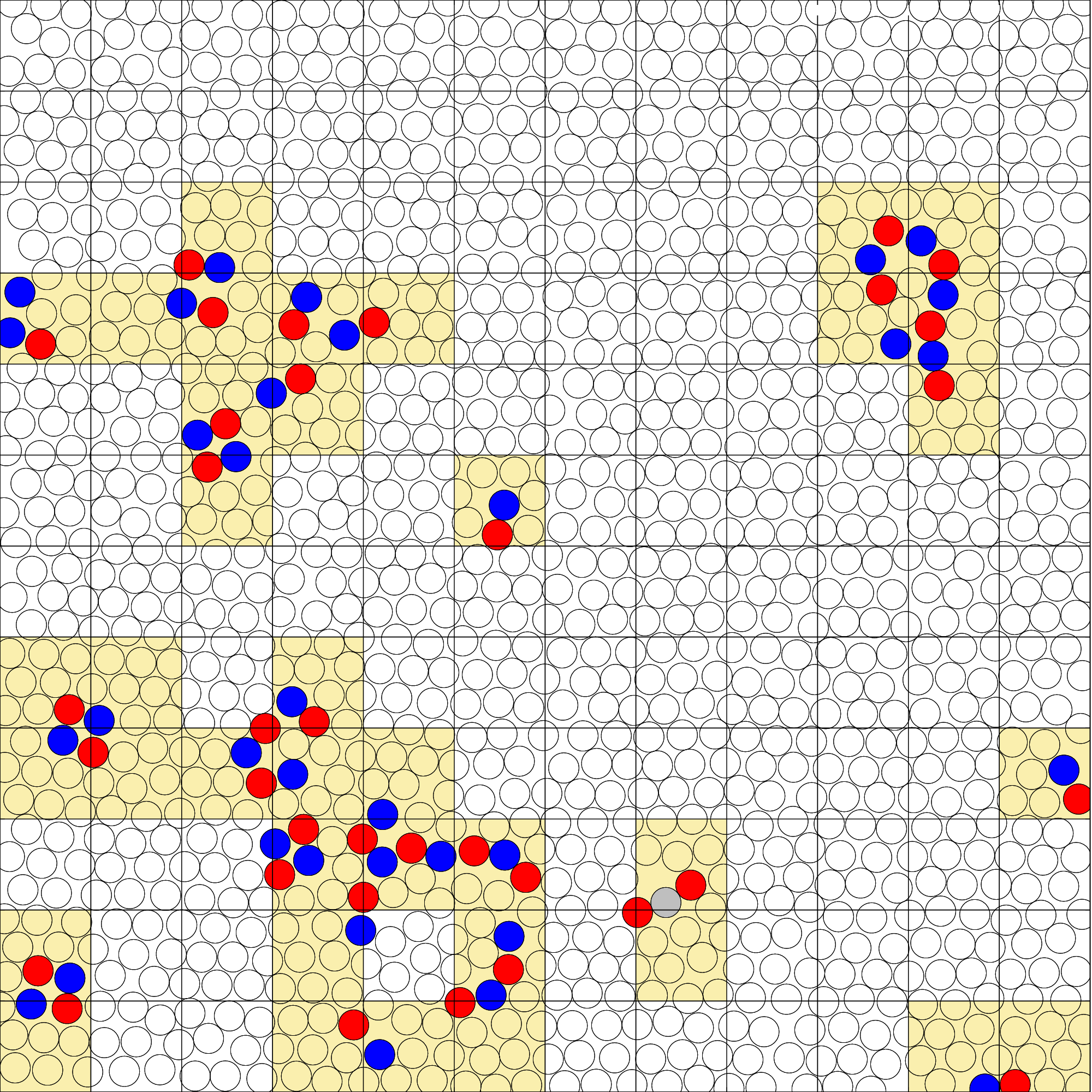}
\caption{Example of the coarse-graining procedure applied to the defect clusters with $d_s=3\sigma_d$.
Disks in red and blue are the particles with 5 and 7 neighbors in the Voronoi construction. Space
is divided in rectangular cells painted yellow whenever the center of  a defect particle falls in it. 
Coarse-grained clusters are built by joining such cells when they share a boundary.  
}
\label{fig:coarse-grained}
\end{figure}

Grain boundaries appear as chains of closely spaced defects 
(see Fig.~\ref{fig:string}), though the latter are not fully 
connected at the single Voronoi cell scale. We fill the microscopic gaps with a coarse-graining procedure routinely applied to study 
gelation~\cite{coniglio2004percolation}. The procedure is demonstrated
in Fig.~\ref{fig:coarse-grained} with an example. The 
thin straight lines separate space into $d_x\times d_y$ rectangular cells, with different linear lengths 
$d_x$ and $d_y$ proportional to a length scale $d_s$. This choice guarantees the covering of the rectangular
$L_x\times L_y$ simulation box. 
More concretely, we used $d_{x,y}=L_{x,y}/\text{int}(L_{x,y}/d_s)$, where int is the integer part function.
A cell is painted yellow
whenever the center of a defected particle lies within it. In this way a particle contributes to, at
most, a single cell. Coarse-grained clusters of defects are built by joining together neighboring cells with at least one defect, that is, those painted yellow. 
Proceeding in this way, the microscopic gaps are filled. 
We used this coarse-graining method in the analysis of the geometry and statistics of the defect clusters,
focusing on the coarse-graining length $d_s=3\sigma_d$. (A study of the effect of this parameter 
can be found in App.~\ref{app:coarse-graining}.) 

We typically found  $\sim10^4$ defected cells in each configuration in the hexatic phase, {later}{not too close to}
the hexatic-liquid transition. 
Once classified in the different classes that we have just defined, we 
counted about $300$ dislocations, 10 disclinations, $1000$ vacancies (which on average include from 3 to 6 cells), and 
$100$ clusters (of $\sim 10$ cells each). These numbers increase when we get closer to the hexatic-liquid transition.
This gives an idea of the number of data points that we use in our statistical analysis.

Of interest for the geometrical characterization of the defect clusters, see Eq.~(\ref{eq:Rg}), 
we identified the centers of mass of the clusters in the periodic box with the help of the method 
explained in Ref.~\cite{Bai08}. 

\subsection{Percolation}
\label{subsec:percolation0}

Percolation occurs when a connected cluster of defects wraps around the periodic boundaries of the system.
In order to identify whether a cluster of defects is percolating we applied the method designed by Machta
 et al.~\cite{Machta_wrapping,Ziff-Newman01}. Accordingly,  while running the algorithm to organize defected 
 cells into connected clusters, for each cluster we keep track of the paths connecting all the sites to the first defected cell we identified (i.e. the root of the algorithm). 
 The wrapping condition for a given cluster is fulfilled whenever the difference between the paths (both along the x- and y-direction) 
 of two different sites belonging to the cluster equals the system size.

Continuous percolation refers to the cases in which the concerned objects are placed in continuous space in 
contrast to the better understood lattice problem in which they occupy the sites of a regular structure. 
Percolation on a lattice is a specially appealing phenomenon since it is well understood theoretically and presents 
critical and universal features, with independence of the microscopic details \cite{StaufferBook}. 
In contrast, continuous percolation~\cite{Heyes89a,Heyes89b,Skvor09}, as it arises, for example, in liquids, is not as well characterized analytically.
Here, by mapping the continuous problem on a lattice one we get closer to the better understood discrete setting.

A way to locate the critical parameters 
for percolation in a system with periodic boundary conditions is to monitor 
the probability of occurrence of a wrapping cluster. Such probability is usually denoted $P^\infty$ and, for finite size systems, it 
smoothly varies from 0 (no percolation) to 1 (complete percolation). $P^\infty$ satisfies finite size scaling and, in the infinite size limit, 
it detaches from zero at the critical control parameter. The best way to estimate the latter in a finite 
size system is, therefore, to perform a finite-size scaling analysis:  
the curves $P^\infty(\phi)$ obtained for different sizes should cross at a common point, 
which we identify with the onset of percolation, $\phi_P$.  

Several characteristics of the clusters in a percolating problem have been defined and analyzed.
An essential aspect of  critical percolation is their fractal morphology. 
A way to study the clusters' geometry is to relate their size, $n_{\mathcal{C}}$, giving the number of cells that constitute the cluster, to a length scale, e.g.,  their radius of gyration 
\begin{equation}
{R_{\mbox{\it g}}}_\mathcal{C}=
\left[
\frac{1}{n_{\mathcal{C}}} 
\sum_{i\in \mathcal{C}} 
({\bold{r}}_i-{\bold{r}}_{\mathcal{C}})^2 
\right]^{\frac12}
\; ,
\label{eq:Rg}
\end{equation}  
where the sum runs over all 
cells in the cluster and ${\bold{r}}_{\mathcal{C}}$ is the position of its
center of mass. The relation between the size and gyration radius defines the fractal dimension~$d_\text{f}$: 
\begin{equation}
n_\mathcal{C}\sim {{R_{\mbox{\it g}}}_\mathcal{C}}^{\!\!d_\text{f}}
\; .
\label{eq:fractal}
\end{equation} 

Another key ingredient of percolation theory is 
the  probability distribution function of cluster sizes, $P(n)$, which takes the 
form \cite{StaufferBook}
\begin{equation}
P(n) \simeq n^{-\tau_n} \, e^{-n/n^*}
\; ,
\end{equation}
where the so-called Fisher exponent, $\tau_n$, is related to the clusters'  fractal dimension via
\begin{equation}
\tau_n = 1+d/d_{\rm f}
\; , \label{eq:Fisher}
\end{equation}
and $n^*$ diverges at the percolation critical point.

 \section{The KTHNY picture: defect unbinding
 }
 \label{sec:point-like}

\begin{figure}
\includegraphics[width=\columnwidth]{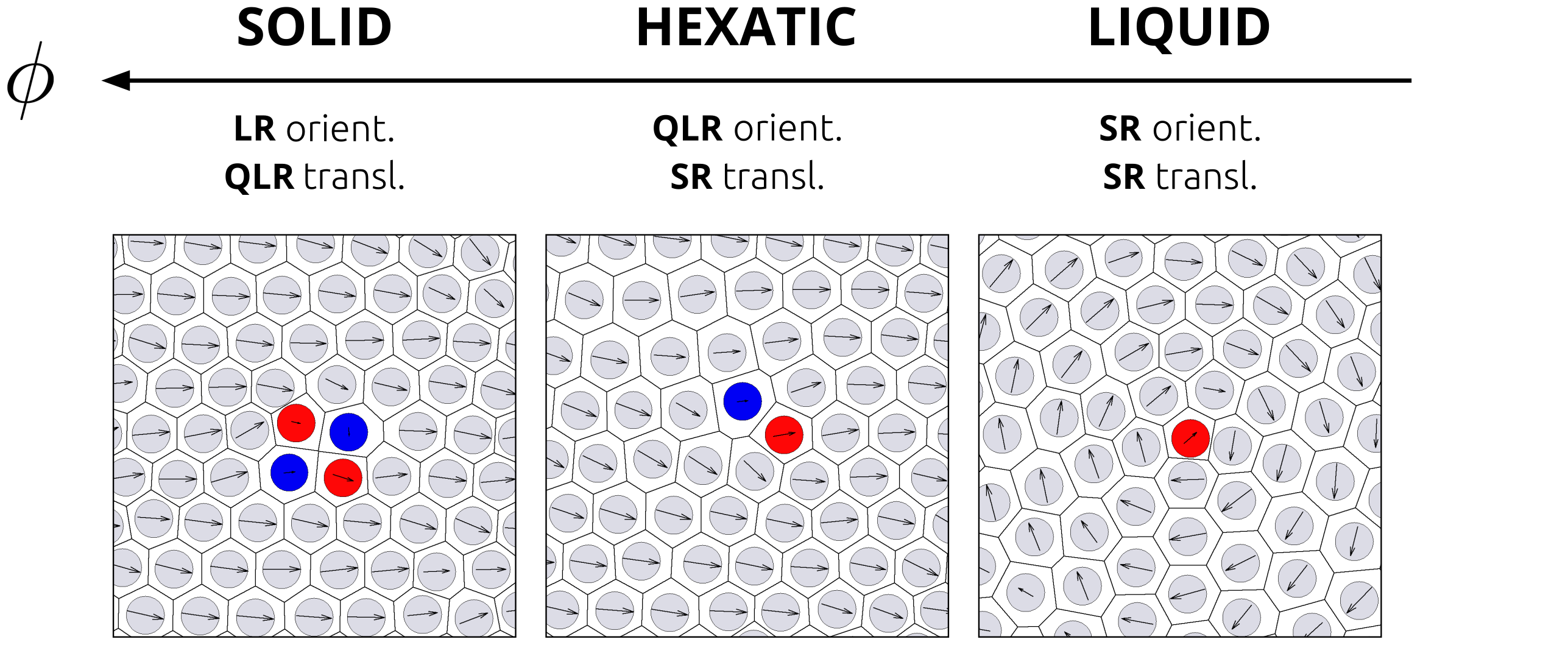}
\caption{Unbinding of dislocations and disclinations: the mechanism behind the solid-hexatic and hexatic-liquid
phase  transitions in the KTHNY theory. As in previous figures, disks in red and blue are 5-fold and 7-fold defected 
particles in the Voronoi construction. The arrows represent the local hexatic order parameter, see Eq.~(\ref{eq:local-hexatic}), attached
to each particle and are used to quantify the bond orientational order. Close to the $\phi$ axis, the nature of the orientational 
and translational order characterizing the three phases  are indicated: Long-Range (LR), Quasi-Long-Range (QLR) or Short-Range (SR).
}
\label{fig:sketch-transitions}
\end{figure}

In the KTHNY picture, melting occurs in two steps, driven by the 
unbinding of dislocations and disclinations. 
Lowering the global packing fraction from a closed packed state, a first transition takes place when groups of 
four point-like defected particles, with five and seven neighbours in the Voronoi tesselation of 
space, separate in pairs of two. The latter are the so-called dislocations, which are formed by two connected particles with 
five and seven neighbours. Decreasing further the global packing fraction, a second transition takes place when dislocations dissociate into two 
isolated defected particles, or disclinations, which are then free to move within the full sample. This scenario is 
represented in Fig.~\ref{fig:sketch-transitions}. 

The BKT-like transitions predict that the number density  of free dislocations and 
disclinations, defined in Eq.~(\ref{eq:def-rhod}), decay at the corresponding critical points $\phi_c$ as
\begin{equation}
\rho_d \sim a \ \mbox{exp} \biggl \{ -b \bigl[\phi_c/(\phi_c-\phi) \bigr]^\nu \biggr \} \mbox{,}
\label{eq:rhod}
\end{equation}
with the parameters $a$ and $b$ and the critical packing fraction $\phi_c$ 
being non-universal, and the exponent $\nu$ taking the values 
\begin{eqnarray}
\nu \;\; \left\{
\begin{array}{ll}
\simeq 0.37 & \qquad \mbox{at the solid-hexatic}
\\
= 0.5 & \qquad \mbox{at the hexatic-liquid}
\end{array}
\right.
\label{eq:nu-theory}
\end{eqnarray}
transitions. These are the results derived
by Nelson \& Halperin~\cite{Nelson1979} and Young~\cite{Young1979}, following 
the development by Kosterlitz~\cite{Kosterlitz1974} for the 2DXY model.  

In this Section we will study in depth the number density of dislocations and 
disclinations, and how they detach from zero at the solid-hexatic and hexatic-liquid
transitions, putting special emphasis on the comparison to the predictions of the KTHNY theory. We complement this study with an analysis of the 
spatial organization of defects.

 \subsection{Number densities}
 \label{subsec:solid-hexatic}

We use the classification introduced in Sec.~\ref{subsec:classification} to identify 
the isolated dislocations and disclinations~\cite{Pertsinidis}, namely, the isolated
miscoordinated pairs of particles or single particles, respectively. In this analysis we 
associate them to free point-like defects. 
Unless otherwise stated, the data in the plots are sampled over 50 independent instantaneous 
configurations selected at sufficiently spaced times in the stationary regime.

\begin{figure}[h!]
\centering 
\includegraphics[width=8.2cm]{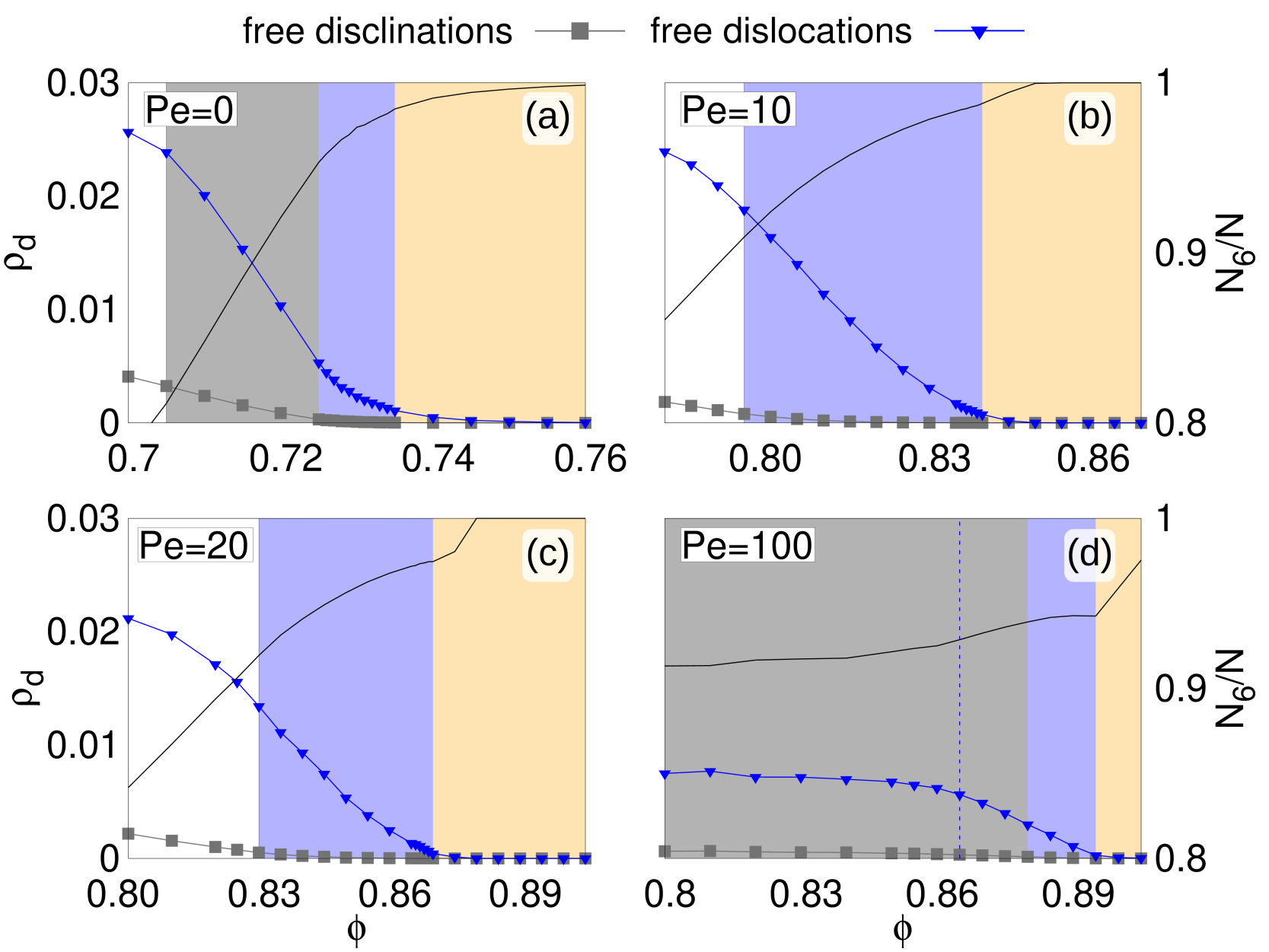}
\caption{Number densities of isolated disclinations and dislocations 
as a function of $\phi$ for passive hard disks (a), ABP at Pe = 10 (b), ABP at Pe = 20 (c) and ABP at Pe = 100 (d). Black solid lines represent the number density of 
non-defected particles, following the scale on the right vertical axis. 
We will discuss  the nature of the defected particles which are not dislocations or disclinations in Secs.~\ref{sec:MIPS} and \ref{sec:clusters}.
The solid, hexatic, phase-coexistence and liquid regions are shown in orange, 
blue, gray and white backgrounds, respectively. The values of the critical parameters separating solid, hexatic and liquid phase 
 are taken from the analysis of the correlation functions and the equation of state presented in Ref.~\cite{PRLino}
 and recalled in Tables~\ref{table:fits-dislocations} and \ref{table:fits-disclinations}. 
 The dashed blue 
vertical line inside the MIPS region (d) indicates the $\phi$ above which local orientational correlations are scale-free. 
All quantities are obtained after averaging over $\simeq 50$ independent configurations sampled from $3$ independent simulations. See Sec.~\ref{subsec:densities-correlations} for the typical number of defects observed in each independent run.
}
\label{fig:population}
\end{figure}

The number densities of free dislocations and disclinations, see Eq.~(\ref{eq:def-rhod}), 
are shown in Fig.~\ref{fig:population} in four representative cases: (a) passive hard disks, Pe = 0; (b) ABP at Pe = 10;  (c)  ABP at Pe = 20; (d)  ABP at Pe = 100.
For equilibrium hard disks the liquid-hexatic transition is discontinuous, showing a coexistence region between the pure hexatic and liquid phases (gray area). In cases (b) and (c) the liquid-hexatic transition is continuous and (d) exhibits MIPS~\cite{PRLino}. In the MIPS coexistence region we only count defects belonging to the dense phase.  The values of the critical parameters separating solid, hexatic and liquid phase are taken from the analysis of the 
 correlation functions and the equation of state presented in Ref.~\cite{PRLino} for the hard disk model, see Tables~\ref{table:fits-dislocations}
 and \ref{table:fits-disclinations}. 

The first noticeable fact in all panels in Fig.~\ref{fig:population} 
 concerns the transition between the solid and hexatic phases.  
 At the vicinity of the hexatic phase (blue background), the number of free dislocations increases sharply, 
 indicating that they break positional QLRO and mediate the solid-hexatic transition at all Pe.  This leads us to the following statement:
\begin{itemize}
\item The solid-hexatic transition in ABP interacting via a hard repulsive potential is consistent, 
at least qualitatively, with the KTHNY scenario.
\end{itemize}

Figure~\ref{fig:population} also informs us about the dissociation of dislocations into disclinations close to the hexatic-liquid  transition.
The density of free disclinations (gray squares) is very close to zero in the hexatic phase and detaches 
significantly from this value when the liquid appears, either as a pure phase, in panels (b) and (c), 
or co-existing with the hexatic, in panels (a) and (d). We see that:
\begin{itemize}
\item The unbinding of disclinations arises in the liquid component of passive and active systems.
\end{itemize}

The data shown in Fig.~\ref{fig:population} correspond to systems with $N=512^2$ particles. 
We have performed a similar analysis in systems with less particles, e.g. $N=256^2$,
and found comparable results. For this range of particle numbers, we have not observed 
strong finite size effects. We argue why this is so when presenting a more specific data analysis
of both dislocation and disclination number densities below. See also App.~\ref{app:finite-size}.


In Fig.~\ref{fig:population} we also plot, with thin black lines, the $\%$ of non-defected particles, 
that is to say, those which  are 6-fold connected, with the scale 
given in the right vertical axis. Note that, as in  Ref.~\cite{Tang89,Tang90}, there is 
quite a lot of 6-fold order even in the liquid phase, not too far from the hexatic-liquid transition,
with approximately 80\% of the particles being of this kind.

 \subsection{Solid-hexatic transition: dislocations}
 \label{subsec:solid-hexatic}

A precise quantitative analysis of the way in which  the density of free dislocations departs 
from zero at the solid-hexatic transition is notably hard even in the passive case. 
For example, the 
curves for the free dislocation density shown in Fig. 3a in~\cite{QiDijkstra} are not accompanied by a fit. 
Still, we have performed 
a convincing analysis of our data points, which exhibit the behaviour predicted by the BKT singularity.

\begin{figure}[h!]
\centering 
\includegraphics[width=8.2cm]{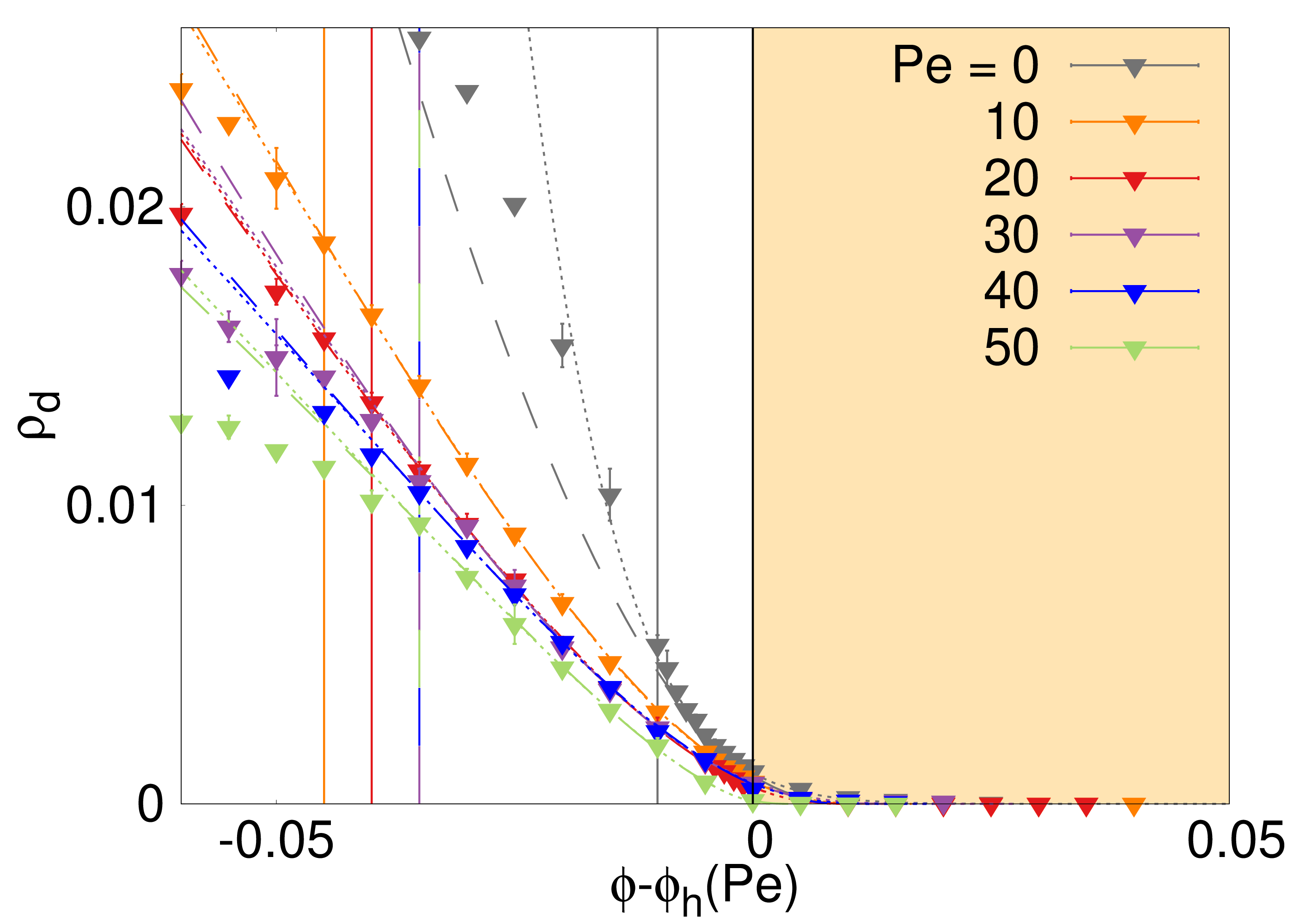}
\caption{Number density of dislocations close to the solid-hexatic transition at different Pe given in the key. 
The values of the off-set $\phi_h$(Pe) in the horizontal axis are the ones found with the methods 
in~\cite{PRLino}, recalled in Table~\ref{table:fits-dislocations}. The region painted orange corresponds to the solid phase.
Dotted and broken lines show three ($a, b, \phi_c$) and four (also $\nu$) 
parameter fits, respectively, to the form in Eq.~(\ref{eq:rhod}). 
In one fit the exponent is fixed to the KTHNY value $\nu=0.37$
and the values of the fitting parameters and the $\chi^2/\mbox{ndf}$ 
are reported in Table~\ref{table:fits-dislocations}.
In the other fit $\nu$ is also an adjustable parameter, see App.~\ref{app:four} for 
details. All fits are performed over the data in the hexatic and solid phases only, 
delimited by the vertical dashed lines to the left of  the origin (of different color for each Pe).
}
\label{fig:BKT-scaling-solid-hexatic}
\end{figure}

We fit the free dislocation number densities shown in Fig.~\ref{fig:BKT-scaling-solid-hexatic} using Eq.~(\ref{eq:rhod})
close to the solid-hexatic transition  for  various values of the P\'eclet number.  
The fits were performed over the data in the hexatic and solid phases only:
the fitting intervals for the different Pe values are delimited with vertical color lines. 
For  all Pe the fitted curves deviate from the data as soon as these cross the hexatic-liquid transition.

Equation~(\ref{eq:rhod}) has four parameters and we followed different fitting procedures to determine them.
We first fixed the value of the 
exponent $\nu$ to the one predicted by the KTHNY theory in the passive limit, 
$\nu=0.37$~\cite{Nelson1979,Young1979}, leaving $\phi_c$, $a$ and $b$ as fitting parameters. 
 In App.~\ref{app:improved-fits} we give more details on this fitting procedure and we estimate 
the confidence interval of the parameters $a$ and $b$.
Alternatively, we considered $\nu$ as a forth adjustable parameter, 
see App.~\ref{app:four}.
We present the analysis as a function of $\phi-\phi_h$(Pe) with 
$\phi_h$(Pe) the solid-hexatic transition value obtained from the study of 
correlation functions and local probability densities~\cite{PRLino}. 

\begin{table}[t!]
\begin{tabular}{|c|c|c|c|c|c|c|c|}
\hline
\;\; Pe  \;\; & \;\; $\nu $ \;\; & \;\; $ a$ \;\; & \;\; $ b $ \;\; & \;\; $\phi_c$ \;\; &  \;\; $\phi_h $ \;\; & \;\; $\chi^2/$ndf \;\;
\\
\hline
\hline
0 & 0.37 & 8 & 2 & 0.75 & 0.735 & 1.61
\\
\hline
10 & 0.37 & 1.5 & 1.61 & 0.853 & 0.840 & 2.76
\\
\hline
20 & 0.37 & 1.2 & 1.59 & 0.883 & 0.870 & 1.34
\\
\hline
30 & 0.37 & 2 & 1.9 & 0.897 & 0.880 & 2.08
\\
\hline
40 & 0.37 & 0.81 & 1.47 & 0.898 & 0.885 & 0.791
\\
\hline
50 & 0.37 & 0.38 & 1.17 & 0.895 & 0.890 & 0.493
\\
\hline
\end{tabular}
\caption{Dislocation unbinding at the solid-hexatic transition. Analysis of the fitting parameters 
in Eq.~(\ref{eq:rhod}) for the density of free dislocations plotted in Fig.~\ref{fig:BKT-scaling-solid-hexatic}.
The exponent   is fixed to the HNY value, $\nu=0.37$. The values of $\phi_h$(Pe) are the ones estimated 
from the analysis of the correlation functions and probability densities in~\cite{PRLino}, while $\phi_c$(Pe) are extracted 
from the fit.
}
\label{table:fits-dislocations}
\end{table}

Although  the performance of the two fitting procedures may seem similar,  a better
judgement  of their relative quality  is  gained from the comparison of the fitted $\phi_c$(Pe) and
$\phi_h$(Pe), see Table~\ref{table:fits-dislocations} and App.~\ref{app:four} (a similar strategy was applied to the 2DXY model
in~\cite{Batrouni98}). 
The two values are closer when the exponent $\nu$ is fixed to $\nu=0.37$, the HNY value.
The fitted values of $\phi_c$  at different Pe are around 1-2\% off the ones 
determined with the measurements of correlation functions and probability density profiles~\cite{PRLino}, providing 
an alternative way to locate the solid-hexatic transition. Indeed, in our ABP model $\phi_c$ is only slightly above $\phi_h$  in the five active 
cases considered when $\nu=0.37$ (e.g. $\phi_c=0.853$ vs. $\phi_h=0.840$ at Pe = 10).
A similar weak deviation between the critical points was measured by Han et al.~\cite{Han08} in their experiments with a 
passive microgel. {We note, however, that close to $\phi_h$, these authors found around 10 times more dislocations 
than we do. Far from $\phi_h$ the percentage of dislocations in~\cite{Han08} and here are similar. The different form of   the
fitting curve in~\cite{Han08} compared to ours is due to the excess in $\rho_d(\phi_h)$. The increase around $\phi_c$ 
is controlled by the parameter $b$ in the fitting curve, which should take a larger value in~\cite{Han08} (not reported) than in our
fit. The proportion of particles in dislocations and the form of the curve $\rho_d$ are similar to ours in other simulations,
see e.g.~\cite{QiDijkstra,Anderson17}, though no quantitative analysis of the critical behavior was performed in these references.}

One may wonder how robust the results in Table~\ref{table:fits-dislocations} are against  coarse-graining.
The coarse-graining procedure explained in Sec.~\ref{subsec:coarse-graining}
builds aggregates, like the ones shown in Fig.~\ref{fig:coarse-grained}, by gathering 
together the single defects that are closer than the coarse-graining length $d_s$.
After this construction, one computes the number density of all defect species with the classification scheme
of Sec.~\ref{subsec:classification}. 
Quite naturally, the number of dislocations and disclinations is diminished after coarse-graining since many of them get attached to larger clusters. 
In App.~\ref{app:coarse-graining}, see Fig.~\ref{fig:coarse-dislocation-density},  we study the effect of the coarse-graining length $d_s$ 
on the critical behaviour of the number density of dislocations close to $\phi_h$. The three parameter fit, with fixed $\nu=0.37$, yields values of 
$\phi_c$ that steadily approach (from above) the $\phi_h$ measured in~\cite{PRLino} for $d_s$ increasing from 
$2$ to $5$. (In studies of the vortex pair unbinding at the BKT transition of quantum gases, the filtering induced via coarse-graining also has the effect 
of approaching the vortex unbinding temperature to the critical one~\cite{Foster10}.)

We conclude that the dislocation 
data are consistent with the HNY exponent $\nu=0.37$ all along the solid-hexatic transition, and 
we therefore state that: 
\begin{itemize}
\item
The solid-hexatic transition in ABP interacting
via a hard repulsive potential complies with  the KTHNY scenario even 
quantitatively. Our measurements suggest that the exponent
$\nu$ is independent of Pe on this critical line.
\end{itemize}

An example of a typical configuration in the hexatic phase with several dislocations
is displayed in Fig.~\ref{fig:disclinations_hexatic}(b). One can appreciate the spatial arrangement
of dislocations which will be analyzed in Sec.~\ref{subsec:radial-dist}. 

 \subsection{Hexatic-liquid transition: disclinations}
 \label{sec:hexatic-liquid}
 
In the KTNHY scenario the hexatic-liquid transition is driven by the dissociation of dislocations into free disclinations. 
We now investigate whether this is indeed what occurs for passive and active Brownian disk systems.
 
\begin{figure}[h!]
\vspace{0.25cm}
\includegraphics[width=8cm]{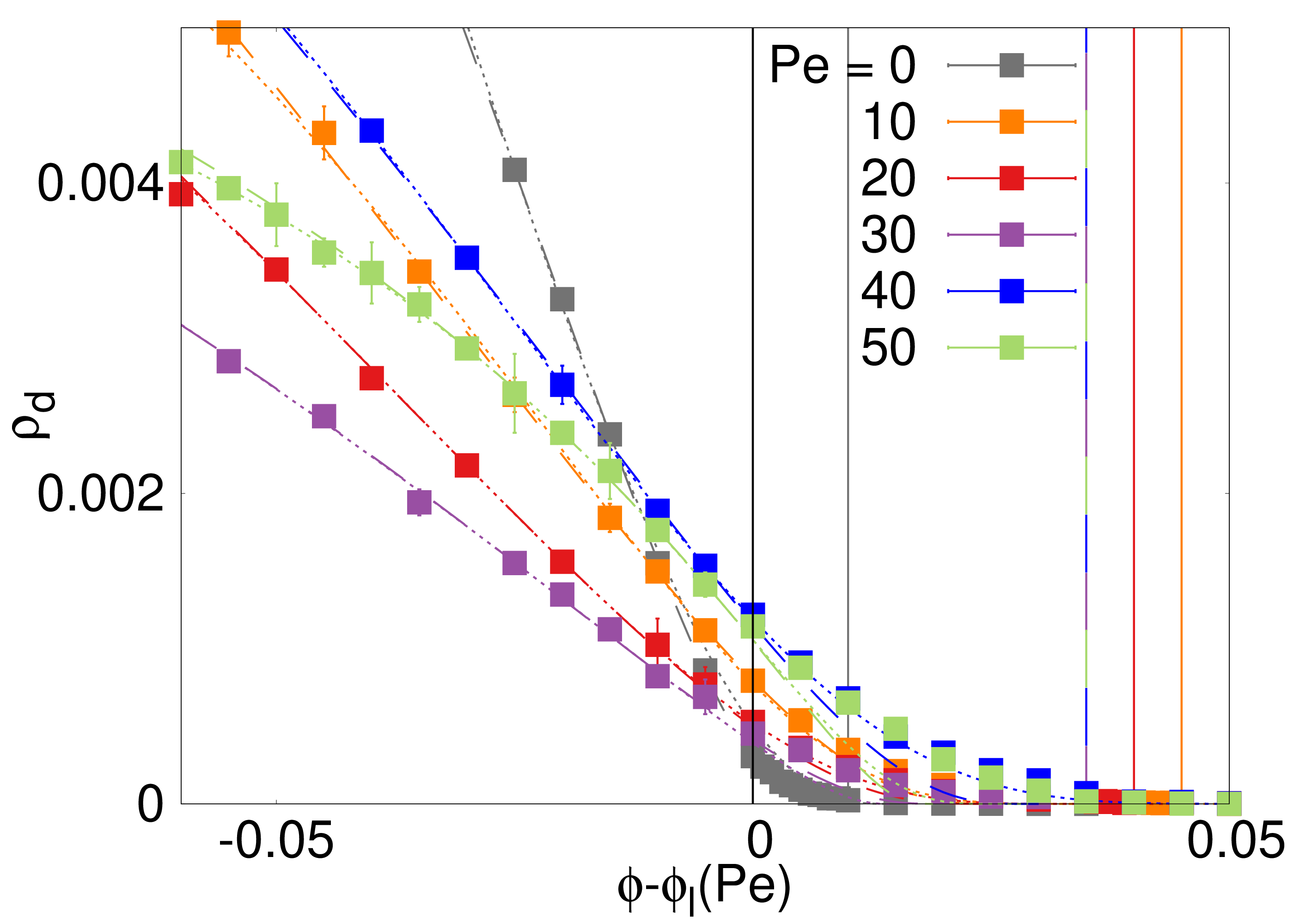}
\caption{Number density of disclinations close to the hexatic-liquid transition at different Pe.
The values of the off-set $\phi_l$(Pe) in the horizontal axis are the ones in Table \ref{table:fits-disclinations}. 
Dotted and broken lines show three ($a, b, \phi_c$) and four (also $\nu$) parameter fits, respectively, to the form in Eq.~(\ref{eq:rhod}). 
In Table~\ref{table:fits-disclinations} we give the values of $a$, $b$, $\phi_c$, $\phi_l$(Pe) and the $\chi^2/\mbox{ndf}$ for the three parameter fit. Details on the four parameter fits are given in App.~\ref{app:four}.  All fits are performed over the data in the liquid phase only. Colored vertical lines show the location of the 
solid-hexatic transitions $\phi_h(Pe)$ for all the Pe values considered.
}
\label{fig:disclinatios-HN}
\end{figure}

Figure~\ref{fig:disclinatios-HN}  displays KTHNY fits, Eq.~(\ref{eq:rhod}), 
to the number density of disclinations close to the packing fraction where  they detach from zero. 
We follow the procedure already used to analyze dislocations, and consider both a protocol where $\nu$ is fixed to its expected theoretical value $1/2$, 
{see also App.~\ref{app:improved-fits}}, 
and  an alternative approach where $\nu$ is a fourth fitting parameter.
We  compare the quality of these fits according to two criteria: how the critical $\phi_c$(Pe) 
extracted compares to the $\phi_l$ previously determined~\cite{PRLino}, and whether the values of the fitting exponent are reasonable.

Concerning $\phi_c$  we see that the values found with the four-parameter fit are higher and possibly too 
close to the $\phi_h$ of the solid-hexatic transition, see Table~\ref{table:four-parameter-disclination} in App.~\ref{app:four}.
Nonetheless, this is not  enough to exclude this fit since, all in all, the distance 
between the fitted $\phi_c$ and the critical values estimated in~\cite{PRLino} are similar in the two protocols, 
see Table~\ref{table:fits-disclinations}.

The second test, that is, whether the values of the fitted $\nu$ follow a reasonable trend as a function of Pe,
suggests, though, that the four parameter fit is less reliable: the fitted $\nu$ reported in Table~\ref{table:four-parameter-disclination} 
varies strongly and in a non-monotonic 
way with Pe and, moreover, for some Pe takes unreasonable values. Therefore, one would conclude that the three-parameter 
fit with $\nu=0.5$ is more consistent and quite satisfactory.

Nevertheless, going back to Fig.~\ref{fig:disclinatios-HN} and looking at the raw data
one notices that the trend of the curves, organized from top to bottom as Pe increases from 0 to 30, is 
inverted for larger Pe. Moreover,  the density of disclinations for $\phi>\phi_l$(Pe) is considerably 
larger than zero at high Pe, which obscures the interpretation of the data and the subsequent fits. 

\begin{table}[b!]
\begin{tabular}{|c|c|c|c|c|c|c|c|}
\hline
\;\; Pe  \;\; & \;\; $\nu $ \;\; & \;\; $ a$ \;\; & \;\; $ b $ \;\; & \;\; $\phi_c$ \;\; &  \;\; $\phi_l $ \;\; & \;\; $\chi^2/$ndf \;\;
\\
\hline
\hline
0  & 0.5 & 0.072 & 0.62 & 0.734 & 0.725 & 0.430
\\
\hline
10 & 0.5 & 0.06 & 0.81 & 0.823 & 0.795 & 1.09
\\
\hline
20 & 0.5 & 0.05 & 0.8 & 0.857 & 0.830 & 0.710
\\
\hline
30 & 0.5 & 0.025 & 0.64 & 0.866 & 0.845 & 0.895
\\
\hline
40 & 0.5 & 0.053 & 0.71 & 0.880 & 0.850 & 0.809
\\
\hline
50 & 0.5 & 0.016 & 0.41 & 0.874 & 0.855 & 0.233
\\
\hline
\end{tabular}
\caption{Disclination unbinding at the hexatic-liquid transition. Analysis of the fitting parameters in 
Eq.~(\ref{eq:rhod}) for the density of disclinations plotted in Fig. \ref{fig:disclinatios-HN}.
The exponent $\nu$ is
fixed to the KTHNY value $\nu=0.5$. The values of $\phi_l$(Pe) are the ones estimated 
from the analysis of the correlation functions and probability densities in~\cite{PRLino}, 
while $\phi_c$(Pe) are extracted from the fit. 
{At Pe = 0, $\phi_l$ is the upper limit of the co-existence region.}
}
\label{table:fits-disclinations}
\end{table}

\begin{figure}[h!]
\vspace{0.25cm}
\includegraphics[width=7cm]{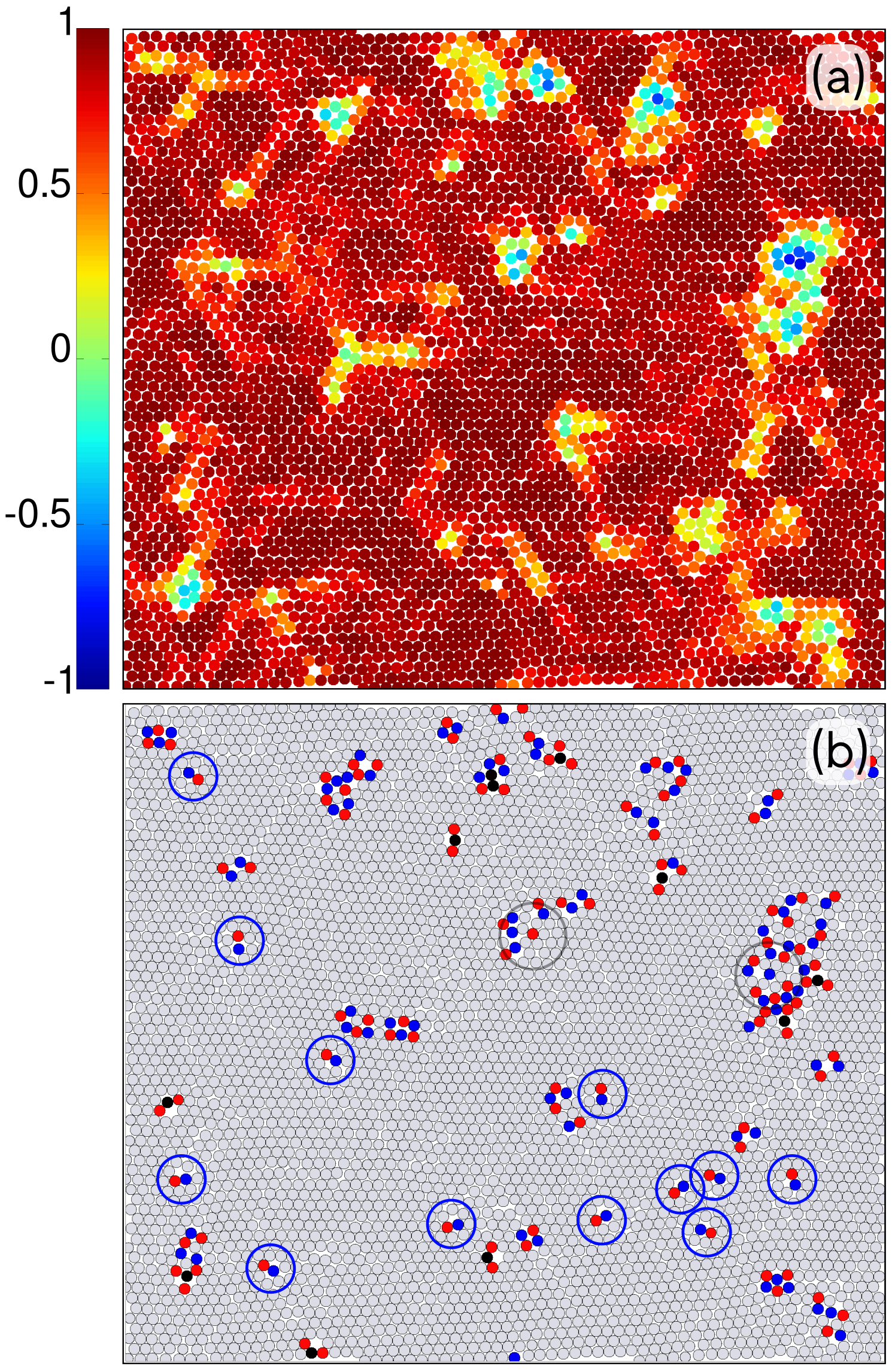}
\caption{Local hexatic order and defects for  $\phi$ = 0.805 and Pe = 10. 
Since $\phi_l$ = 0.795 and $\phi_h$ = 0.840 at Pe = 10, the case shown lies within the hexatic phase,
close to the melting transition.  
(a) Each disk is colored according to the projection of the local hexatic order parameter (\ref{eq:local-hexatic})
along the horizontal axis.
(b) The defected particles in the same configuration are shown in red and blue. 
Blue and gray circles enclose dislocations and disclinations, respectively. 
}
\label{fig:disclinations_hexatic}
\end{figure}

A better understanding of what happens close to the hexatic-liquid transition is gained from the 
visualization of the local hexatic order parameter, Eq.~(\ref{eq:local-hexatic}), and the defects configurations. An 
example is shown in Fig.~\ref{fig:disclinations_hexatic} for  parameters such that the 
system is in the hexatic phase, above but close to $\phi_l$(Pe), where a few disclinations  yield a non-vanishing (but small)
$\rho_d$. 
The first observation is that disclinations are not free to move apart: they are very close to 
ensembles of many defects.  This can be appreciated in panel (b), where two gray
circles highlight two defected particles that count as free disclinations. These disclinations are located 
in regions  where  the local hexatic order deviates from the majority (red) orientation, see Fig.~\ref{fig:disclinations_hexatic} (a).  
On top, these disclinations cannot freely move around
and they do not destroy the quasi-long-range orientational  order of the hexatic phase. Finally, the local character of these 
point-like defects also explains why we do not see strong finite size effects in the measurement of $\rho_d$ 
in this range of parameters: there is no characteristic length associated to them {which} could be in competition with the 
linear system size (see App. \ref{app:finite-size}).
More  generally, Fig.~\ref{fig:disclinations_hexatic} demonstrates that close to the hexatic-liquid transition
defects are not just point-like: they aggregate forming large objects that invade the sample in a way that we discuss in 
Sec.~\ref{sec:clusters}. 

In App.~\ref{app:coarse-graining}, see Fig.~\ref{fig:coarse-disclination-density},  we show the results of counting the number of disclinations 
after appliying the coarse-graining procedure explained above,  Sec.~\ref{subsec:coarse-graining}. Contrary to what happened with the 
dislocations, as soon as $d_s\geq 3$ we remove practically all disclinations 
and end up with an almost vanishing associated number density. The analysis of the 
pair correlation of disclinations in Subsec.~\ref{subsec:radial-dist} gives further support to their spatial proximity  to 
other defects
and to the fact that {they can} be erased by {an even} mild coarse-graining.

In conclusion, the behavior of the disclination number density close to the hexatic-liquid transition 
is more complex than the one of dislocations at the solid-hexatic threshold. 
\begin{itemize}
\item
The analysis of the number density of disclinations is not incompatible with the KTHNY fits with 
$\nu=0.5$. However, the proximity of the solid-hexatic transition interferes with the analysis of the hexatic-liquid one.  
More importantly, other features conspire against the simple unbinding  interpretation.  
Disclinations do not appear to be free but are notably close to large defected structures.
This suggests that some other mechanism may
also be at work at the hexatic-liquid transition.
\end{itemize}

\vspace{0.25cm}

\subsection{Radial distributions: spatial organization}
\label{subsec:radial-dist}

We analyze the spatial organization of particles, dislocations and disclinations
by measuring their spatial correlation functions defined in Eq.~(\ref{eq:radial}).

 \begin{widetext}
 
\begin{figure}[h]
\centerline{
\includegraphics[width=\textwidth]{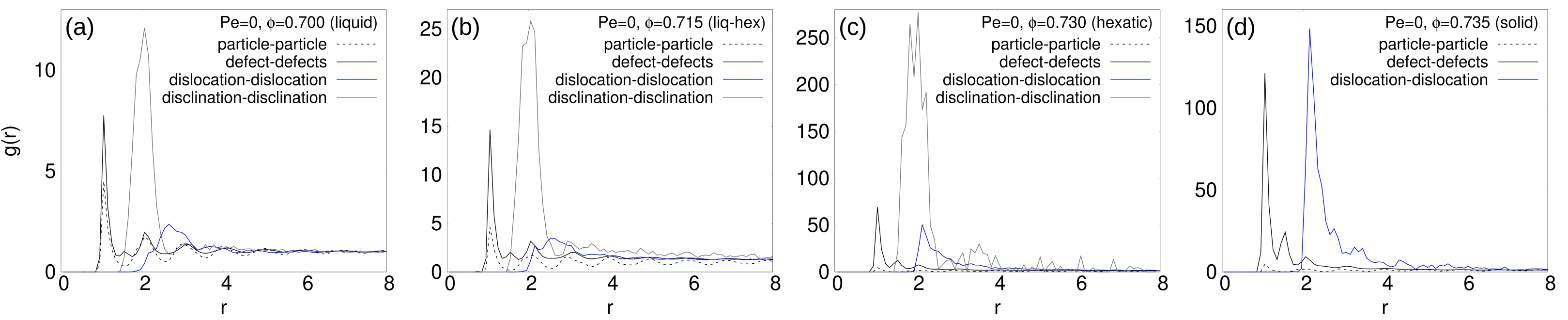}
}
\caption{Radial distribution functions defined in Eq.~(\ref{eq:def-rhod}) in the various equilibrium phases. The 
curves represent the data for  all particles (dashed black), all defected particles (solid black), 
dislocations (blue), and disclinations (gray). {The global packing fraction increases from left to right and is 
indicated in the keys together with the} corresponding phases are indicated in the keys. 
In the solid phase, panel (d), disclinations are absent. 
}
\label{fig:corr-fcts}
\end{figure}

\end{widetext}

The radial distribution function for different species is shown in Fig.~\ref{fig:corr-fcts} for passive systems in the liquid, co-existence region, hexatic and solid phases.
The peak structure of the particle-particle radial distribution (dashed curves) 
reproduces the  well-known behavior of a hard-disk system at the value of surface fraction considered~\cite{Jegge20,Poncet21}.
The location of the peaks in the defect pair correlation is very close to 
the one of the particles themselves because defected particles are immersed in the dense disks
 layer. Therefore, except for small fluctuations in their position, they copy the same underlying structure of the whole sample. 
The dislocation-dislocation radial distribution function (blue curves) 
has a relatively wide bump departing from zero at $r\approx2$ and centered at distances 
$r \approx3$. 
It represents the probability density of separation between dislocations, like the ones highlighted in Fig.~\ref{fig:disclinations_hexatic}.

The disclination $g(r)$ presents a first 
peak at  $r\approx2$ which is much more pronounced in the 
hexatic and solid than  in the liquid. 
This supports the expectation that point-like defects are  
more tightly bonded in the ordered phases. 
Still, 
\begin{itemize}
\item
the isolated 
disclinations are likely to stay close to each other even in the liquid, where one would have expected  them (accordingly to the KTHNY scenario) to 
separate the most.
\end{itemize} 

This  is confirmed by the fact that, upon coarse-graining,  
the number of disclinations drastically diminishes when using a coarse-graining length larger than $r=2$, see~App.~\ref{app:coarse-graining}.
These features are similar to  the results reported  for  the Gross-Pitaevskii classical field equation for Bose gases,  
in which positive and negative vortices show a pairing correlation that does not disappear above $T_{\rm KT}$~\cite{Foster10}.

The conclusion is that 
\begin{itemize}
\item
dislocations do not pack as closely as disclinations 
do; they have more freedom, with the consequence of broadening the first peak of their $g(r)$. 
\end{itemize}

This conclusion 
 is consistent with the analysis of number densities of dislocations and disclinations, reported in 
 Figs.~\ref{fig:BKT-scaling-solid-hexatic} and \ref{fig:disclinatios-HN}. 
 The number of dislocations grows in agreement with the KTHNY theory at the vicinity of the solid-hexatic transition, while the analysis of 
 disclinations close to the hexatic-liquid transition is less conclusive. 
 As we will see in Sec \ref{sec:clusters}, extended structures of topological defects are the key objects allowing to further 
 characterize the hexatic-liquid transition. 

\begin{figure}[h!]
\begin{center}
\includegraphics[width=\columnwidth]{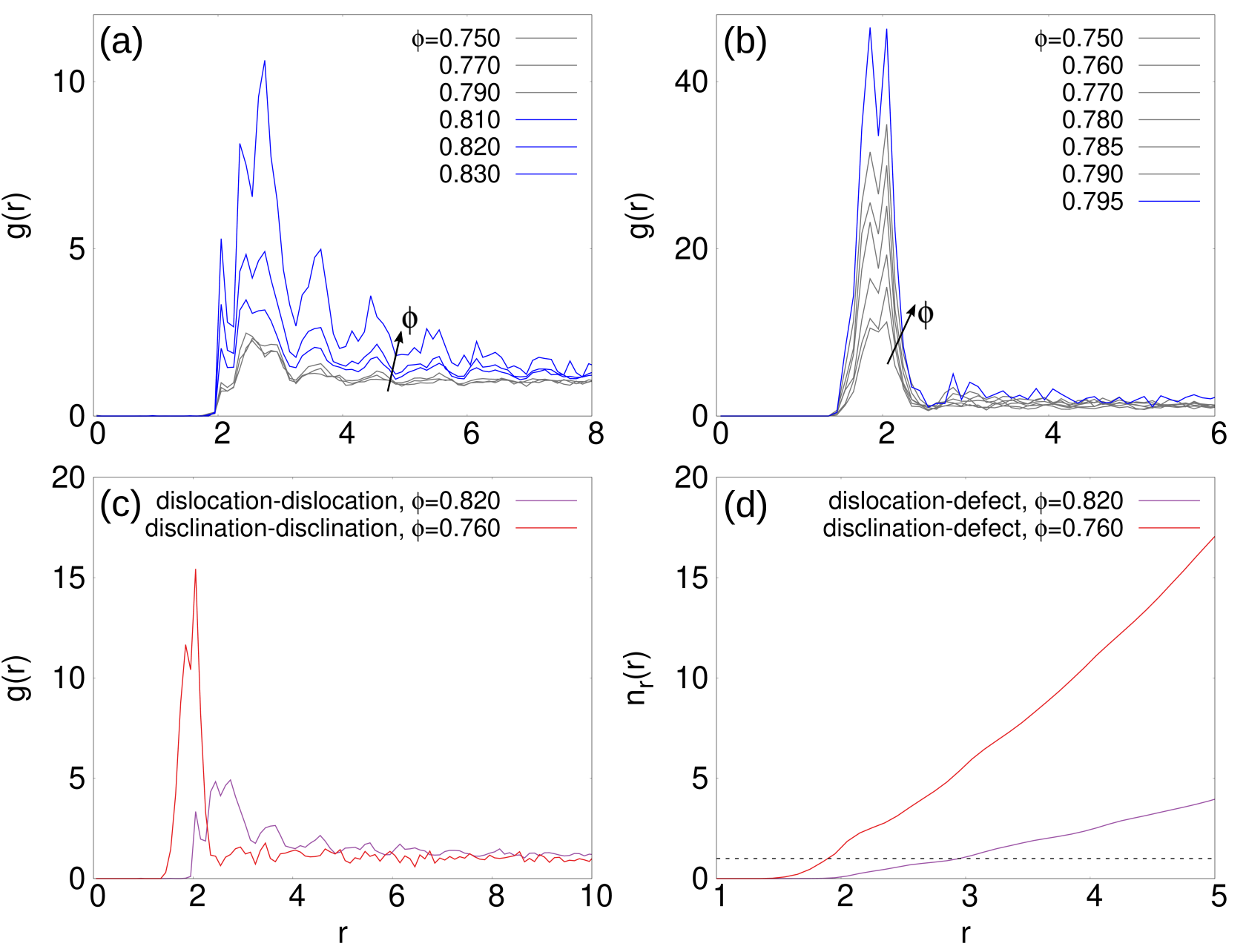}
\end{center}
\vspace{-0.5cm}
\caption{Radial distribution functions of dislocations  (a) and  disclinations (b) 
at Pe = 10, and the packing fractions given in the keys. Blue curves are in the hexatic and black in the liquid phase. 
The normalization of $g(r)$ with the density of defects contributes to the growth of the first peak for higher $\phi$. 
(c) Radial distribution functions at Pe = 10 for dislocations at $\phi=0.820$ (hexatic phase, violet curve) and disclinations at $\phi=0.760$ (liquid phase, red curve). 
The surface fractions were chosen so that dislocations and disclinations have a similar number density, $\rho_d \approx 0.003$. 
(d) Average  number of defected particles at a distance $r$  of a dislocation (red curve) and disclination (violet curve) 
for the $\phi$'s in the key. 
}
\label{fig:corr-Pe10}
\end{figure}

\begin{figure}[h!]
\centerline{
\includegraphics[width=\columnwidth]{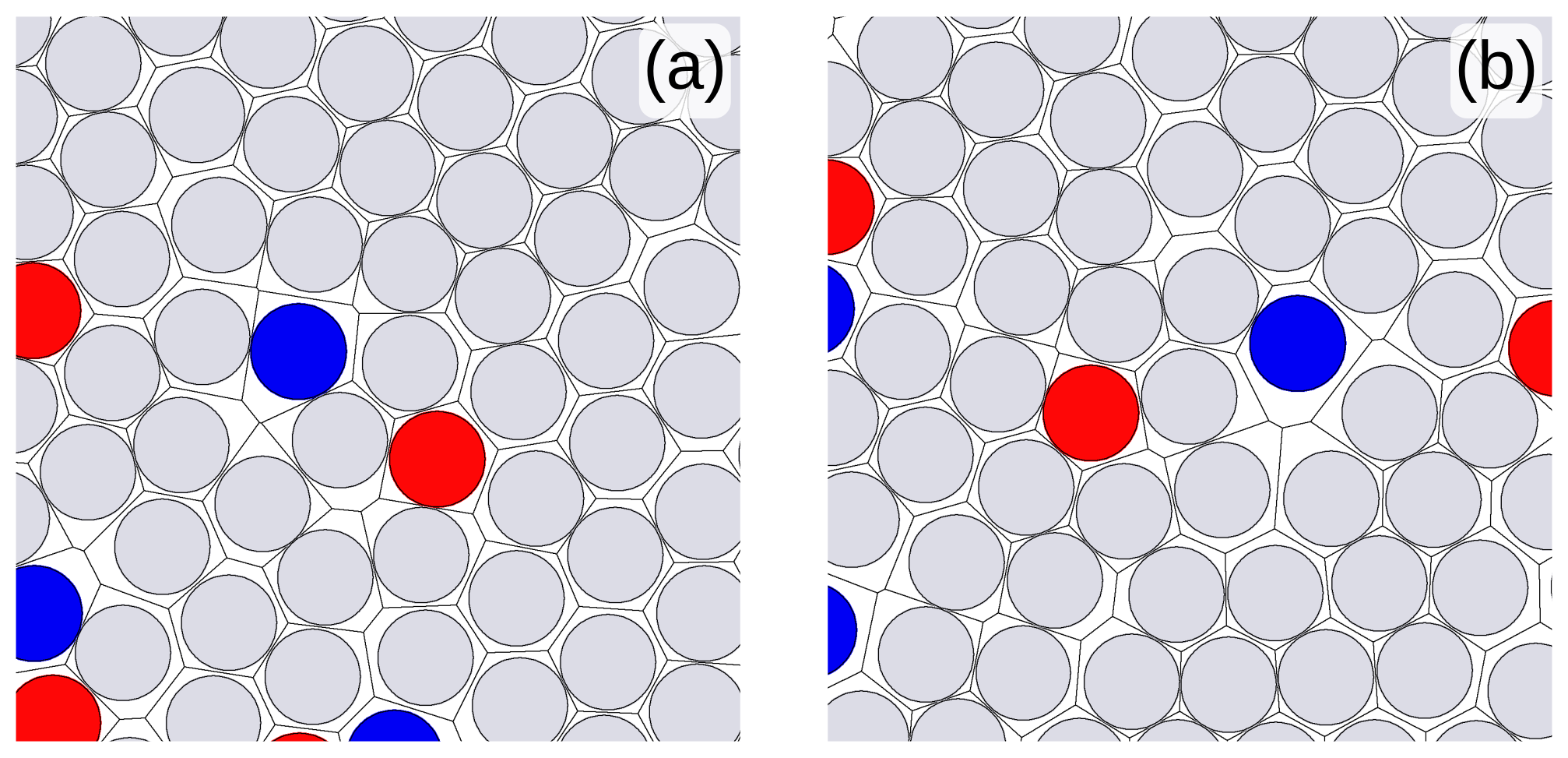}
}
\caption{Configurations at Pe = 10 and $\phi=0.75$ with two nearby disclinations. A segment joining the 
centers of the two defected particles  is almost perfectly aligned with the 
boundary of a Voronoi cell in (a) and it crosses two boundaries of Voronoi 
cells  in (b). The length of the segment is very similar in the two panels, but not identical, and such 
difference gives rise to the two sub-peaks in the disclination-disclination $g(r)$, around 
$r\approx 2$, see Fig.~\ref{fig:corr-Pe10}~(b).}
\label{fig:snapshots-defects}
\end{figure}

Let us now turn to the active case.
Figure~\ref{fig:corr-Pe10} shows the $g(r)$ of dislocations (a) and disclinations (b) at  Pe = 10, and 
for different densities across melting. As the density increases, as expected, the spatial organization of dislocations gets magnified 
because the overall number of defects decreases. Moreover, the denser the system, the sharper the first peak located at a distance $r\approx 3$ (as in equilibrium), 
with respect to the others. For disclinations, on the contrary, their is not much change in the form of $g(r)$
while increasing $\phi$ from the liquid to the hexatic. 
We show only one case in the hexatic phase, $\phi=0.795$, very close to the critical point, 
in order to reveal the behavior of residual tightly bonded disclinations in the ordered phase.
Furthermore, we observe  that the first peak develops a fine structure, composed of two sub-peaks evidenced in Fig.~\ref{fig:corr-Pe10} (b). 
Such fine structure is also present in equilibrium, as shown in Fig. \ref{fig:corr-fcts} (c). Figure~\ref{fig:snapshots-defects} 
illustrates the local configurations that give rise to the double first peak.
The  disclinations are separated by one single lattice layer, while the ones belonging to the second sub-peak have one dislocated lattice layer in between. The difference between the two configurations can also be understood from the inspection of the Voronoi tesselation close to the defected particles, see the caption of Fig.~\ref{fig:snapshots-defects} for more details.
We clarify the difference between the pair distribution of dislocations and disclinations with a direct comparison in 
Fig.~\ref{fig:corr-Pe10} (c) of two cases where the number density of the two species are similar. While for dislocations 
we have $g(r)>1$ over a significative extent beyond $r\approx 2$, for disclinations the distance probability density is 
extremely concentrated around the minimum value $r\approx 2$. 
In addition, the total number  of defects of any species at a distance $r$, $n_r$,  grows much faster for disclinations that for dislocations. Moreover, 
for disclination,  it reaches $n_r=1$ at $r\approx 2$ (see Fig.~\ref{fig:corr-Pe10} (d)), further confirming that, even deep in 
the liquid phase, every 
disclination has on average one neighboring defected particle at a distance smaller {than}~$2$.


\section{MIPS and finite length strings}
\label{sec:MIPS}

In this Section we focus on parameters in the  MIPS region of the phase diagram
(for some recent studies of this phase see~\cite{capriniPRL2020, Hermann21,Valeriani21,Dittrich21,Maggi21})
where the system separates into a macroscopic dense  and a dilute phase, and we investigate the
defects in the dense component only. In other words, 
we do not consider the particles in the dilute phase nor in the bubbles within the dense one~\cite{Caporusso20},
 even though  they are most likely mis-coordinated. For the same reason, 
we do not count the particles sitting on the boundaries between the dense phase (see App.~\ref{app:particle-clustering} for the method used for
its identification)
 and the gas, see
Fig.~\ref{fig:snap-construction}.

In short, we start by evaluating the number density of point-like defects and we then 
study more complex structures made by nearby topological defects.
We  will find an extended, though finite-length, network of defects which separates 
orientationally ordered micro-domains~\cite{Caporusso20}.

\subsection{Point-like defects}

\begin{figure}[h!]
\centerline{
\includegraphics[width=\columnwidth]{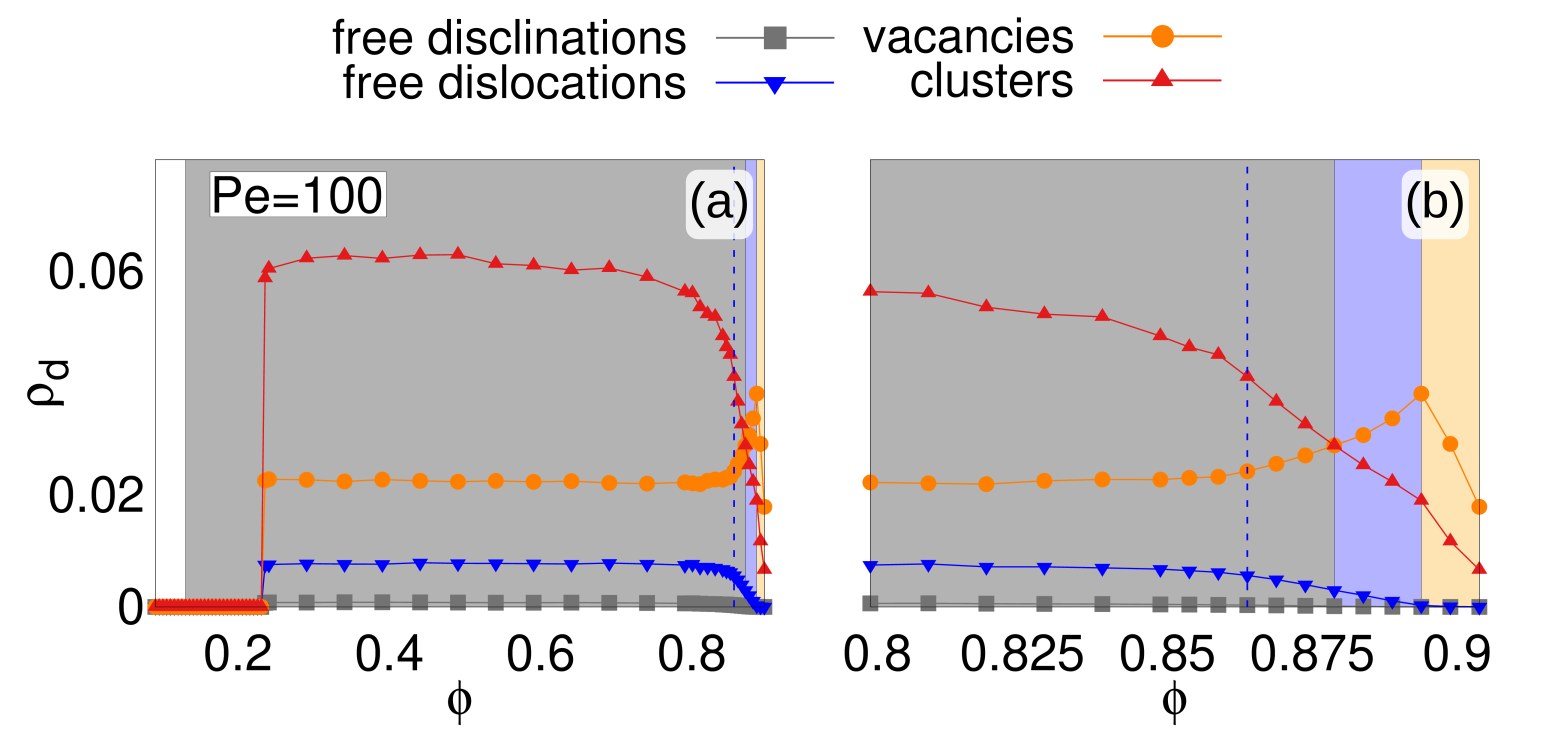}
}
\caption{Number density of all kinds of defects at a fixed high Pe and varying $\phi$. The gray zone is the MIPS 
sector of the phase diagram and the dashed blue line inside it indicates the $\phi$ above 
which local orientational correlations are scale-free. Panel (b) zooms over the high $\phi$ region. The number density of 
vacancies  has a net peak at the solid-hexatic transition. The decrease in number to the left of the 
peak is concomitant with  the increase in dislocation number. The number density of vacancies is 
$\phi$ independent in MIPS below the hexatic dotted line. {The same conclusion applies to the density of 
defected particles in clusters.}
}
\label{fig:number-density-MIPS}
\end{figure}

The parameter dependence of the number density of point-like defects across the MIPS coexistence {sector} 
is displayed  in Fig.~\ref{fig:number-density-MIPS}. 
The number density of each kind of defect remains practically constant 
below the packing fraction that we identified as the end of the QLR hexatic order~\cite{PRLino} and 
until the low density spinodal~\cite{Joan, PRLino}.
Therefore, 
 \begin{itemize}
 \item
 the bulk of the dense phase generated through MIPS is characterized by a density of  point-like defects set by Pe and not $\phi$. 
 \end{itemize}
 This is another aspect of MIPS that behaves as in equilibrium phase-separation: 
 in the $NVT$ ensemble,  the parameter that triggers the phase separation, here Pe (would be the temperature 
 $T$ in an equilibrium system of attractive particles), controls the nature of the two co-existing phases, 
 hence the density of point-like defects, while the mean particle 
 density sets the fraction of the system belonging to each of them.  
 

Besides dislocations and disclinations, vacancies are another kind of localized defect that displays an interesting parameter dependence within MIPS.  In the way we defined 
them, vacancies include bounded dislocation pairs~\cite{Pertsinidis}, they do not break positional QLRO, and can thus be present in the solid phase. 
As shown in Fig.~\ref{fig:number-density-MIPS}, the number of vacancies increases as the density is decreased in the solid regime, but rapidly decays as we get into the 
hexatic {phase}, leaving a peak at the transition. Such decay is concomitant with an increase in the number of dislocations, 
since some vacancies can be though of as two bounded dislocations. 
Besides a high $\phi$ region, where the solid and hexatic phases penetrate MIPS, the number density of vacancies also remains constant within MIPS.

\vspace{0.25cm}

\begin{figure}[h!]
\centerline{
\includegraphics[width=\columnwidth]{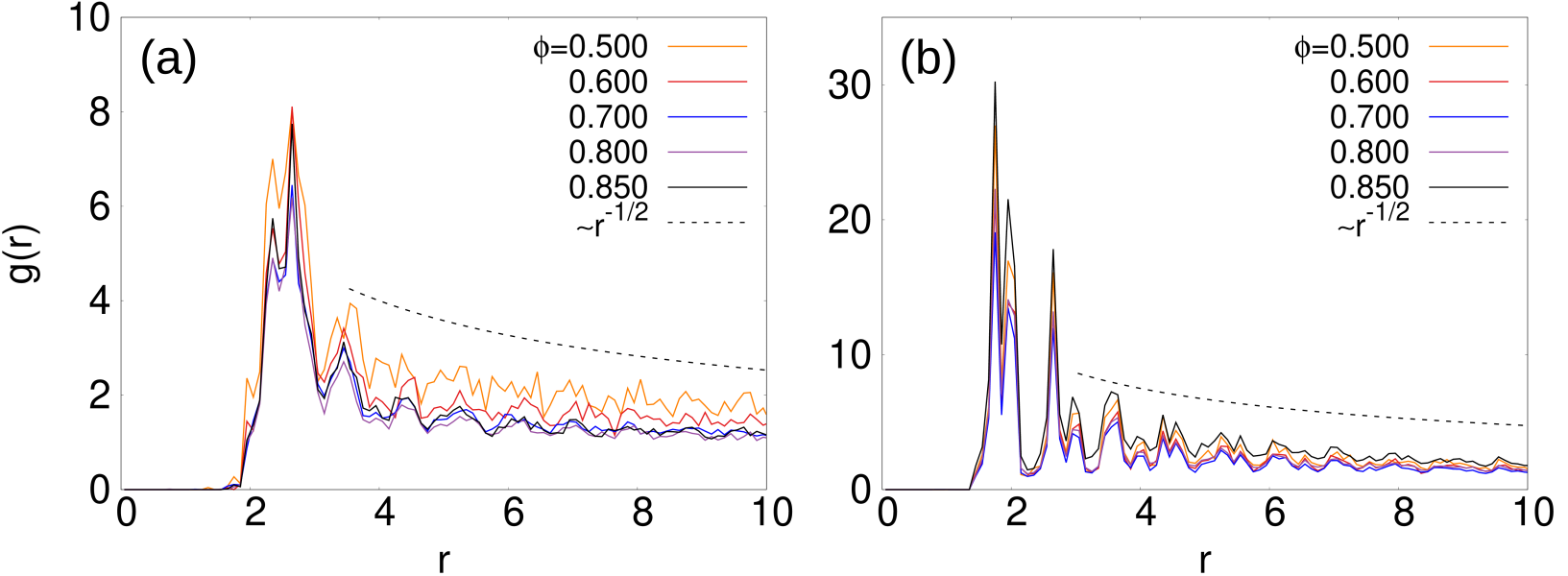}
}
\caption{Variation with $\phi$ of the radial distribution functions of dislocations (a) and 
disclinations (b) in the MIPS phase of the active system at Pe = 100. 
The quasi-periodic peaks in both panels have an
 envelope which  falls off algebraically and is well represented by $r^{-1/2}$. 
}
\label{fig:pair-corr-Pe100}
\end{figure}

Figure~\ref{fig:pair-corr-Pe100} displays the pair correlation of dislocations and disclinations in the 
dense phase of MIPS at different global packing fractions. The curves have multiple {well-defined} peaks 
with decreasing height at farther distances, and they remain almost unchanged when varying $\phi$. 
This is again consistent with the 
fact that the actual density of the dense phase is set, approximately, by the 
upper limit of MIPS and is hence independent of $\phi$. 
The spatial distribution of disclinations displays a sharper peak structure, indicating that disclinations are 
more correlated in space than dislocations,  forming a structure with a higher degree of order. 
Note that the two sub-peak structure of the disclination $g(r)$ at very short $r$, already reported in Fig. \ref{fig:corr-Pe10}, 
is also present in MIPS. Finally, compared to the correlations in the passive system or for weak activity, 
there are many more sharp peaks here. Interestingly enough, these peaks have an envelop that is well approximated by 
the algebraic decay $r^{-1/2}$. This decay is much slower than what we saw in homogeneous phases and we will argue that 
it is due to the string-like organization of defects in the dense phase built via MIPS that we discuss below.

\subsection{A network of  defects}\label{sec:stringMIPS}

Up to now we have focused on point-like defects: dislocations, disclinations and vacancies. 
However, most mis-coordinated particles in the MIPS dense phase cannot be identified as such, but appear in 
clusters comprising several defected particles. 

Figure~\ref{fig:number-density-MIPS} also shows (with red triangles joined by lines) the number density of particles 
belonging to clusters of topological defects. Apart from a 
relatively sharp increase close to the upper limit of the MIPS region, this curve is approximately $\phi$ independent
all the way until the lower (spinodal) limit of MIPS. 

\begin{figure}[h!]
\centerline{
\includegraphics[width=\columnwidth]{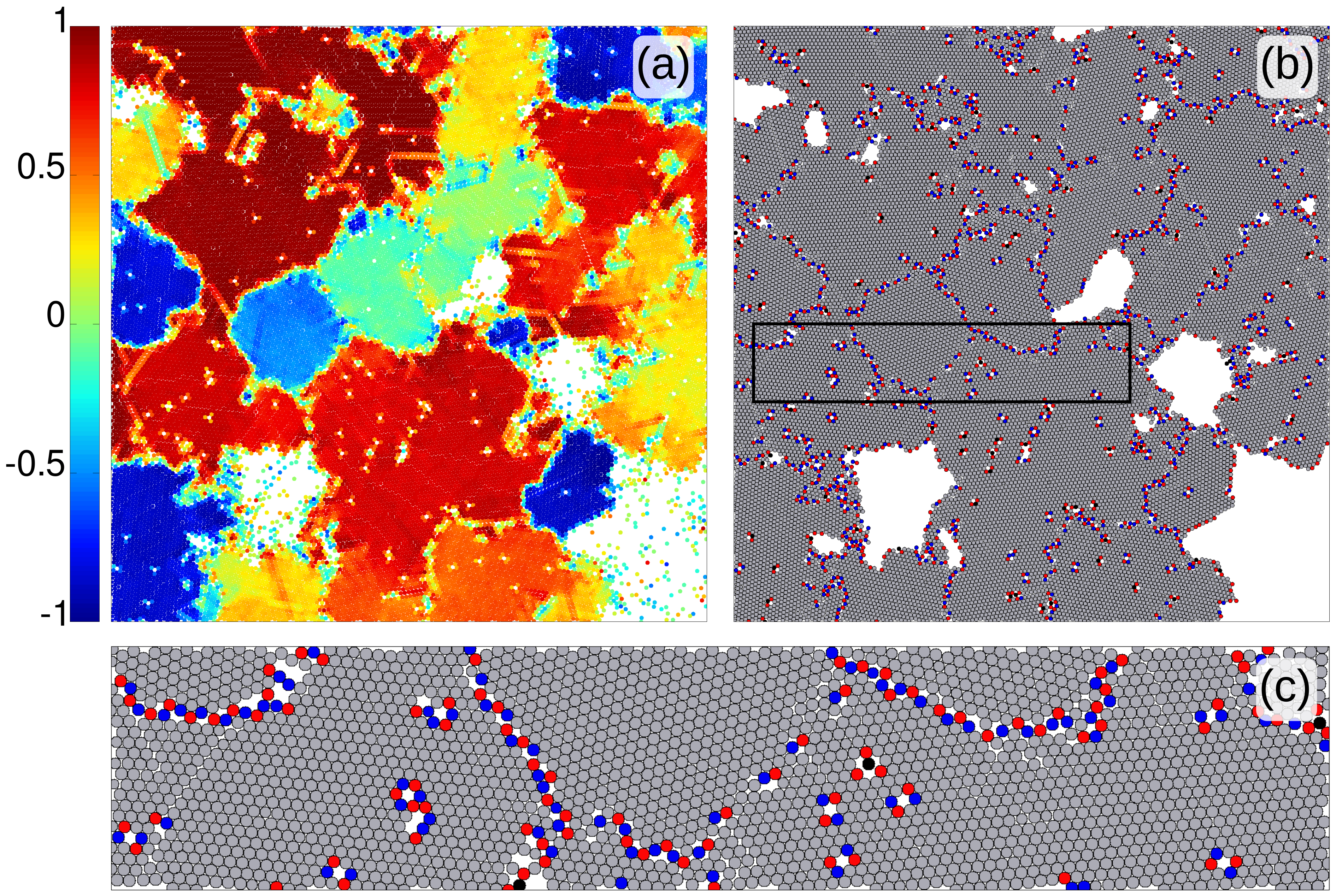}
}
\caption{A zoom over a region of the dense phase in the MIPS regime.
(a) The map of the local hexatic order {parameter} projected on the horizontal direction. 
(b) Defects represented in red (particles with 5 neighbors) and blue 
(particles with 7 neighbors)  in the background of 6-fold coordinated particles (in gray) 
and the gas (in white).
(c) Enlargement of the zone surrounded by a rectangle in (b).
}
\label{fig:MIPS}
\end{figure}

The structure of these clusters can be appreciated in the image in 
Fig.~\ref{fig:MIPS}.  We show in (a)
the local hexatic order color map with finite-size hexatic domains.
The right panel (b) displays the same configuration with 6-fold coordinated particles painted in  gray,
 the ones with 5 neighbors  in red, those with  7 neighbors in blue,
and the ones with  less than 5 or more than 7 neighbors in black. The particles in the 
gaseous phase are eliminated and do not appear in the image.
An enlargement of the zone within the rectangle  is
displayed in (c), and demonstrates a nearly linear arrangement of defects 
along an interface.

\vspace{0.25cm}

The analysis in~\cite{Caporusso20} showed that the MIPS dense phase is  a patchwork of micro
domains with different orientational order separated by rather sharp interfaces. In stationary conditions 
their radii $\tilde R_H$ are exponentially distributed according to
\begin{equation}
P(\tilde R_H) \simeq R^{-1}_H \; e^{-\tilde R_H/R_H}
\; . 
\label{eq:PRH}
\end{equation}
The characteristic length-scale $R_H$ coincides with the averaged one, $\langle \tilde R_H\rangle =
\int d\tilde R_H \, \tilde R_H \, P(\tilde R_H)$, and it follows
$R_H \simeq$ $ \ln ({\mbox{Pe}-a}) +b$ as
estimated from a two parameter ($a,b$) fit, see the inset in Fig.~4~(b) in \cite{Caporusso20}. $R_H$ is weakly dependent 
on $\phi$. As the hexatic micro-domains are compact, from $R_H$ one can estimate an averaged hexatic 
area, $A_H = \pi R_H^2$. The values 
taken by $A_H = \langle \tilde A_H \rangle$ at three Pe values are given in Table~\ref{tab:MIPS}.

The dense phase is also populated by gas bubbles~\cite{Caporusso20,Shi20} with algebraically distributed radii
\begin{equation}
P(\tilde R_B) \simeq {\tilde R}_B^{-\tau_B} \, e^{-\tilde R_B/R_B^*}
\label{eq:PRB}
\end{equation}
with a Pe-independent exponent $\tau_B \simeq 2.2$ and 
$R_B^*$ increasing with Pe~\cite{Caporusso20}.  The exponent $\tau_B$ is also independent of 
$\phi$, at least at high $\phi$ where the bubble statistics can be reasonably well sampled. 
The lower the packing fraction, the larger the cutoff $R_B^*$, since the total gas-phase increases and a few
bigger bubbles appear~\cite{Caporusso-private}.
In Table \ref{tab:MIPS} we also report the mean area of the bubbles, {$\langle \tilde A_B\rangle$,} for the same three Pe values.

The enlargement of the dense component in Fig.~\ref{fig:MIPS}, with the hexatic domains and gas bubbles,  
illustrates the features just described. The configurations are 
smooth in the sense that the areas of both hexatically ordered patches and gas bubbles 
are not fractal but rather  scale as $\tilde A_H \propto {\tilde R}^d_H$
and $\tilde A_B \propto {\tilde R}^d_B$ with $d=2$.

The crossover between the averaged area of the hexatically  ordered patches and the bubble size
from $\langle \tilde A_H \rangle <  \langle \tilde A_B\rangle$ to $\langle \tilde A_H \rangle >  \langle \tilde A_B\rangle$
as Pe increases reported in Table  \ref{tab:MIPS}, 
can also be appreciated in the snapshots in Fig.~\ref{fig:snapshots-MIPS}. These two scales, 
$\langle \tilde A_H \rangle$ and  $\langle \tilde A_B\rangle$, are shown as red (hexatic) and black (bubbles) 
filled disks above the first row.

The dense phase formed via MIPS is thus plagued with defects forming elongated clusters, 
or strings, associated to the formation of grain boundaries delimiting regions with different orientation. 
One can also clearly see from Fig.~\ref{fig:MIPS}  that the defect strings tend to form a network but are also 
interrupted by the gas bubbles in cavitation~\cite{Caporusso20,Shi20}.
These chains of closely spaced defects are not fully 
connected at the single cell scale, as evidenced in Figs.~\ref{fig:string}~(e)
and \ref{fig:MIPS}~(b). 
We therefore treated the configurations with the coarse-graining procedure explained in 
Sec.~\ref{subsec:coarse-graining} with $d_s=3\sigma_d$. Clusters of defects obtained with this technique
are shown in Fig.~\ref{fig:snapshots-MIPS}.  (The boundaries with the gas bubbles are also occupied by 
defected particles but we recall that we do not count them.)  All in all, we identified in this way the 
coarse-grained clusters of defects that we next analyze with the tools devised to understand the morphology and statistics
of percolating objects, {recalled in Sec.~\ref{subsec:percolation0}}.

\begin{figure}[h!]
\centerline{
\includegraphics[width=0.9\columnwidth]{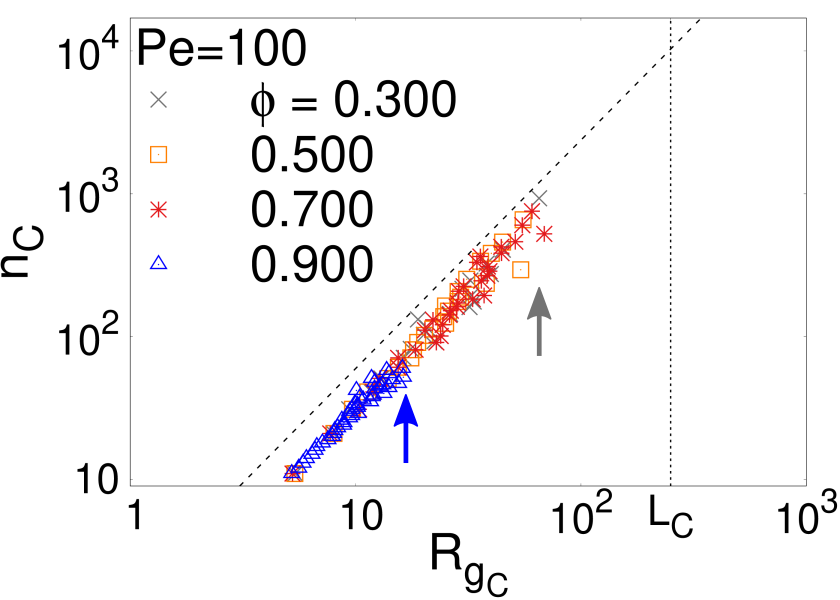}
}
\caption{Scatter plot of the size of the defect clusters against their radius of gyration for parameters at  high activity, 
in the MIPS region of the phase diagram.
The blue and gray vertical arrows point towards the largest $n_C$ for $\phi$ in the hexatic phase and in MIPS, respectively.
From $n_{\mathcal{C}} \simeq {R_{\mbox{\it g}}}_\mathcal{C}^{d_{\rm f}}$, shown with a dashed line close to the data, 
the estimated fractal dimension is  $d_{\rm f} = 1.60$ (error bars are given in Table~\ref{tab:MIPS}).The vertical dotted line indicates the approximate linear size of the dense droplet in MIPS (for a system with equal area of dense and gaseous phases for which $L/2 \approx 250$).
}
\label{fig:MIPS-Rg}
\end{figure}

A measure of the length scale of a cluster is given by its radius of gyration $R_g$.
The data in Fig.~\ref{fig:MIPS-Rg} are scatter plots for the 
cluster size, $n_{\mathcal{C}}$, against the radius of gyration, ${R_{\mbox{\it g}}}_\mathcal{C}$, in systems with different packing fraction but all 
at the same Pe = 100. Again, within our numerical accuracy, 
all data fall on top of each other yielding a single estimate for the fractal dimension 
$d_{\rm f} \simeq 1.60$. The error bars for this measurement and the 
ones at other Pe are given in Table~\ref{tab:MIPS}.
Data for other Pe values also show $\phi$ independence (within 
our numerical accuracy). The outcome of the evaluation of $d_{\rm f}$ are reported
in Table~\ref{tab:MIPS}, see also the gray data-points in Fig.~\ref{fig:fractal-dimension}.
The results  show a monotonic {weak} decay with Pe varying, roughly, from 
$1.65$ at Pe = 50 to $1.57$ at Pe = 200.  In short, we see 
\begin{itemize}
\item
a weak decay of $d_{\rm f}$ with increasing Pe, which suggests that the cluster network 
gets thinner as Pe increases.
\end{itemize}

\begin{figure}[h!]
\centerline{
\includegraphics[width=\columnwidth]{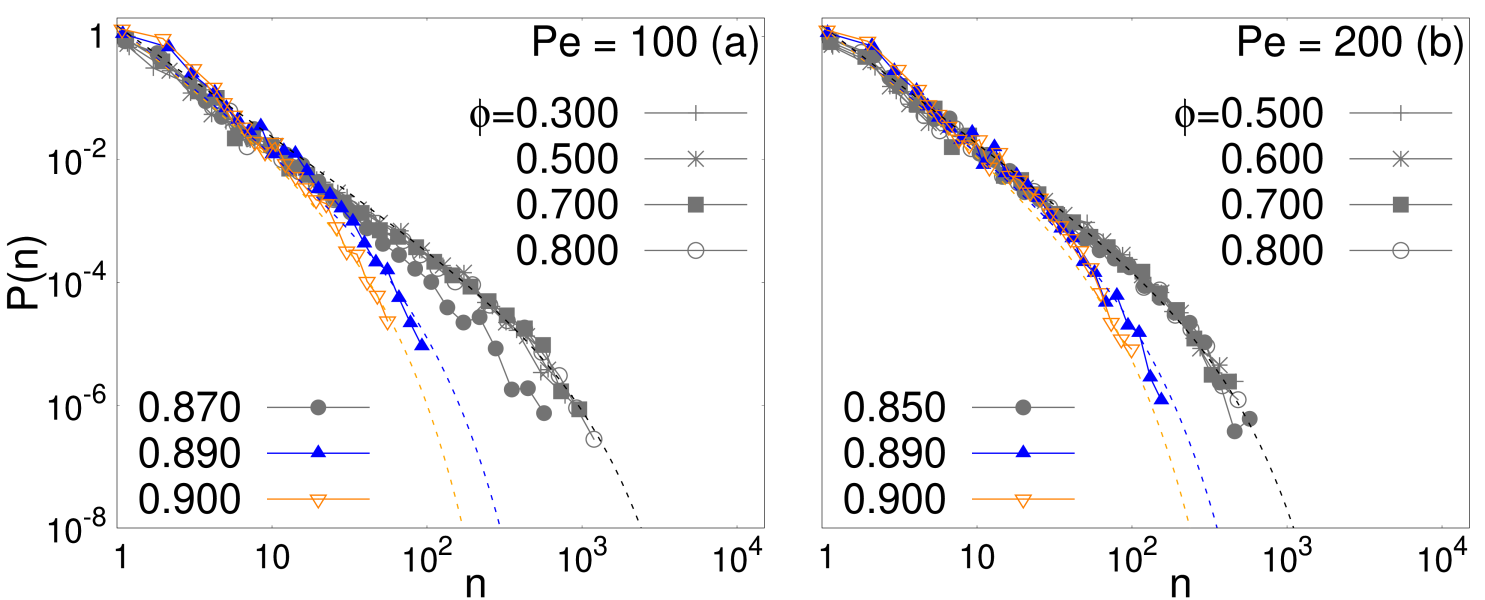}
}
\caption{Probability distribution of defect cluster sizes at  Pe~=~100 (a) and Pe = 200 (b) and several packing fractions, across the different phases. 
The data represented in gray lie in the MIPS region, while the blue and orange in the hexatic and solid, respectively. 
The outlier gray data in (a) are for parameter within MIPS, in between the lower limit of the hexatic phase and the end of MIPS.
Decays of the type  $n^{-\tau_n} e^{-n/n^*}$  with $\tau_n \simeq 2.2$  are shown with 
dotted lines ($n^*\simeq 30$ in the solid, $n^*\simeq 50$ in the hexatic for Pe = 100 and Pe = 200 alike, 
$n^*\simeq 500$ for Pe = 100 and $n^*\simeq 200$ for Pe = 200 in the MIPS region). 
The data in the MIPS phase with hexatic order are compatible with the $2.2$ exponent, though with a shorter cut-off than the rest.
This is due to the presence of a macroscopic single-color hexatic domain (implying the divergent hexatic correlation length) 
and small regions of others colors. Thus, the domain boundaries are also shorter. 
}
\label{fig:MIPS-fig}
\end{figure}

The cluster size distribution $P(n)$ is shown in  Fig.~\ref{fig:MIPS-fig}. It was 
obtained by counting the number of coarse-grained clusters made of $n$ defected cells
at Pe =  100,  200 and various values of $\phi$ specified in the keys. The data in the solid (orange) and hexatic (blue) 
are separated from all curves within the MIPS region of the phase diagram (gray), that fall on each other
(apart from one outlier which is in the hexatic region within MIPS).
Thus, 
\begin{itemize}
\item
the defect cluster size distribution and the morphology of the clusters remain largely invariant within MIPS at fixed Pe,
\end{itemize}
similarly to what we saw in Fig.~\ref{fig:population} (d) and with the analysis of the 
number density of all kinds of defects. 
These results are in line with the absence of packing fraction dependence of other observables at fixed Pe within MIPS.

Solid and hexatic data are well represented by exponential decays $P(n) \propto e^{-n/n^*}$. 
For instance, at Pe = 100 we found that the scales in the 
exponential decay are $n^*\simeq 30$ in the solid and $n^*\simeq 50$ in the hexatic. 
The master curve for the MIPS data shows a much slower decay, 
suggesting a first power law cut-off at large sizes, 
\begin{equation}
P(n) \simeq n^{-\tau_n} f(n/n^*)
\; ,
\label{eq:Pn}
\end{equation}
with $f(x\gg 1) \to 0$.
Although we lack a precise knowledge of the scaling function of the cluster size distribution, which renders the estimation of $\tau_n$ quite ambiguous, 
from the data in Fig.~\ref{fig:MIPS-fig}, we can infer $\tau_n\approx 2.2$ for the two Pe.
The values of the average $\langle n\rangle \equiv \int_0^\infty dn \, n \, P(n)$
are also reported in Table~\ref{tab:MIPS} and decrease with 
increasing Pe.  From the two panels in Fig.~\ref{fig:MIPS-fig}, where we used an exponential form for the 
function $f$,  one sees that, consistently,  the cut-off length $n^*$ is 
a decreasing function of Pe. 

\vspace{0.25cm}

\begin{table}[h!]
\begin{tabular}{|c|c|c|c|c|c|}
\hline
\;Pe \; & \; $\langle \tilde A_H \rangle $ \;& \; $\langle \tilde A_B\rangle$ \; &  \;\; $\langle n \rangle$  \;\; & \;\ $d_{\rm f}$ \;\; 
\\
\hline
\hline
50  &  68.9  & 296.328 & 28.590 & $1.74 \pm 0.04$  
\\
\hline
100 &  151.7  & 378.894 &  26.360 & $1.66 \pm 0.06$
\\
\hline
200 &  795.0 & 442.757 &  22.271 & $1.53 \pm 0.08$  
\\
\hline
\end{tabular}
\caption{ 
Geometric parameters in MIPS. 
The averaged $\tilde A_H$ and $\tilde A_B$ were computed from the numerical data. 
Within our numerical accuracy $\tau_B \simeq 2.2$~\cite{Caporusso20}
and $\tau_n \simeq 2.2$ at the three Pe and quite independently of $\phi$ (at least not too close to the lower
spinodal).
The average cluster size $\langle n \rangle$ is computed from the cluster size distribution $P(n)$ {displayed 
in Fig.~\ref{fig:MIPS-fig}} and the fractal dimension is extracted 
from the analysis of the radius of gyration of the clusters, {in Fig.~\ref{fig:MIPS-Rg}}. 
}
\label{tab:MIPS}
\end{table}

\begin{figure}[h!]
\includegraphics[width=8.8cm]{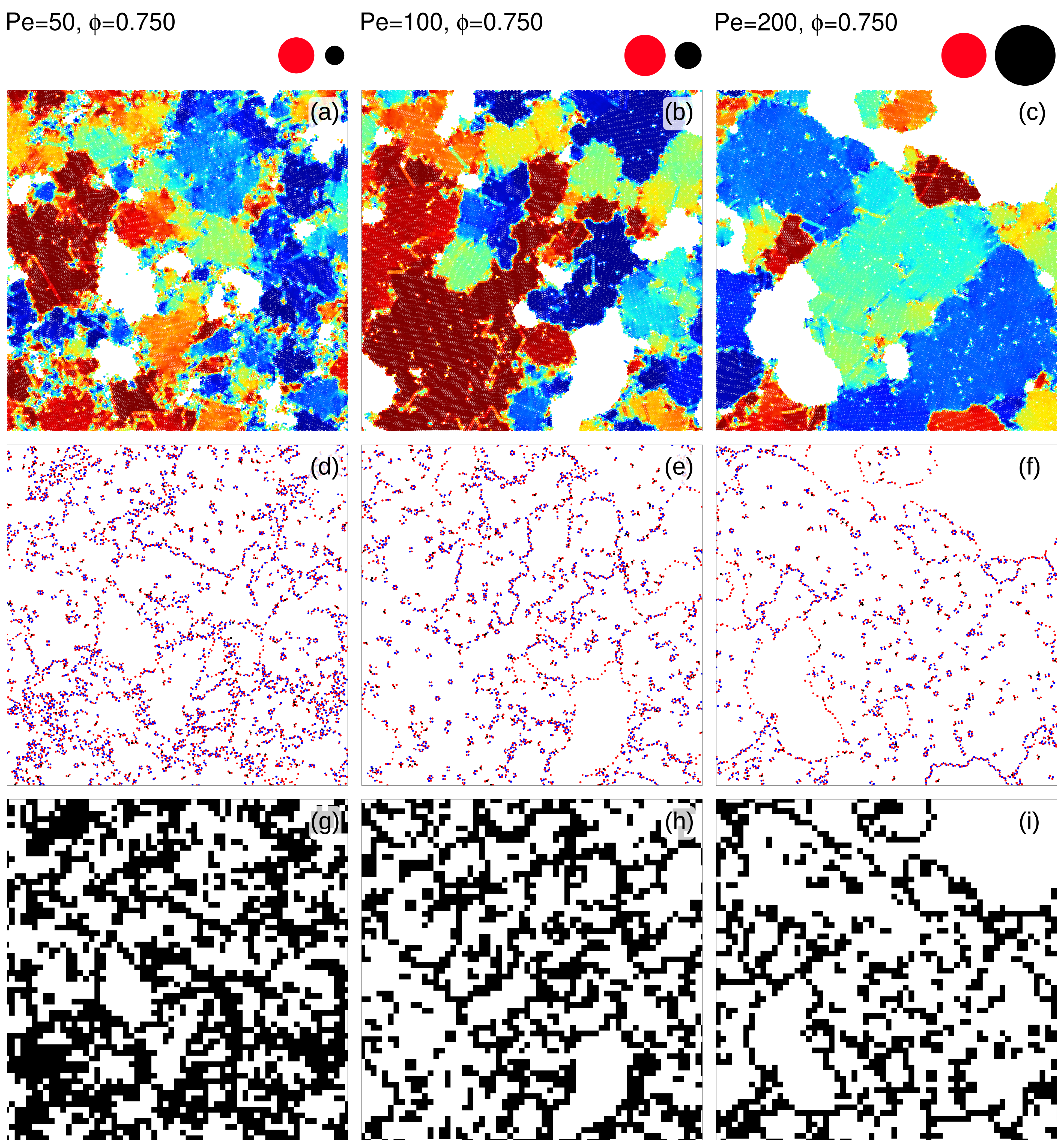}
\caption{Snapshots showing the influence of Pe on the topological defects (support to the data in Table III).
In the three columns, Pe = 50, 100, 200 from left to right, all at $\phi=0.75$. 
We  eliminated the particles in the gas and we used a white background for the gas regions. 
First row:  color map of the local hexatic order parameter projected on the horizontal axis. 
Second row: only the defected particles are represented, showing the string-like structures located at the 
boundaries between domains with different orientation. Third row:  clusters of defects after coarse-graining
with $d_s=3\sigma_d$. As Pe increases, the network structure gets sharper. The red and black 
disks on top of the plots represent the scales $\langle \tilde A_B\rangle$ and $\langle \tilde A_H\rangle$, 
respectively.
}
\label{fig:snapshots-MIPS}
\end{figure}

In Table \ref{tab:MIPS} we report the  average area of the hexatic domains, $\langle \tilde A_H\rangle $, gas bubbles, $\langle \tilde A_B \rangle$, 
and average cluster size, $\langle n \rangle$, together with the fractal dimension of the clusters of defects, 
$d_{\rm f}$. A first observation is that while $\langle \tilde A_H\rangle $ grows,  $\langle n \rangle$ decreases with Pe, which is 
compatible with the idea that the defect clusters are mostly located along the boundaries between hexatic patches.
A second  remark is that $d_{\rm f}$ {weakly} decreases with Pe, suggesting that clusters get more string-like as Pe increases. 
This is in agreement with the inspection of snapshots, see, for instance, Fig. \ref{fig:snapshots-MIPS}.

We stress here that the way in which we count the defects, excluding the boundaries
between dense and gaseous regions, inhibits the possible percolation of 
defect {clusters} within MIPS. Indeed, if the network of grain boundaries within the
MIPS drop percolated, we would observe different distributions at different
densities concomitant with the growth of the {dense} drop {size} with $\phi$. However, this 
does not happen, see Fig.~\ref{fig:MIPS-fig}. Instrumental to this result is the 
appearance of gas bubbles~\cite{Caporusso20,Shi20} within the droplet {which} cut 
the defect network. 

Finally, 
\begin{itemize}
\item
in the MIPS regime, we  see hexatic domains, leaving behind a network of grain-boundaries, 
that can become large but remain finite. The topological defects are mainly located along these boundaries
{and the network is cut by the gas bubbles.}
\end{itemize}

 \section{Melting and percolation of clusters of defects}
 \label{sec:clusters}
 
In Sec.~\ref{sec:point-like} we studied the densities of point-like defects, how they compare to the 
 standard KTHNY theory of melting in passive systems, and their behavior in the presence of self-propulsion.
 Next, in Sec.~\ref{sec:MIPS}, we focused on MIPS and  we characterized 
 the aggregation of defects into a ramified non-percolating network. Now,
  we investigate in depth the collective behavior of defects.

For parameters close to the hexatic-liquid transition, most mis-coordinated cells cannot be identified as 
disclinations or dislocations, but appear in clusters comprising defects of alternating 
topological charge, see Fig.~\ref{fig:string} and Fig.~\ref{fig:percolation-cluster}. 
These objects, when sufficiently large, lie beyond the KTHNY theory. Thus, quite naturally, we 
need to understand which role they play in the melting of the hexatic. 

A related purpose of this Section is to investigate whether defect clusters show any qualitative difference depending on the  
first order (for very small  Pe) or second order  (for larger Pe) nature of the hexatic-liquid transition. For this reason, we also 
study Brownian particles with a soft interacting potential for which the melting  transition in the passive 
limit has been shown to be continuous~\cite{KapferKrauth} (see also App.~\ref{app:soft-disks}).

\subsection{Number densities}

We start by comparing the number density of particles belonging to the different kinds of defects
in the passive and active hard disk system, as well as in the soft particle passive one.
\begin{figure}[h!]
\centering 
\includegraphics[width=8.2cm]{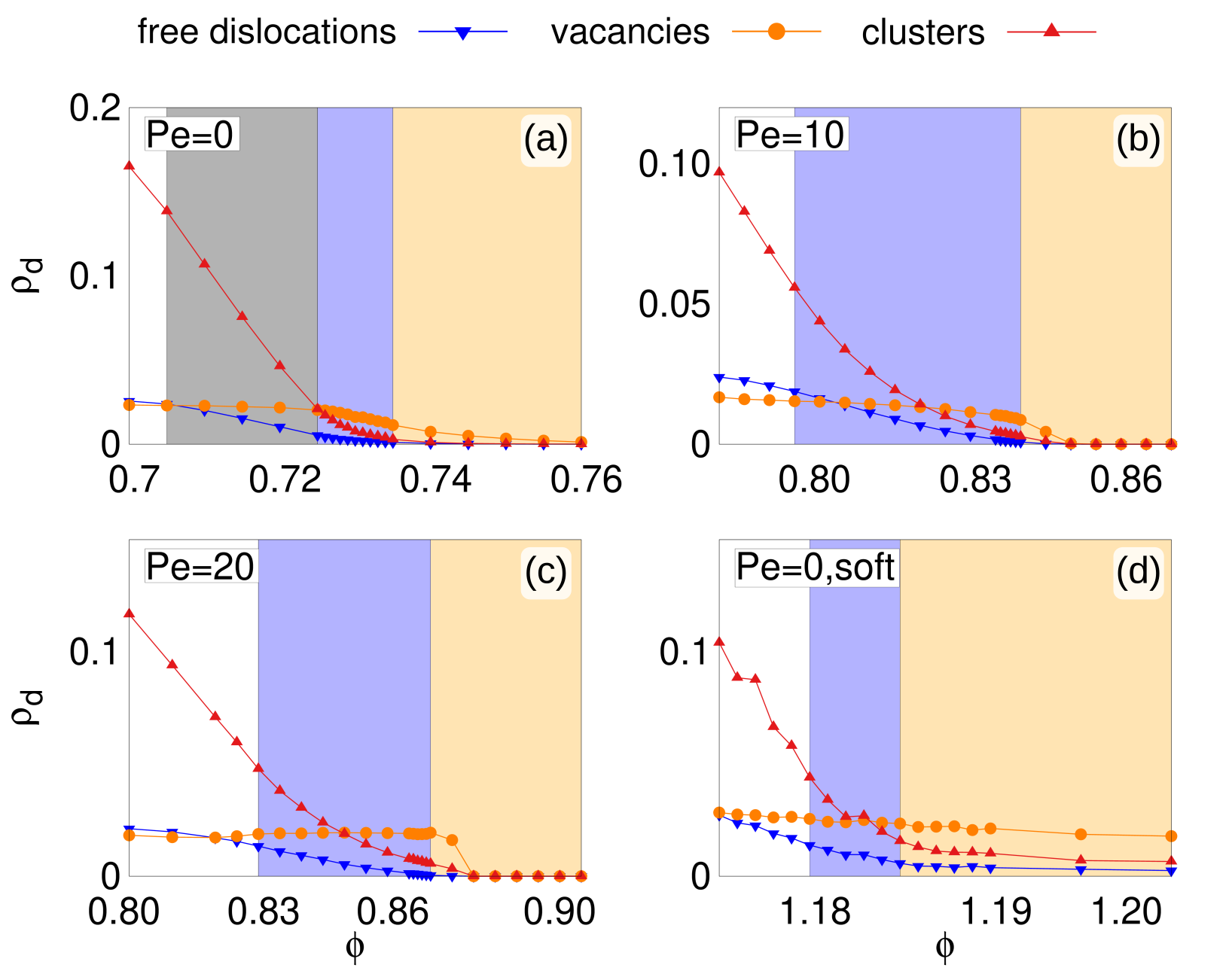}
\caption{Normalized number density of dislocations, vacancies and clusters, 
as a function of $\phi$ for passive hard disks~(a), ABP at Pe = 10 (b) and  Pe = 20 (c), and 
passive soft disks (d).
The solid, hexatic, phase-coexistence and liquid regions are shown in orange, blue, gray and white, respectively. 
}
\label{fig:population1}
\end{figure}
Figure~\ref{fig:population1} shows that  
\begin{itemize}
\item
the number density of particles in clusters dominates the distribution of defects in the hexatic and liquid phases, 
proving that topological excitations are collective rather than localized in this sector of the phase diagram.
\end{itemize}
This also suggests that their proliferation might play an important role in the mechanism driving the melting of the hexatic, instead of (or at least in combination with) 
the unbinding of disclinations. 
Clusters of defects in hard-disk systems have been associated to the formation of grain boundaries 
delimiting regions of different hexatic order~\cite{QiDijkstra, KapferKrauth}, which could drive an alternative 
first-order melting mechanism in 2D~\cite{Chui1982,Chui1983, Kleinert1983}.  
For passive soft-disks and active hard disks at Pe = 10, 20, there is no evidence for a 
first-order transition, and yet the density of defected particles in clusters at the liquid-hexatic transition  is very close to 
the passive hard-disk value ($\approx 0.05$). This suggests, as we show below, that the 
proliferation of clusters might be generic and not responsible for the first-order character of the hexatic-liquid transition of passive hard disks. 

\begin{figure}[b!]
\includegraphics[width=\columnwidth]{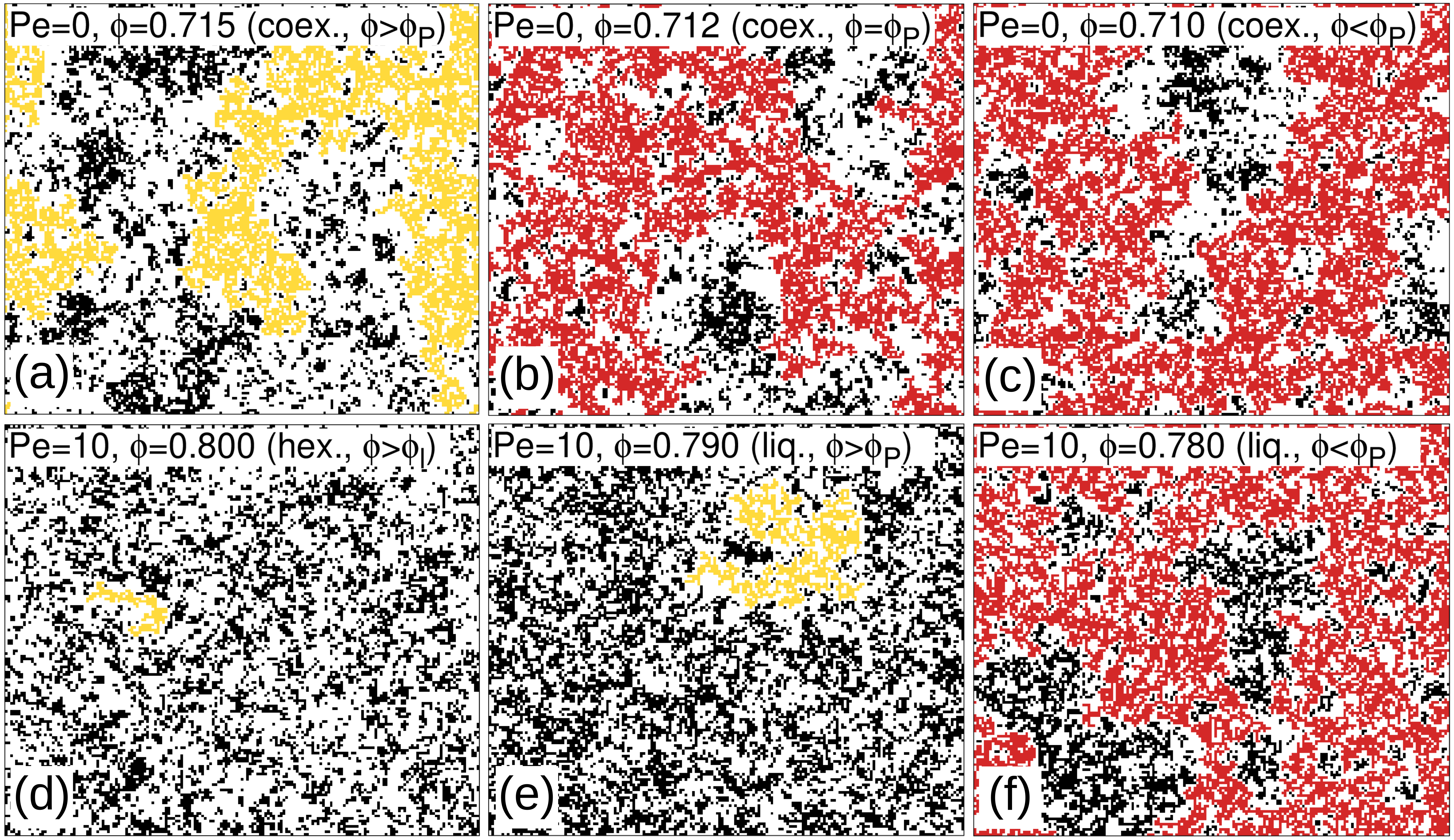}
\caption{Coarse-grained clusters of defects across the hexatic-liquid transition. Finite size clusters are painted in black.
The largest cluster in the system is painted red or yellow, whether it percolates or not.
The first row corresponds to equilibrium hard disks at packing fractions 
(a)  $\phi=0.715$ (in the coexistence region, above percolation), (b)  $\phi=0.712$ (in the coexistence region, at the onset of percolation $\phi_P$) and (c) $\phi=0.710$ (in the coexistence region, below percolation), spanning the coexistence region between hexatic and liquid.
In the second row, we show ABP at Pe = 10, with (d) $\phi=0.800$ (in the hexatic), (e) $\phi=0.790$ (in the liquid, above percolation) and (f) $\phi=0.770$ (in the liquid, below percolation). 
We recall that we use $d_s=3 \sigma_d$ as coarse-graining linear length.}
\label{fig:percolation-cluster}
\end{figure}

For soft particles, we found that the number density of {isolated} dislocations is different from zero at all $\phi$ (even in the solid),  but 
its variation is faster below the transition, see Fig.~\ref{fig:population1}~(d). 
Such behavior departs from a recent study reporting that the 
solid phase melting in a system of even softer Active Brownian particles  
is not driven by the proliferation of dislocations, as a KTHNY scenario would predict,  
opening the possibility of breaking orientational LRO in the absence of any defect~\cite{PaliwalDijkstra}.
For very hard disks (with thus very small mutual overlaps) the solid cannot melt in the absence of free 
dislocations, as our analysis clearly shows. However, 
particle softness influences strongly the nature of melting~\cite{KapferKrauth}, as also shown by the 
results in App.~\ref{app:soft-disks}. Under activity, the competition 
between the potential stiffness and self-propulsion is likely to affect even further the melting 
scenario. 
Although it is hard to make a quantitative comparison between 
the results in~\cite{PaliwalDijkstra} and ours, because of the large difference in the potential 
considered, it is however surprising that these authors did not find a proliferation of 
defects in their solid phase.

\subsection{The numerical percolation curve}
\label{subsec:percolation}

In the study of defects in MIPS, we already treated the configurations with the coarse-graining procedure explained in  Sec.~\ref{subsec:coarse-graining}.  We apply here the same procedure to all defect clusters.
Details on the effect of the coarse-graining linear length $d_s$ are given in App.~\ref{app:coarse-graining}.

Figure~\ref{fig:percolation-cluster} displays, on the first row, configurations (after coarse-graining) 
within the hexatic-liquid  coexistence region of equilibrium hard-disks, with the packing fraction decreasing 
from left to right. In the second row, we show configurations for Pe~=~20 across the continuous hexatic-liquid phase transition.
The largest coarse-grained defect clusters highlighted in yellow are finite, while the red ones percolate across the system size.
In order to identify whether  a cluster percolates in our periodic lattice 
(made of empty or occupied cells after coarse-graining),  we applied the method designed by Machta et al.~\cite{Machta_wrapping,Ziff-Newman01}.
%
We then use the framework of percolation theory \cite{StaufferBook}, see Sec.~\ref{subsec:percolation}, to analyze the statistics of clusters of defects and argue whether critical percolation-like behavior emerges in the system. 

\begin{figure}
\includegraphics[width=\columnwidth]{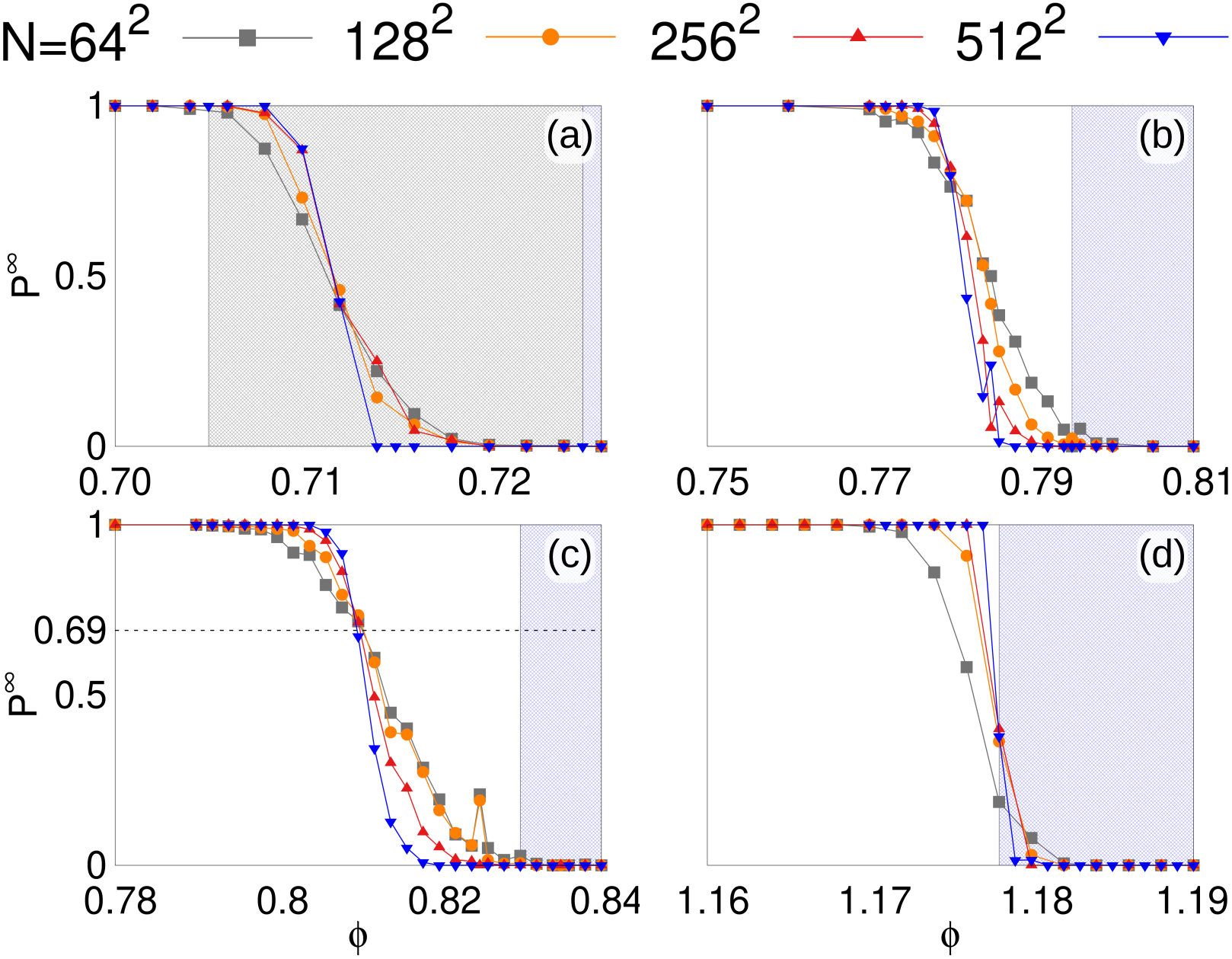}
\caption{Percolation probability $P^\infty$ against the global packing fraction
for different system sizes and fixed Pe~=~0 (a), 10 (b), 20 (c) and Pe~=~0 soft-disks (d). 
The curves cross at a point $\phi_P$(Pe) that is identified as the percolation transition. The analytical prediction of the percolation probability at the critical point \cite{pinson1994} is indicated by a horizontal dotted line. 
}
\label{fig:Heyes}
\end{figure}

After the identification of percolating clusters in the snapshots, the aim now is to locate a percolation curve, $\phi_P$(Pe),  
in the phase diagram (see the red symbols in Fig.~\ref{fig:phasediagram}). 
Figure~\ref{fig:Heyes} shows a finite size analysis of the {percolation} probability $P^{\infty}$ 
as a function of the packing fraction for fixed Pe.
All curves smoothly increase from 0 to 1 when $\phi$ decreases from high to low values,  in the vicinity of the 
hexatic-liquid transition. As the system sizes increase the {variation} of $P^{\infty}$ {with $\phi$} becomes {steeper}  
and {approaches} a step function in the thermodynamic limit.  
The curves for different system sizes are thus expected to cross at a single point that we identify as the percolation critical 
point $\phi_P$~(see~\cite{Heyes89a,Heyes89b} for a similar investigation of particle systems). The value of $P^{\infty}$ at critical percolation on a square lattice with PBC has been computed in \cite{pinson1994}, yielding the value indicated by a horizontal dotted line in Fig. \ref{fig:Heyes} (c). Such prediction matches noteworthy our numerical data for Pe=10, and remains consistent with the data  in the other two cases, for which more data points around the onset of percolation would be needed to put the comparison into quantitative test. 

The full analysis proves that 
\begin{itemize}
 \item
 clusters of defects percolate at a critical packing fraction $\phi_P$(Pe). In equilibrium conditions,  
 $\phi_P$ is located in the hexatic-liquid coexistence region for hard-disks and at the liquid-hexatic transition for soft-disks. 
 In the active case, $\phi_P$(Pe) lies below, but rather close to, the liquid-hexatic transition point.
 {These conclusions are reached within 
 our numerical accuracy,}
\end{itemize}

More precisely, while for passive hard-disks $\phi_P$ is located at $\phi \simeq 0.712$, in the middle of the coexistence region, for Pe~=~10, 20 we find $\phi_P \simeq 0.780, \ 0.810$ respectively. The latter values are smaller than $\phi_l$(Pe) 
(see the colored regions in Fig.~\ref{fig:Heyes} for a reference) and are thus located within the {co-existence region (Pe $<3$) 
or} in the liquid phase {(Pe $>3$)}. 
For the passive soft-disks, percolation coincides with the hexatic-liquid critical point up to our numerical accuracy.

\begin{figure}
\includegraphics[width=\columnwidth]{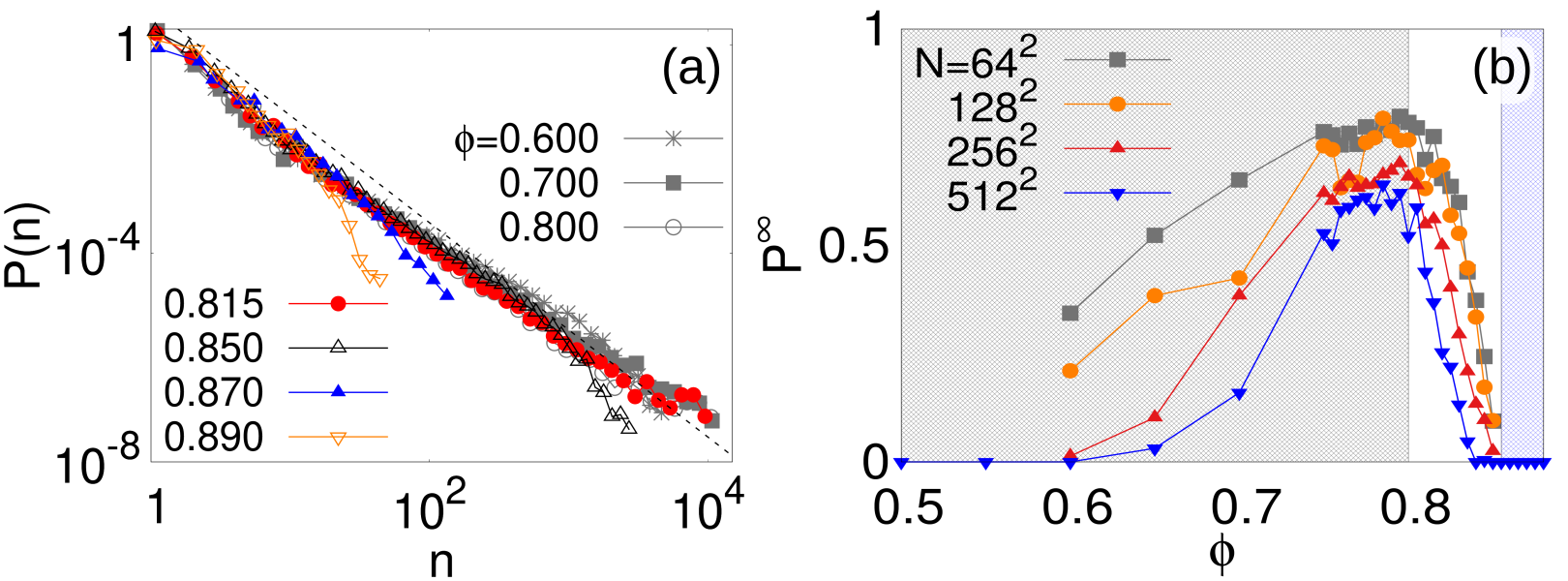}
\caption{(a) Size distribution of the defect clusters at fixed Pe~=~50 and different values of the global packing fraction given in the key across the solid (orange symbols), hexatic (blue symbols) and liquid (black symbols) phases above the upper limit of MIPS. The dotted line corresponds to an algebraic decay with exponent $\tau_n=187/91$. Red symbols correspond to the case $\phi=0.815$ for which we observe a power-law distribution.
(b) Percolation probability at Pe~=~50 and different system sizes. Again, the blue region indicates the hexatic phase, 
the white one the liquid and the gray the coexistence region, here due to MIPS. 
}
\label{fig:percolation_pe50}
\end{figure}

For larger Pe values, the system demixes by MIPS over a range of packing fractions, as shown in Fig.~\ref{fig:phasediagram}.
 For Pe values in between the MIPS critical point~\cite{Partridge19,Maggi21} (located at around Pe $\approx 35$) and the intersection of the liquid-hexatic 
 critical line with the MIPS coexistence region, percolation can also occur. As shown in Fig.~\ref{fig:percolation_pe50} (a), the cluster size distributions 
 $P(n)$ for Pe = 50 display distinctive features of percolation phenomena. Starting from the solid phase, where clusters have relatively small sizes and are exponentially distributed, the distribution broadens as $\phi$ is decreased and eventually becomes algebraic at $\phi=0.815$, in between the hexatic-liquid transition, at $\phi_l=0.855$, and the MIPS high density branch. 
The analysis of $P^{\infty}$ is consistent with such behavior of $P(n)$. As shown in Fig.~\ref{fig:percolation_pe50} (b), the probability of a percolating cluster increases fast as we enter the liquid phase from the hexatic, were $P^{\infty}$ vanishes. However, $P^{\infty}$ does not reach its saturating value  $P^{\infty}=1$ as it occurs in the absence of MIPS (see Fig \ref{fig:Heyes}). The growth of $P^{\infty}$  is interrupted by MIPS.  As anticipated in Sec. \ref{sec:stringMIPS}, MIPS prevents the percolation of clusters of defects via the formation of bubbles that are constantly reshaping.   However, the system displays a power law decay $P(n)\sim n^{-\tau_n}$, with an exponent close to the one of 2D site percolation $\tau_n=187/91 {\sim 2.05 = \tau_n^*}$
{at a precise $\phi$ which we identify as} the percolation point  
$\phi_P$ ({We use this criterium here,} rather than  the finite-size analysis of $P^{\infty}$ as we did for smaller Pe values.) 
See App.~\ref{app:df_Pe50} for a detailed analysis of the structure of defect clusters at the percolation threshold for Pe~=~50.

\subsection{Radius of gyration and fractal dimension}
\label{subsec:radius-gyration}

\begin{figure}[h!]
\vspace{0.25cm}
\centering
\includegraphics[width=0.4\textwidth]{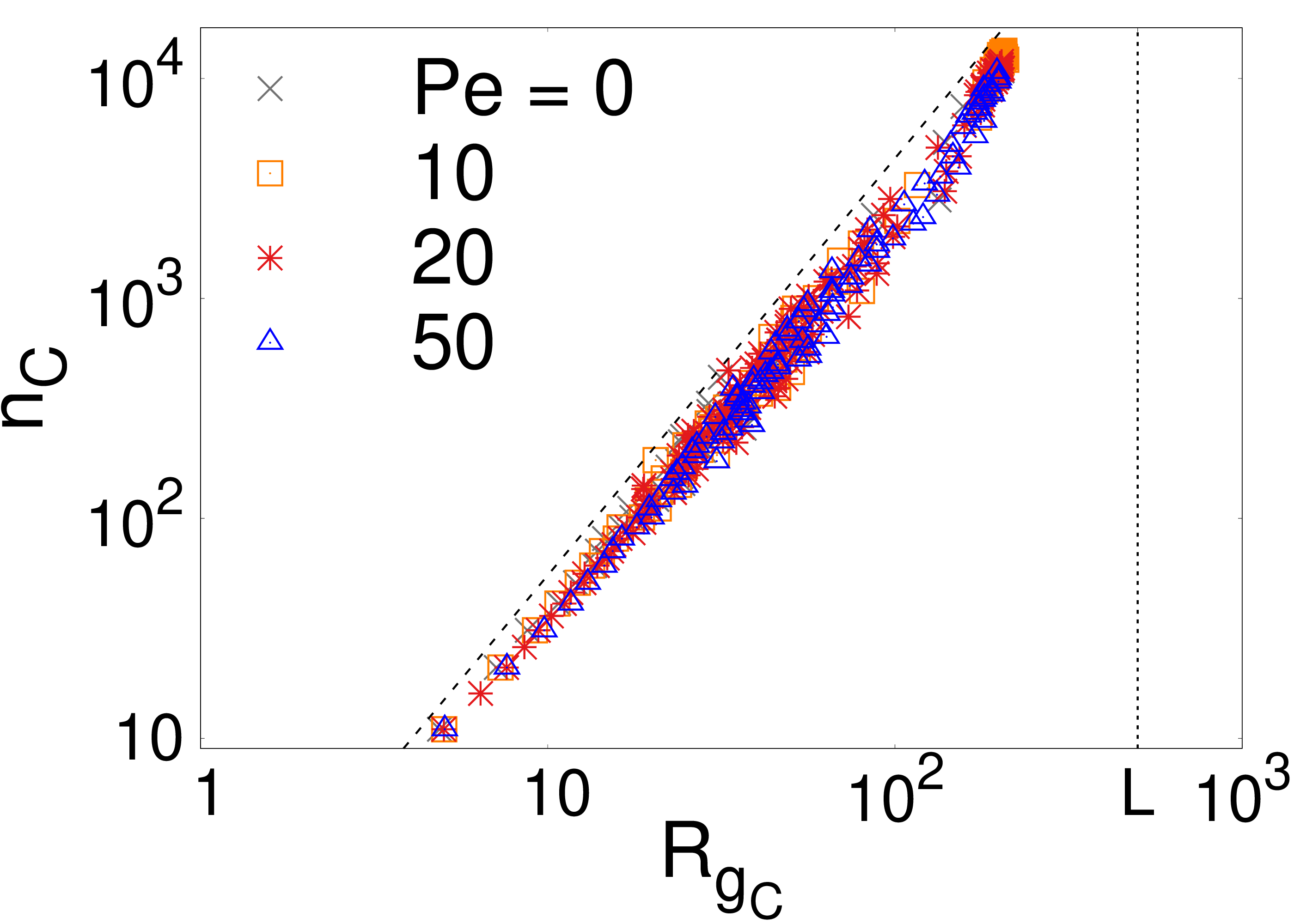}
\caption{Scatter plots of the size $n_\mathcal{C}$ against the radius of gyration of the clusters ${R_{\mbox{\it g}}}_\mathcal{C}$. 
The different data-points (colors) correspond to the 
different values of Pe given in the key. In all cases, the packing fraction is $\phi_P(\text{Pe)}$. 
The black dashed line corresponds to  $n_{\mathcal{C}} \simeq {R_{\mbox{\it g}}}_\mathcal{C}^{d_{\rm f}}$, with $d_{\rm f}=91/48$.  
The vertical dotted line corresponds to the linear size $L\approx 500$ of the system.
}
\label{fig:Rg}
\end{figure}

\begin{figure}[h!]
\vspace{0.25cm}
\centering
\includegraphics[width=\columnwidth]{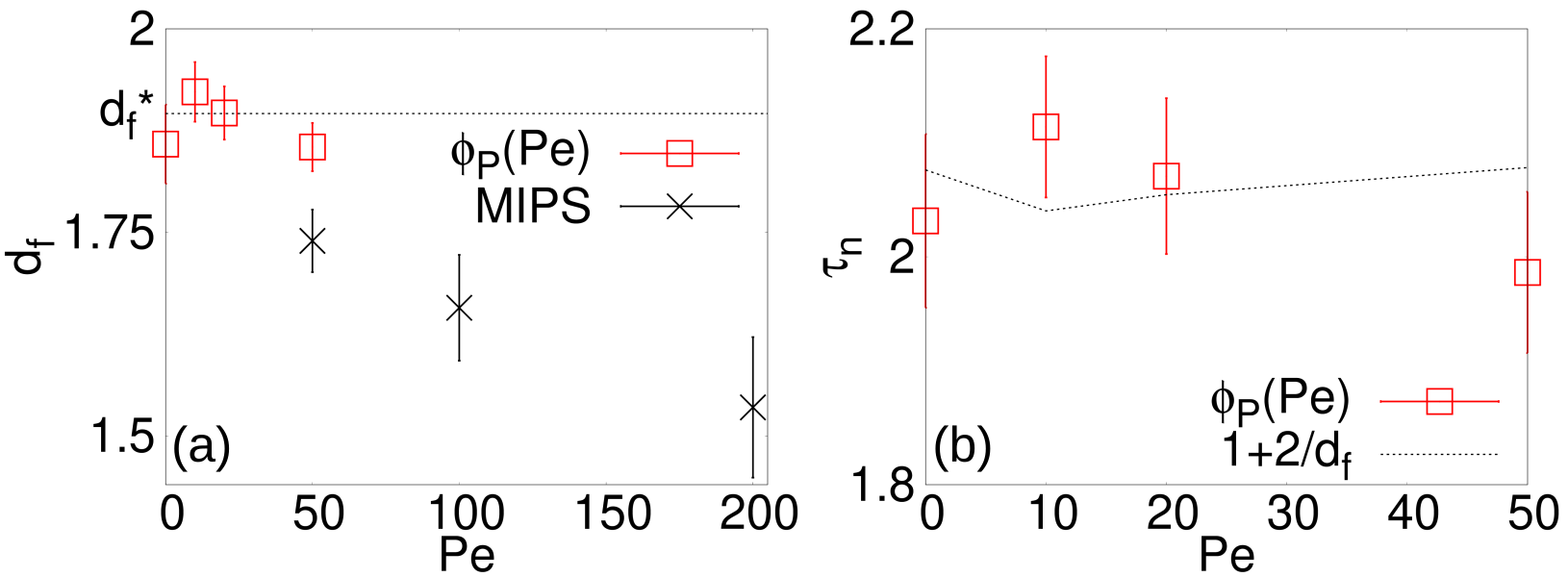}
\caption{(a) The fractal dimension $d_{\rm f}$ of the defect clusters as a function of Pe for 
global densities along the $\phi_P$(Pe) curve away from MIPS (in red), and within MIPS {at} $\phi=0.5$ (in black).
The dotted line is {at} $d^*_{\rm f}=91/48 \simeq 1.90$, {the value of}  uncorrelated site percolation on a 2D lattice~\cite{StaufferBook}.
(b) The exponent $\tau_n$ {was} evaluated from the algebraic decay of the cluster size distribution at the percolation critical point, 
{$\phi_P$}, away from MIPS.
The black dotted line is the value of $\tau_n$ extracted from the scaling relation {in} Eq.~(\ref{eq:Fisher}), 
using the values of $d_{\rm f}$ reported in (a).
}
\label{fig:fractal-dimension}
\end{figure}

We turn now to the characterization of the (possibly fractal) geometry of {the} clusters. 
Typical scatter plots of $n_\mathcal{C}$ against ${R_{\mbox{\it g}}}_\mathcal{C}$ are shown in Fig.~\ref{fig:Rg} for the hard disk system. 
The data-points are taken at different Pe values and  fixed packing fraction $\phi_P(\text{Pe})$, the 
percolation {curve} determined in Sec.~\ref{subsec:percolation}.
The relation between the mass and radius of gyration is in reasonabl{y good} agreement with the expectations from standard 
2D critical percolation, for which $d_{\rm f}=91/48 \sim 1.90 {= d^*_{\rm f}}$. 
The numerical estimates of the fractal dimension $d_{\rm f}$ extracted from such scatter plots for different Pe values at $\phi=\phi_P$ are reported in red, and
{at} $\phi=0.5$ in MIPS in black, in Fig.~\ref{fig:fractal-dimension}~(a).
{The data points are consistent with constant behavior on the curve $\phi_P$(Pe) (red) while they have a weakly decreasing dependence on 
Pe in MIPS (black).}

\subsection{Cluster size distribution}
\label{subsec:cliuster-size}

The full analysis of clusters, in view of percolation theory, leads us to carefully reconsider their size distributions $P(n)$. 
These are obtained by measuring the number of coarse-grained clusters made of $n$ defected cells, excluding the spanning one. 

\begin{figure}[h!]
\vspace{0.75cm}
\centering 
\includegraphics[width=8.5cm]{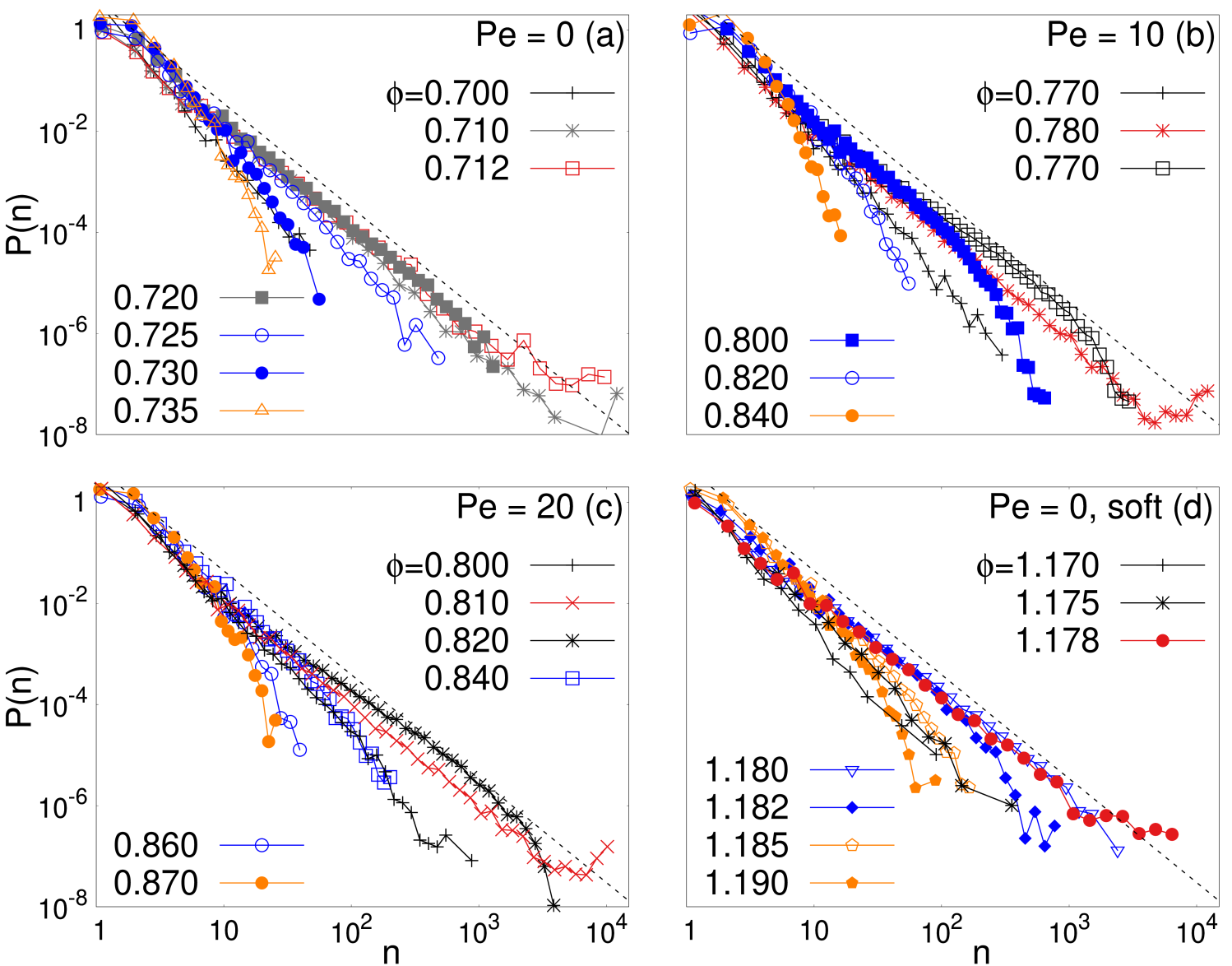}    
\caption{Cluster size distribution  for (a) passive hard disks, 
(b) passive soft-disks, (c) ABP at Pe = 10 and (d) ABP 
at Pe = 20. The different sets of data correspond to different $\phi$ given in the keys. 
The curve that we identify with  critical  percolation behavior is represented in red.
An algebraic decay $P(n)\propto n^{-\tau_n}$ with $\tau_n=187/91$ is plotted with dotted black lines
{in all panels}.
At criticality, the Fisher exponent $\tau_n$ is related to space dimension and the cluster fractal dimension via Eq.~(\ref{eq:Fisher}),
 see Fig.~\ref{fig:fractal-dimension}. All the panels represent the data on the same scale on both the vertical and horizontal axis.
}
\label{fig:Pn}
\end{figure}

In Fig.~\ref{fig:Pn} we show $P(n)$ for passive hard disks (a),  ABP at Pe = 10 (b), Pe = 20 (c), and passive soft disks (d). 
In all cases, we show results for different $\phi$ across all phase transitions.  Starting from the solid (orange symbols), the effect of decreasing $\phi$ is clear, as the 
distribution broadens to include larger clusters. The data in the hexatic phase (in blue) also follows this trend. 
The key feature to highlight here is that, in all cases -- for passive disks, both hard and soft, and also active ones -- $P(n)$ 
becomes \emph{scale free} at a packing fraction slightly below $\phi_l$, the value  below which orientational correlations decay exponentially (red data). This is the hallmark of 
percolation. Consistently, the packing fraction at which the distributions become scale-free coincides with the value 
$\phi_P$ extracted from the finite-size analysis of $P^\infty$. As already 
mentioned, percolation is observed: (i)  in the middle of the coexistence region for equilibrium hard disks (corresponding to the intersection point of the Binder cumulant associated to 
hexatic order obtained for different system sizes \cite{PRLino}), (ii) at the hexatic-liquid transition for equilibrium soft disks, (iii) slightly below the hexatic-liquid transition for ABP. 

At a critical percolation point, the scaling relation $\tau_n=1+d/d_{\rm f}$, must hold. 
As our analysis {demonstrates}, 
our coarse-grained clusters of defects at $\phi_P$ follow this scaling. 
The algebraic decay $n^{-\tau_n}$, with $\tau_n=187/91$ -- 
the exponent corresponding to random site percolation in 2D, for which $d_{\rm f}=91/48$ -- 
captures the behavior of $P(n)$ at $\phi_P$, see Fig.~\ref{fig:Pn}.
Thus,
 \begin{itemize}
 \item
 at $\phi_P$(Pe),
the statistics of the coarse-grained clusters is well described by critical percolation, 
irrespective of the discontinuous (for passive hard disks~\cite{BernardKrauth}) or continuous 
(for passive soft disks~\cite{KapferKrauth} and ABP at Pe = 10, 20~\cite{PRLino}) character of the 
hexatic-liquid transition. 
\end{itemize}

At $\phi_P$(Pe) the measurement of $d_\text{f}$ provides a consistent prediction for the 
Fisher exponent $\tau_n$ characterizing the distribution of 
cluster sizes $P(n)$. This is confirmed by the data presented in  Fig. \ref{fig:fractal-dimension} (b)-(c). The Fisher exponent extracted from the decay of $P(n)$ at $\phi_P$ is reasonably 
close (considering the difficulty of such numerical measurements) to the value predicted by the scaling relation~(\ref{eq:Fisher}), 
using the value of $d_{\rm f}$ obtained independently 
from the {mass}-radius of gyration {relation}. 
In the MIPS sector, though, such scaling is likely to be lost. First, clusters do not percolate inside MIPS. 
Secondly, as already discussed, the numerical determination of $\tau_n$ from the size distributions is somehow ambiguous. 
{The algebraic decay of $P(n)$ expands on rather short scales only and the exponent seems to be 
independent of Pe while $d_{\rm f}$ appears to decay with Pe.}

The percolation of coarse-grained clusters is illustrated in Fig.~\ref{fig:percolation-cluster}.
For hard-\-disks  in their equilibrium hexatic phase, the largest cluster (yellow) does not span the system.
As we decrease $\phi$, the size distribution broadens until a percolating cluster arises (red), right in the middle of the liquid-hexatic coexistence region \cite{PRLino}.
The spatial location of the percolating structure is correlated with the one of the liquid as we explain in the next Subsection.

\subsection{The emergence of the liquid}
\label{subsec:clusters-liquid}

The correlation between the location of the defect clusters and the emergence of the liquid 
is proven by {its comparison to} the maps of the local density 
\begin{equation}
\phi_i=\phi(\bold{r}_i) = \frac{\pi \sigma^2}{4A_i}
\; , 
\end{equation}
with $A_i$ the area of the Voronoi cell attached to the $i$th particle, 
and the maps of the local hexatic order parameter defined in Eq.~(\ref{eq:local-hexatic}). 

\begin{widetext}

\begin{figure}[h!]
\centering
\includegraphics[width=\textwidth]{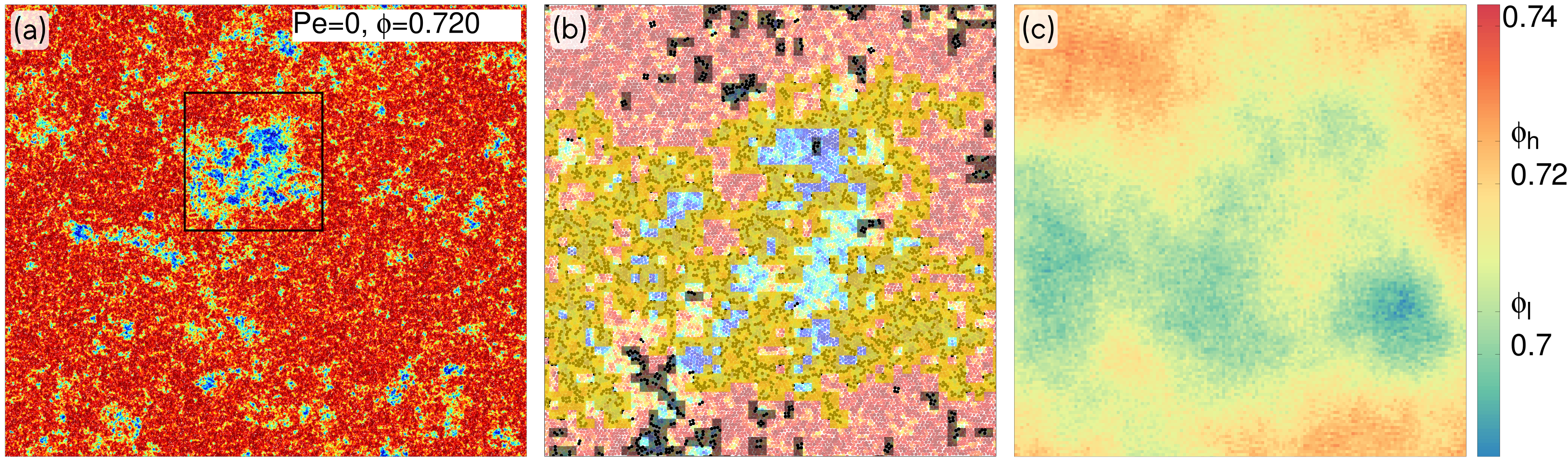}
\caption{
(a) Color map of local hexatic order parameter (see the main text and~\cite{PRLino} for the color code), for a hard disk system with Pe = 0 and $\phi=0.720$, within the hexatic-liquid coexistence region and close to the high-density binodal. The red area is oriented along the mean orientation of the sample and corresponds to its hexatic region, in coexistence with the liquid one. (b) Detailed view of the zone surrounded by  the black square in (a), 
showing a drop of liquid coexisting with the hexatic phase that largely surrounds it. Black cells are finite- size defect clusters. Yellow cells belong  to the biggest (although finite) cluster  of defects  (see also Fig.~\ref{fig:percolation-cluster}). 
(c) Local density $\phi_i$ corresponding to (b)  
then coarse-grained over a length of $20\, \sigma_d$. The liquid and hexatic coexistence densities, 
appearing in the scale  on the right are $\phi_l = 0.705$ and $\phi_h = 0.725$. A clear correlation between the location of the biggest cluster in yellow in panel (b) and the appearance of the liquid in panel (c).
}
\label{fig:liquid-drop}
\end{figure}

\end{widetext}

\begin{figure}[t!]
\includegraphics[width=\columnwidth]{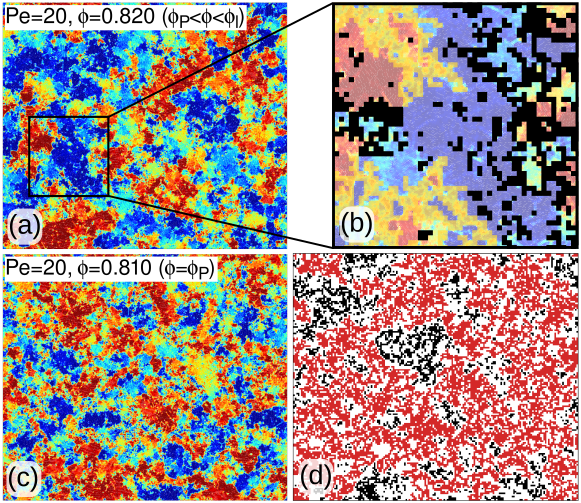}
\caption{(a) Map of the local hexatic order parameter $\psi_6(\bf{r})$ projected on 
the mean orientation of the system for active, Pe = 20, hard disks
close to the hexatic-liquid transition ($\phi_l=0.830$), within the liquid phase.
(b) Zoom over the region within the 
black square in  (a) with the defect cluster shown in black on top of the $\psi_6(\bf{r})$ background.
(c) Map of the $\psi_6(\bf{r})$ projection at $\phi_P$(Pe = 20)$=0.810$. (d) Clusters of defects in the configuration 
in (c) with finite ones in black and the percolating one in red. 
}
\label{fig:morphology}
\end{figure}

In Fig.~\ref{fig:liquid-drop} we {exhibit}  how the growth of the liquid within 
the hexatic occurs in correspondence with the proliferation of clusters of defects, 
in a passive system with a global packing fraction within the coexistence regime. 
Panel (a) displays the map of the local hexatic order parameter. {In this plot} we use the following convention: red particles are oriented along the horizontal direction, blue the opposite one, and the intermediate directions follow the code in panel (a) of Fig.~\ref{fig:morphology}.
The square box selects a region of the system that is zoomed over in (b). In this sub-region a 
liquid drop co-exists with the hexatically ordered surrounding. The black cells are small size coarse-grained defect clusters
while the yellow ones belong to the largest  defect cluster in this region. 
The correlation between the location of the largest defect cluster and the 
liquid zone is confirmed in (c). In this latter case,  the map of the local density is plotted according to 
the color code on the right bar,  also indicating the values of the liquid and hexatic coexisting packing fractions.

We conclude that 
\begin{itemize}
\item
For parameters in the co-existence region of the equilibrium hard-disks system,
the emergence of a percolating defect cluster can be attributed to the percolation of the liquid domain.
\end{itemize}

In Fig.~\ref{fig:morphology} we present further evidence for such correlation in an active
system at Pe = 20, close to the hexatic-liquid transition. 
Panel~(a) presents the color map of the local hexatic order parameter, with red indicating the 
averaged, and also the majority, orientation in the system. 
As evidenced in the selected view Fig.~\ref{fig:morphology}~(b), {which} zooms over the box in (a),
most defects are located at boundaries between different hexatic orders, where the hexatic melts
{\it cfr.} Fig.~\ref{fig:percolation-cluster}~(b) and Fig.~\ref{fig:morphology}~(a).
In short, Fig.~\ref{fig:morphology}~(a)-(b) suggests 
that
\begin{itemize}
\item
at the continuous hexatic-liquid transition at intermediate Pe, coarse-grained clusters can be identified with grain-boundaries
between regions with different orientational order.
\end{itemize} 
Percolation in this case is associated with the emergence of a system-spanning cluster of defects, 
along which the liquid phase develops.
For Pe = 20, there is no liquid-hexatic coexistence, and yet there is percolation slightly below the hexatic-liquid transition.
 For equilibrium soft-disks,  the  liquid-hexatic transition is continuous and  accompanied by the percolation of clusters of defects. 

\subsection{Is critical percolation of defects driving melting?}
\label{subsec:critical-percolation}

Experience teaches us that one  has to be extremely careful before claiming that there is a strict relation between a thermodynamic phase transition driven by the competition between interactions and fluctuations, such as melting, and a phenomenon of purely geometric nature,  like percolation.
In many {lattice} spin models, the properties of their second order thermal phase transitions can be rigorously 
described in terms of the critical percolation 
of suitably defined Fortuin-Kasteleyn (FK) clusters, that {\it are not}  the obvious geometric ones one could {\it a priori} try to use~\cite{FK1972a,StaufferBook}. 
Moreover, although the FK clusters should not be useful to describe a first order thermodynamic transition, in cases with very long correlation lengths such as the 
2D Potts model with $Q$ \raisebox{-0.1cm}{$\stackrel{>}{\sim}$} 4, one could easily see them percolate close to the transition in finite size samples~\cite{Marco}. 

Our simulations of finite, off-lattice, particle systems, provide clear evidence that active, passive, continuous and discontinuous hexatic-liquid transitions 
are accompanied by the percolation of  {coarse-grained} defect clusters.
However, {we should stress that the particle problem we are dealing with is defined in the continuum and, in addition to the 
{\it provisos} exposed in the previous paragraph, we face  the difficulty  of having to transform the particle configurations into lattice 
ones via coarse-graining.}

We cannot therefore posit that the relevant clusters to describe the transition quantitatively
are the coarse-grained defect ones we used.
Still, albeit we cannot establish a rigorous link 
between melting and the {geometric} transition,
the identification of a {critical} percolation phenomenon sheds new light on the microscopic mechanisms driving 
2D melting, as the liquid  phase does indeed emerge from the defect clusters similarly to the pre-melting of 3D colloidal crystals \cite{Alsayed2005}.

 \section{Conclusions}
 \label{sec:conclusions}

We close the paper with a concluding Section that we divide in two parts. In the first one we recall the 
context in which our work inscribes, that is,  the melting  of spherically symmetric particle systems. 
In the second one we present a short summary and discussion of our results.

\subsection{Context}
\label{subsec:discussion}

In order to consider our results in the right perspective, we first observe  that passive melting is a fundamental 
natural process and that local but also extended defects play a key role in it. 

While the main mechanism for bulk 3D melting, as one in which the liquid pervades the system along grain boundaries between crystalline 
domains, is fully accepted, the one(s) leading to 2D melting are far from being fully understood. 

Three dimensional experiments  demonstrate the pre-melting along grain 
boundaries in the bulk of crystalline passive materials~\cite{Alsayed2005}. They 
thus prove that this is the basic mechanism promoting the direct 
melting transition from the crystal to the liquid.  Besides, thermodynamic 
measurements reveal that the 3D melting transition is a conventional first order one~\cite{Gasser09}.

One of the predictions of the  KTHNY theory for 2D melting~\cite{Halperin1978,Nelson1979,Young1979}, 
the fact that the transition takes place in two 
steps with an intermediate hexatic phase is by now widely accepted. Experiments 
in a host of different materials have established the existence of this intermediate 
 phase with QLRO. Early~\cite{Murray87,Tang89} and more 
recent~\cite{Kusner94,Maret1999colloids,Guillamon09} experiments 
providing the necessary evidence 
are discussed in the review articles~\cite{StrandburgRev,GlaserRev,vonGrunberg07,Gasser09,Gasser10,Wang16}.
It is important to stress, though, that some other materials with different  interactions, conform better~\cite{Wang16} 
to Chui's single first order transition arising through grain boundary melting even in 2D~\cite{Chui1983}.

The KTHNY theory~\cite{Halperin1978,Nelson1979,Young1979} also predicts the order of, and mechanisms for, the two transitions;  basically, 
BKT singularities~\cite{Kosterlitz1973,Kosterlitz1974} due to the unbinding of rather simple defects: 
dislocations in the solid-hexatic and disclinations in 
the hexatic-solid. Admittedly, experimental and numerical 
proofs of these predictions have remained elusive. In the terms of von Gr\"unberg, Keim and Maret ``...the colloid KTHNY literature on
this point is full of irritating and partly contradictory findings"~\cite{vonGrunberg07}.

Experimental studies of the defect mediated scenario for the two step transition
appeared in~\cite{Kusner94} where a substrate-free dipolar system was monitored and in~\cite{Han08},
also reported in~\cite{Wang16}, where a monolayer of a repulsive colloidal system was studied. 
Based on a qualitative analysis of data, Kusner et al.~\cite{Kusner94} claimed that free dislocations appear in the hexatic, and 
free disclinations at the liquid-hexatic transition, with no signature of defect cluster formation.
In~\cite{Han08}, the free dislocation and 
free disclination densities detach from zero at different values of the packing fraction. 
While the latter is in rather 
good agreement with the critical value estimated from correlation and susceptibility measurements, 
the former is displaced towards a higher packing fraction compared to the one deduced from 
the translational susceptibility peak. The density of free dislocations is compatible with the KTHNY 
criticality with $\nu = 0.37$ while there is no quantitative analysis of the density of free disclinations.
Other experimental papers, like the ones by Tang et al.~\cite{Tang89,Tang90}, report the appearance of 
dislocation loops and complex clusters in the free expansion melting of 2D charged colloidal 
micro-spheres and cast some doubt on the hexatic-liquid transition being solely driven by the dissociation 
of dislocations.

Being the collection of clean experimental data so hard, progress in this field 
came from numerical simulations which indicate that, for sufficiently hard disks, the transition 
from the hexatic to the liquid is of first order instead of BKT~\cite{BernardKrauth}. The measurements in this 
reference, and most following articles studying this hexatic-liquid transition, see, 
e.g.~\cite{KapferKrauth,PRLino,Ciamarra20a,Ciamarra20b,Padilla20,Khali21,Marzena21}, 
focused on macroscopic observables which do not provide clues about the mechanisms driving the phase transition. 

Arguments for extended, possibly percolating, defect clusters
{which could} render the transitions first order have been put forward by various authors. This proposal 
is an old debated topic, see e.g.~\cite{Murray87,Bladon95,vonGrunberg07},  which has 
received renewed interest~\cite{KapferKrauth,Qi14,Mazars15,Mazars19,Khali21} after the 
Bernard \& Krauth work~\cite{BernardKrauth}, which differentiates the order of the hexatic-liquid transition from the one of the
solid-hexatic. Although there is some numerical evidence for the proliferation of defects and their clusterization close to the hexatic-liquid transition
of both passive~\cite{Reichhardt03,Mazars19,Khali21} and active~\cite{PaliwalDijkstra} 2D systems,
we have found no quantitative support for, nor evidence against, the percolation of 
defect clusters at, or close to, this transition. 

The geometrical description of phase transitions, 
such as the one provided by percolation of the 
rightly identified objects (be them fully geometrical or also partially statistical) is very appealing. 
Its  search drives research efforts in different fields of material science (jamming, gelation, complex systems, etc.). 
It is therefore quite clear the interest in trying to elucidate whether a
geometric phenomenon is behind the breakdown of  the hexatic phase in 2D particle systems.

 \subsection{Summary of results}
 \label{subsec:summary}

 We provided the first quantitative numerical analysis of topological defects in 2D passive 
and active Brownian particle systems, reaching conclusions that confirm the classical KTHNY scenario
concerning the solid-hexatic transition, but  lie beyond it regarding the hexatic-liquid one. 

Concretely, we characterized the melting of passive and active disks in terms of the statistics and spatial organization 
of point-like topological defects and the statistics and geometry 
of  the clusters they form.

On the one hand, we found that the KTHNY scenario of dislocation unbinding 
gives a qualitative and quantitative description of the solid-hexatic transition 
of the hard disk system for all 
activities. Our measurements point towards the universality of this phase transition in the sense that 
the BKT scaling holds and the exponent $\nu$ is, within our numerical accuracy, equal to 0.37 all along the solid-hexatic 
transition line. {It would be worth proving this statement analytically, possibly extending 
the analysis in~\cite{Zippelius80}  to a model with energy injection. This project goes beyond the scope of this 
publication.}

On the other hand, we generically observed percolation 
of clusters of defects at the vicinity of the hexatic-liquid transition: within the coexistence 
region for equilibrium hard disks, at the transition for soft ones, and below the transition for active hard disks. 
The particles involved in the clusters overwhelm in 
number those participating in the dissociation of dislocations, which also exist as soon as the liquid phase 
appears. The fewer disclinations we identify are always very close to extended defect clusters and are 
not free to move in the sample. The geometry of the percolating cluster appears to be quite independent of the 
activity and very close to the one of conventional uncorrelated percolation in 2D.  

Although defect cluster percolation has been associated to first-order melting scenarios in the past, we found it across both 
first and second-order transitions,  in- and out-of-equilibrium. In all these cases, 
we found that the clusters of defects appear and grow in regions of density depletion and 
where interfaces between {patches} with different orientational order sit. 
The liquid starts permeating the sample from these
spots.

What is the relationship between melting, a phase transition driven by the competition 
between interactions and fluctuations, and percolation, a phenomenon of pure geometric nature? 
In the context of classical spin systems, the connexion between thermal phase transitions and the 
percolation of suitably defined clusters can be rigourously established through, for instance, the Fortuin-Kasteleyn (FK) 
representation, which exactly maps a spin system on a geometric model~\cite{FK1972a,FK1972b,FK1972c}. 
We do not have a way to faithfully represent a system of interacting particles in terms of geometric clusters 
and, therefore, we cannot establish a rigorous connexion between melting and percolation.
However, all the results reported point in the direction of suggesting the existence of such a relation.

 At high Pe, MIPS prevails over the hexatic-liquid transition, and clusters of defects are large and ramified 
 but, in the way we defined and counted them, do not percolate. Indeed, the dense MIPS droplet is also filled 
 with gas bubbles  with an algebraic distribution of sizes  that disconnect the defect clusters at all packing fractions, even beyond 
 the hexatic-liquid transition {inside MIPS}. Then, our defect cluster size probability distribution depends on 
 Pe but not on the global packing fraction.
 
 In short, we presented an unprecedented comprehensive analysis of topological defects in the stationary regime of 
 2D passive and active Brownian particles. We clarified the nature of the phase transitions in a way 
 that sheds new light on the old problem of 2D melting, going beyond the KTHNY picture,  
 and showing that percolation is an underlying phenomenon associated to 2D melting, both in equilibrium and non-equilibrium conditions. 
 With a robust scaling analysis we characterized the statistic and geometric properties of the defect clusters and we found universal features  
 along the critical percolation curve $\phi_P({\rm Pe})$. We also proved the important role payed 
 by the dynamic bubbles in cavitation in the dense MIPS phase. {They} break the
 network of defects which lie mostly on the interfaces between micro-domains with orientational order.
 
\noindent
\section*{Acknowledgments.}
We thank L. Berthier, C.B. Caporusso, C. Miguel, M. Picco, A. Suma, J. Tailleur and F. van Wijland 
for very useful discussions.
We acknowledge access to the MareNostrum Supercomputer at the BSC,  IBM Nextscale GALILEO at 
CINECA (Project  INF16-fieldturb)  under  CINECA-INFN  agreement and  Bari ReCaS e-Infrastructure funded by MIUR 
{\it via} PON  Research  and  Competitiveness  2007-2013  Call~254 Action~I. 
DL acknowledges Ministerio de Ciencia, Innovaci\'on y Universidades MCIU/AEI/FEDER for financial support under grant agreement RTI2018-099032-J-I00.
LFC acknowledges financial support from ANR-THEMA (AAP CE30).
IP acknowledges support from Ministerio de Ciencia, Innovaci\'on y Universidades MCIU/AEI/FEDER for financial support under grant agreement PGC2018-098373-B-100 AEI/FEDER-EU and from Generalitat de Catalunya under project 2017SGR-884, and Swiss National Science Foundation Project No. 200021-175719.
 
\appendix

\section{Soft disks}
\label{app:soft-disks}

In this Appendix we present some details on the behaviour of the passive soft disk system.

\begin{figure}[h!]
\vspace{0.25cm}
\centering
\includegraphics[width=8.5cm]{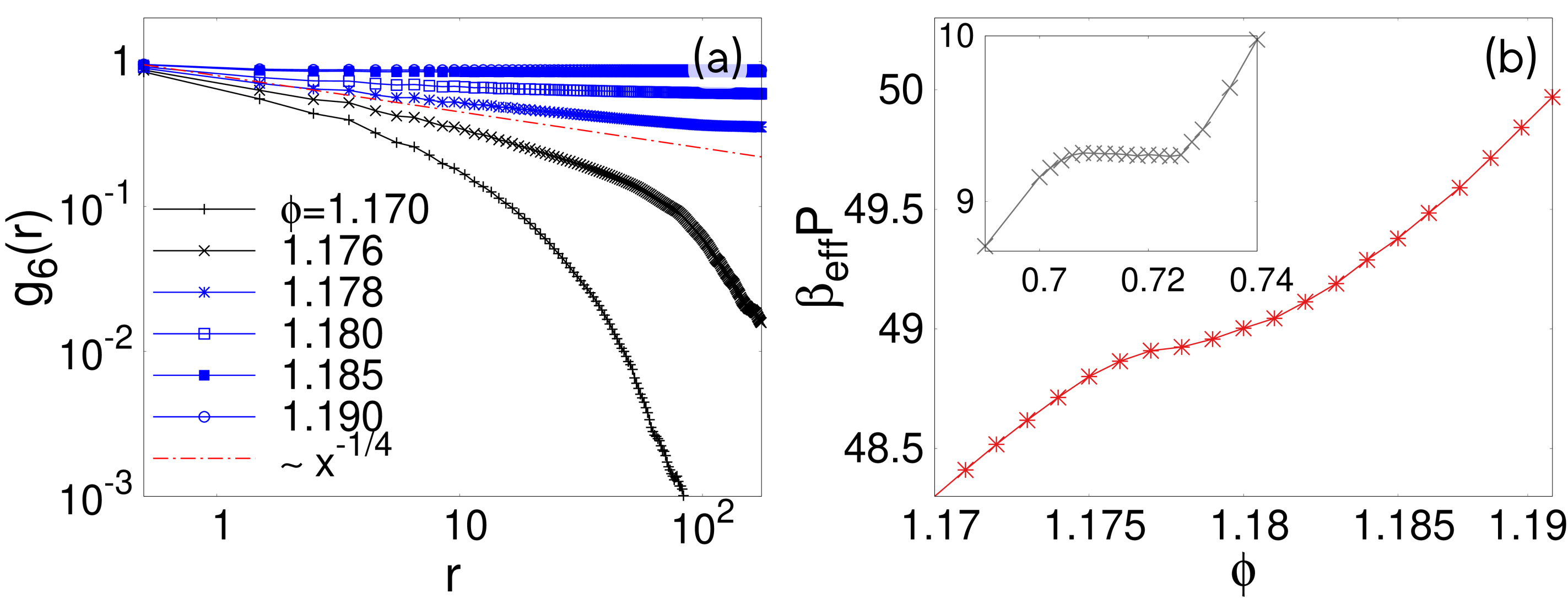}
\caption{Soft disks in equilibrium. (a) Orientational correlation function.
Consistently with the colour code used in the main text, black curves are for parameters in the liquid phase, 
while blue curves are used for the hexatic. The red straight line is the critical power-law predicted by HNY 
for the liquid-hexatic BKT transition~\cite{Halperin1978}.
(b)  Monotonically increasing equation of state in the region across 
the liquid-hexatic transition. In the inset, the equation of state for hard disks at equilibrium, obtained in Ref.~\cite{PRLino}, 
with the loop pointing towards the direction of a first order phase transition {in this case}.
}
\label{fig:corr}
\end{figure}

In Fig.~\ref{fig:corr} (a) we show the orientational correlation function (built with 
the hexatic order parameter measured at 
two points in space separated by a distance $r$). The different colors follow the code 
used in the main text, that is, blue for hexatic and black for liquid. In the main part of panel 
(b) we present the equation of state, $P(\phi)$, for soft disks and we compare it to the 
one of hard disks, displayed in the inset. The absence of a loop in the curve in the main 
panel, replaced by an inflexion point, indicates that the 
hexatic-liquid transition is continuous for soft disks. 
The comparison with the hard-disk $P(\phi)$ in the inset, 
obtained as a detailed view of Fig. 2 (a) in~\cite{PRLino},  
 makes this fact clear.

\section{Particle clustering}
\label{app:particle-clustering}

The identification of a particle cluster in a continuous system is not unique and several proposals 
for ways to do it have been summarized in~\cite{Sator03}.
We adopted a definition based on the proximity between the particles concerned in configurational space (similar to the
Stillinger clusters of liquids). More precisely, we identified particle clusters using the Density-Based Spatial Clustering of 
Applications with Noise (DBSCAN)~\cite{Esler96}. With the chosen 
parameters for the DBSCAN algorithm, one particle is considered to belong to a cluster whenever it has at least 12 neighbors within a radius of $2.1\, \sigma_d$ 
with $\sigma_d$ the particle diameter
(that is slightly above the third shell radius of the triangular lattice). In this way we pinpointed the dense phase in MIPS.


\section{Finite size effects}
\label{app:finite-size}

\begin{figure}[h!]
\vspace{0.25cm}
\centering
\includegraphics[width=\columnwidth]{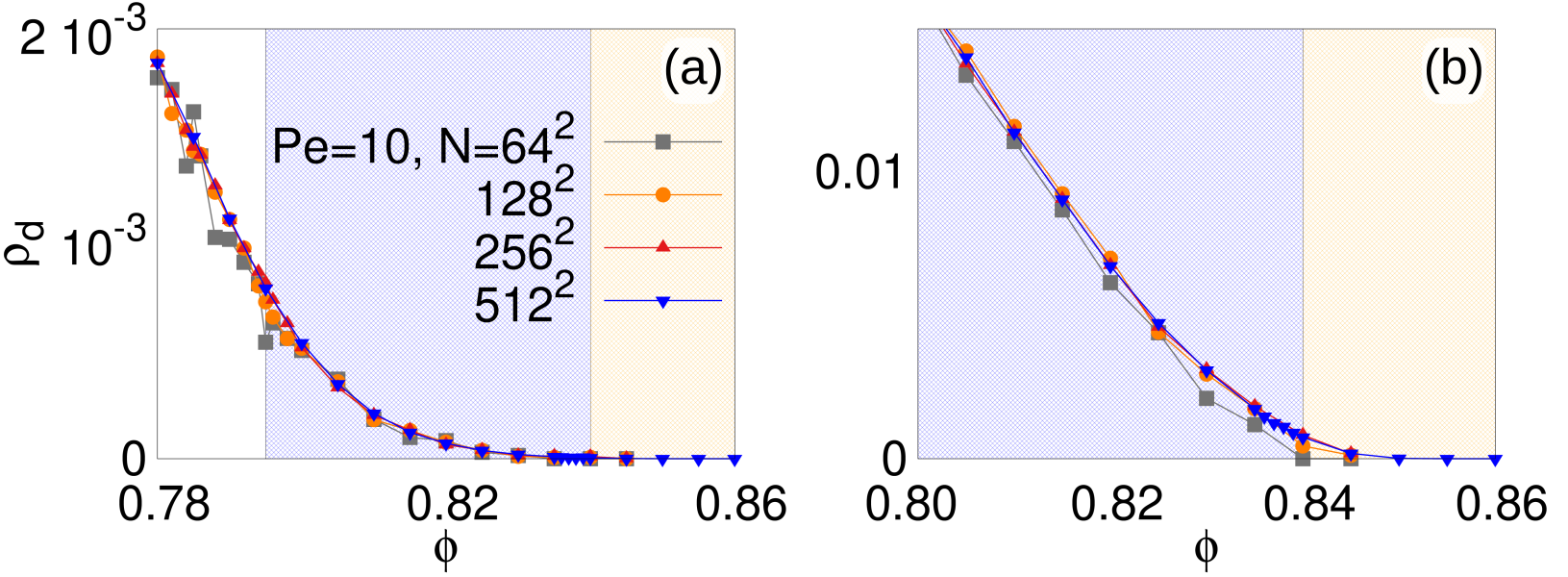}
\caption{Finite size effects on the number densities of point-like defects, as defined in Eq.~(\ref{eq:def-rhod}).
(a) Dislocations.
(b) Disclinations.
The system sizes are given in the key.
}
\label{fig:finite-size}
\end{figure}

In Fig.~\ref{fig:finite-size} we show that the counting of particles belonging to dislocations and disclinations is not much influenced by
finite size effects. Indeed, the number density of dislocations and disclinations do not vary much with the sizes of the 
systems. The measurements were done on independent systems with the sizes given in the keys. Although {\it a priori} quite 
surprising, a similar lack of strong finite size effects was observed in the counting of vortices in the disordered phase of the 
2DXY model~\cite{Jelic11}.

We see a similar size independent in the pair correlation functions  {(not shown here).}


\section{Improved fits of point-like defect densities}
\label{app:improved-fits}

We have shown in Sec.~\ref{sec:point-like} that the behavior  of free dislocations close to the solid-hexatic transition is altogether compatible with the KTHNY unbinding mechanism, so that the number density of free dislocations is strongly related to the spatial divergence of the 
hexatic correlation length predicted by the theory.

Here we provide an additional analysis of the critical behavior of the number density of dislocations, 
in order to determine a reliable uncertainty interval for the non-universal values of the parameters given in Tables~\ref{table:fits-dislocations}
and \ref{table:fits-disclinations}.

\vspace{0.25cm}

\begin{figure}[h!]
\includegraphics[width=\columnwidth]{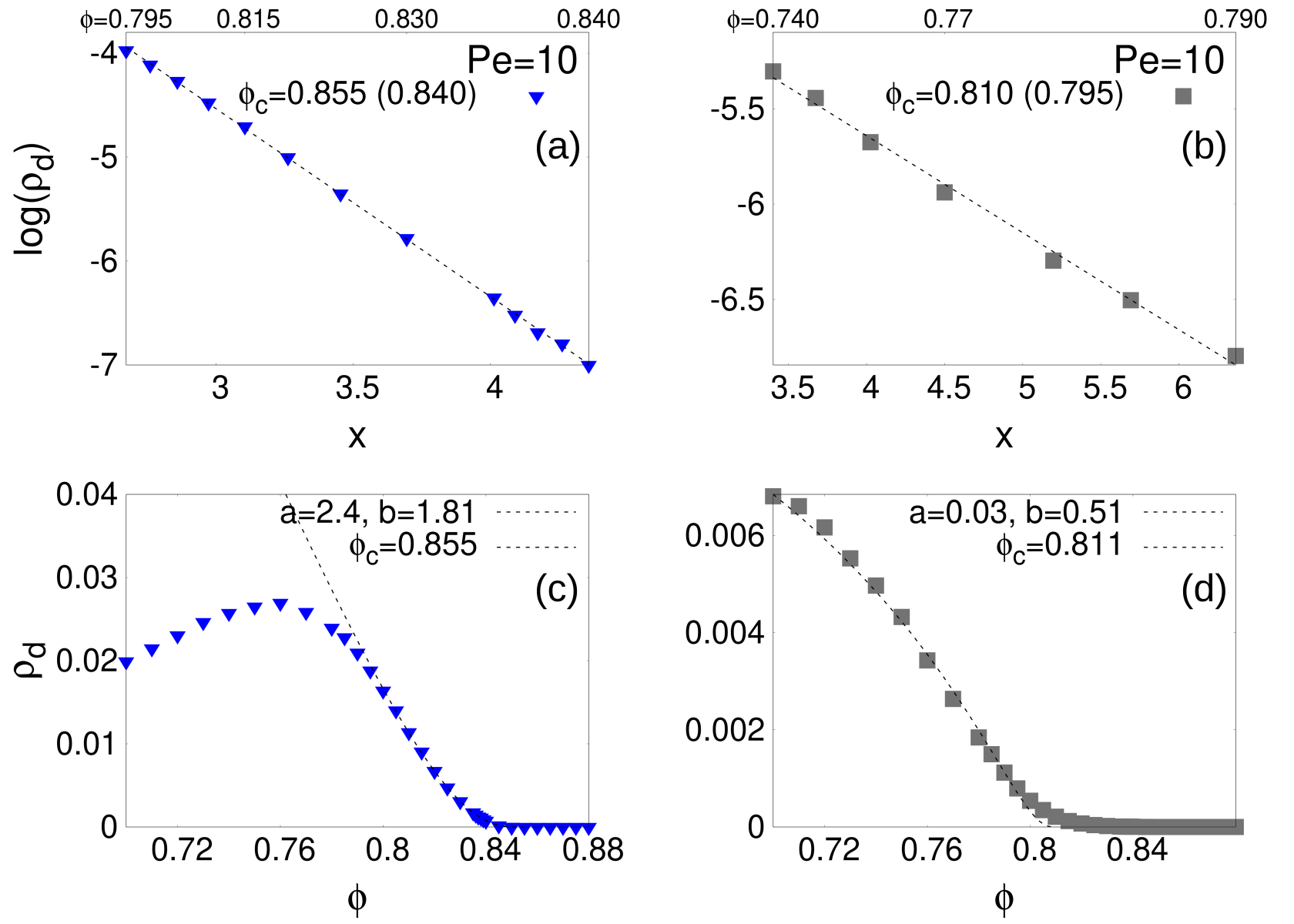}
\caption{
In the first row, the linear representation, Eq.~(\ref{eq:linear_fit}), of the point-like defects at Pe = 10 with  $x=[\phi_c/(\phi_c-\phi)]^{\nu}$. 
(a) Dislocations with $\nu=0.37$ and (b) disclinations with  $\nu=0.5$. $\phi_c$ is chosen to take the values shown as labels in the panels. 
Between parenthesis the values of $\phi_h$ and $\phi_l$ estimated from the study of correlation functions and local densities. In the second row, 
the result of the linear fit fixing the remaining parameters $a$ and $b$.
}
\label{fig:linear_KT_fit}
\end{figure}

Let us first analyze the free dislocation number densities.
Starting from the full BKT critical form in Eq.~(\ref{eq:rhod}), we fix the value of the critical exponent 
$\nu=0.37$ predicted by HNY for the solid-hexatic transition, and we consider the new variable 
$x=[\phi_c/(\phi_c-\phi)]^{\nu}$, so that the logarithm of the number density of dislocation obeys the linear relation:
\begin{equation}
\centering
\log \rho_d=\log a-bx \mbox{.}
\label{eq:linear_fit}
\end{equation}

As shown in {the first row in} Fig.~\ref{fig:linear_KT_fit} for Pe = 10, we find that, interestingly, the numerical data 
{for dislocations} do indeed follow 
a linear behavior if we tune the value of $\phi_c$ around the initial guess $\phi_h$, given from the analysis of the translational correlation 
functions in~\cite{PRLino}, and recalled in Table~\ref{table:fits-dislocations}. Moreover, the optimal value of $\phi_c$ which reproduces the linear prediction is really 
close to the one computed with the fit of the data to the full expression in Eq.~(\ref{eq:rhod}), also given in Table~\ref{table:fits-dislocations}. On the one hand, 
this proves the robustness of the fitting procedure presented in the text. On the other hand, it serves as an alternative procedure 
to obtain an independent estimation of the parameters $a$ and $b$ and their related errors. Moreover, the results of this (linear) evaluation, 
given in the key of Fig.~\ref{fig:linear_KT_fit}, are not affected by the complexity of a full non linear fit of the stretched 
exponential form of Eq.~(\ref{eq:rhod}).

We also show in {the second row in}
Fig.~\ref{fig:linear_KT_fit} the same analysis for the number density of disclinations across the liquid-hexatic transition, 
being now the fixed value of the critical exponent the one of the standard BKT transition $\nu=0.5$. Similar results as the ones discussed 
above for dislocations apply to the disclinations as well. The {parameters} of the full fit are given in Table~\ref{table:fits-disclinations}.


\section{Four parameter fits}
\label{app:four}

In Tables~\ref{table:four-parameter-dislocations} and \ref{table:four-parameter-disclination}
we display the values extracted from four-parameter fits to the dislocation and disclination number 
densities close to the solid-hexatic and hexatic-liquid transitions, respectively. \vspace{0.25cm}
\begin{table}[h!]
\begin{tabular}{|c|c|c|c|c|c|c|c|}
\hline
\;\; Pe  \;\; & \;\; $\nu $ \;\; & \;\; $ a$ \;\; & \;\; $ b $ \;\; & \;\; $\phi_c$ \;\; &  \;\; $\phi_h$ \;\; & \;\; $\chi^2/$ndf \;\;
\\
\hline
\hline
0  & 9 & 13 & 0.002 & 1 & 0.735 & 0.920
\\
\hline
10 & 0.6 & 0.4 & 0.7 & 0.857 & 0.840 & 2.89
\\
\hline
20 & 0.3 & 5 & 3 & 0.881 & 0.870 & 1.39
\\
\hline
30 & 0.8 & 0.2 & 0.3 & 0.909 & 0.880 & 2.08
\\
\hline
40 & 0.7 & 0.2 & 0.4 & 0.90 & 0.885 & 0.924
\\
\hline
50 & 0.2 & 7 & 3 & 0.892 & 0.890 & 0.461
\\
\hline
\end{tabular}
\caption{Dislocation unbinding at the solid-hexatic transition. The fitting parameters 
in Eq.~(\ref{eq:rhod}) for the density of free dislocations  plotted in Fig.~\ref{fig:BKT-scaling-solid-hexatic}.
In the table above, $\nu$ is also a fitting parameter. The values of $\phi_h$(Pe) are the ones estimated 
from the analysis of the correlation functions and probability densities in~\cite{PRLino}.}
\label{table:four-parameter-dislocations}
\end{table}
\begin{table}[h!]
\begin{tabular}{|c|c|c|c|c|c|c|c|}
\hline
\;\; Pe  \;\; & \;\; $\nu $ \;\; & \;\; $ a$ \;\; & \;\; $ b $ \;\; & \;\; $\phi_c$ \;\; &  \;\; $\phi_l$ \;\; & \;\; $\chi^2/$ndf \;\;
\\
\hline
\hline
0  & 0.4 & 0.4 & 2 & 0.7 & 0.725 & 3.24
\\
\hline
10 & 2 & 0.012 & 0.03 & 0.85 & 0.795 & 0.859
\\
\hline
20 & 1 & 0.02 & 0.2 & 0.9 & 0.830 & 0.858
\\
\hline
30 & 0.3 & 0.09 & 2 & 0.86 & 0.845 & 0.965
\\
\hline
40 & 2 & 0.013 & 0.01 & 0.96 & 0.850 & 0.661
\\
\hline
50 & 0.9 & 0.008 & 0.1 & 0.88 & 0.855 & 0.288
\\
\hline
\end{tabular}
\caption{Disclination unbinding at the hexatic-liquid transition. The fitting parameters in 
Eq.~(\ref{eq:rhod}) for the density of free disclinations plotted in Fig. \ref{fig:disclinatios-HN}.
In the table above, $\nu$ is a fitting parameter.The values of $\phi_l$(Pe) are the ones estimated 
from the analysis of the correlation functions and probability densities in~\cite{PRLino}.
}
\label{table:four-parameter-disclination}
\end{table}
In these fits 
we used the BKT scaling form in  Eq.~(\ref{eq:rhod}) and we also let the exponents $\nu$ be determined by the fits.
The fitting intervals and number of data points considered are the same as the ones used 
in the main text and in App.~\ref{app:improved-fits}. The curves resulting from these four parameter fits 
are drawn as broken lines in Fig.~\ref{fig:BKT-scaling-solid-hexatic} and Fig.~\ref{fig:disclinatios-HN} for dislocations and disclinations.
The results in the Tables are not satisfactory for several reasons. First of all, some values of 
$\nu$ are unrealistic (e.g., $\nu = 9$ at Pe = 0 for dislocations and $\nu =  2$ at Pe = 40 for 
disclinations). More generally, they do not show any regular trend. The same lack of regular 
trend can be observed in the fitted  
values of the critical packing fractions $\phi_c$, which show an oscillating behaviour as a function
of Pe in both cases. {Moreover, in some cases they are unrealistically large and close to $\phi_{\rm cp}$.}
The $\chi^2$ normalized by the number of degrees of freedom, is an indicator of the quality of a fit. However, a 
fit with four rather than three parameters will  lead to a better $\chi^2/\mbox{ndf}$ without necessarily implying 
that  it {is} a better fit. 
Indeed, if one looked in more detail into the errors incurred with this kind of 
fit, one would find that they are too large and hence not acceptable. The failure of the fits with too 
many free parameters gives further support to the procedures followed in the main text and in 
App.~\ref{app:improved-fits}, in which we fixed $\nu$ and we searched for $\phi_c$.

As an additional test of reliability of our fits, we also tried a similar fitting procedure in which we fixed
$\nu$ to different values, around the expected HNY critical value. Increasing $\nu$ beyond $0.37$ and $0.5$ for 
dislocations and disclinations
we observed a drift of the fitted $\phi_c$ away from the $\phi_h$ and $\phi_l$, respectively.
Instead, decreasing $\nu$ below these values leads to unstable and unrealistic values for the 
other fitting parameters.

\section{Effects of coarse-graning}
\label{app:coarse-graining}

In Sec.~\ref{sec:topological-defects} we introduced a coarse-graining procedure, with coarse-graining length $d_s$, to 
fill the microscopic gaps between defects and identify the coarse-grained clusters. 

\begin{figure}[h!]
\includegraphics[scale=0.08]{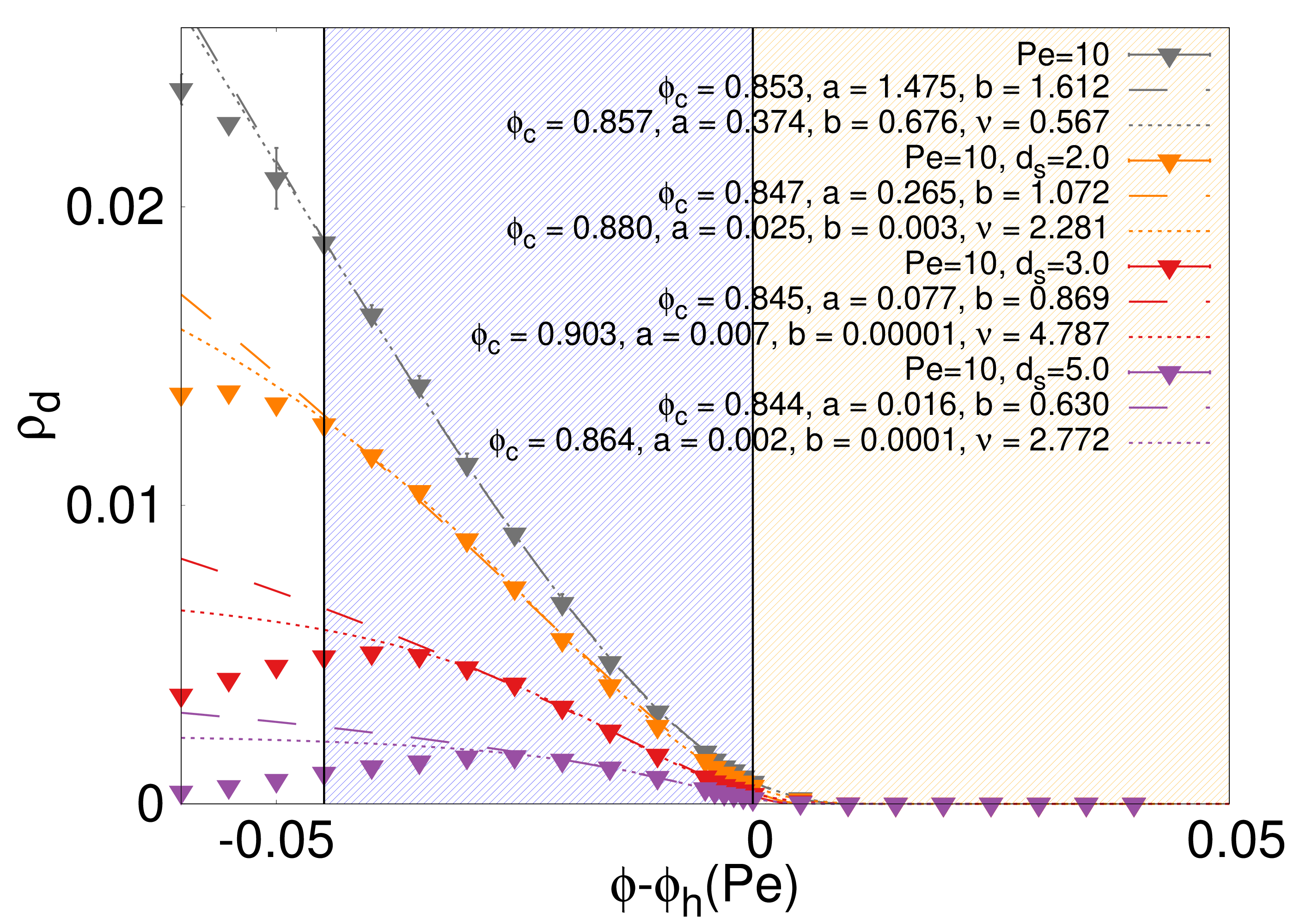}
\caption{Effect of coarse-graining on the critical behavior of the number density of 
dislocations at Pe = 10. Raw data and four (dotted) and three (dashed) parameter fits to the form in Eq.~(\ref{eq:rhod}). 
The coarse-graining length increases from top to bottom, $d_s=0, 2, 3, 5$.
The vertical line on the left is the limit of the fitting interval in the case with no coarse-graining. In the other cases
the fit is done on the data-points to the right of the maximum.
The values of the four and three parameter fits are given in the key.
}
\label{fig:coarse-dislocation-density}
\end{figure}

We first study the effect of such a coarse-graining on the number of unpaired defects. 
The critical behaviour of the number density of dislocations in a system at Pe = 10 for different coarse-graining lengths 
$d_s=0,2,3,5$ (in units of $\sigma_d$), is zoomed over in Figs.~\ref{fig:coarse-dislocation-density}. 
For dislocations, we see that the three parameter fits to Eq.~(\ref{eq:rhod})  with $\nu=0.37$ are very satisfactory 
for all these values of $d_s$. Moreover, the $\phi_c$'s extracted from the fits slowly approach the 
value $\phi_h=0.840$ estimated in~\cite{PRLino} for Pe = 10. Instead, the four parameter fits in which $\nu$ is also 
{an adjustable parameter} yield unreasonable values of $\nu$ and, moreover, no clear trend 
for $\phi_c$. The fitting parameters are
given in the key.

\begin{figure}[h!]
\includegraphics[scale=0.08]{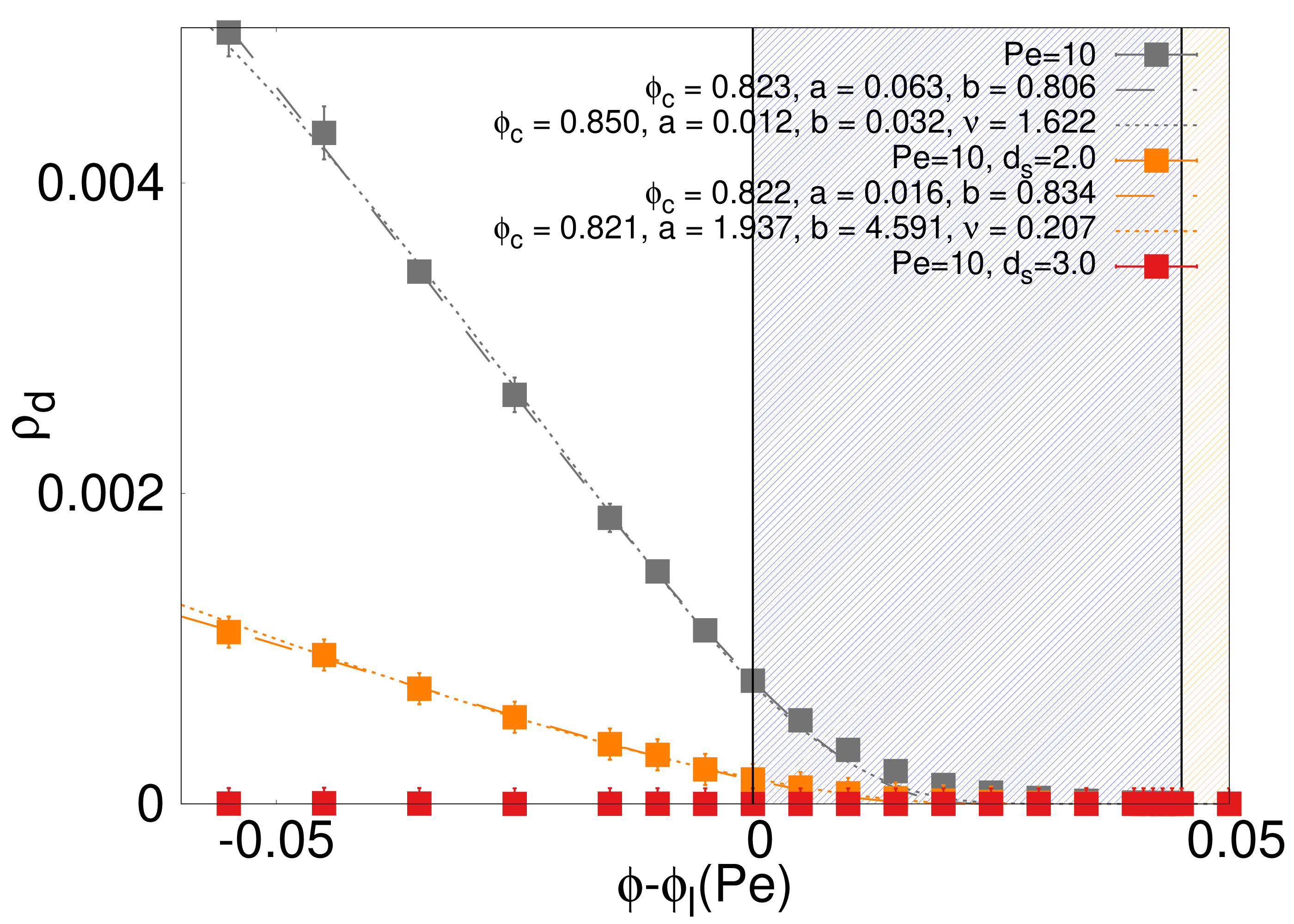}
\caption{Effect of coarse-graining on the critical behavior of the number density of 
disclinations at Pe = 10. Raw data and four (dotted curves) and three (broken curves) parameter fits to the form in Eq.~(\ref{eq:rhod}). 
The values of the four and three parameter fits are given in the key.
}
\label{fig:coarse-disclination-density}
\end{figure}

Instead, for the number density of disclinations displayed in Fig.~\ref{fig:coarse-disclination-density} the 
effect of the coarse-graining length is stronger. Already for $d_s=3$ we have practically 
erased all disclinations. The three parameter fits with $\nu=0.5$ to the two remaining sets of data, with no coarse-graining and using 
$d_s=2$, are {of better quality} than the four parameter ones. 

\begin{figure}[h!]
\includegraphics[scale=0.08]{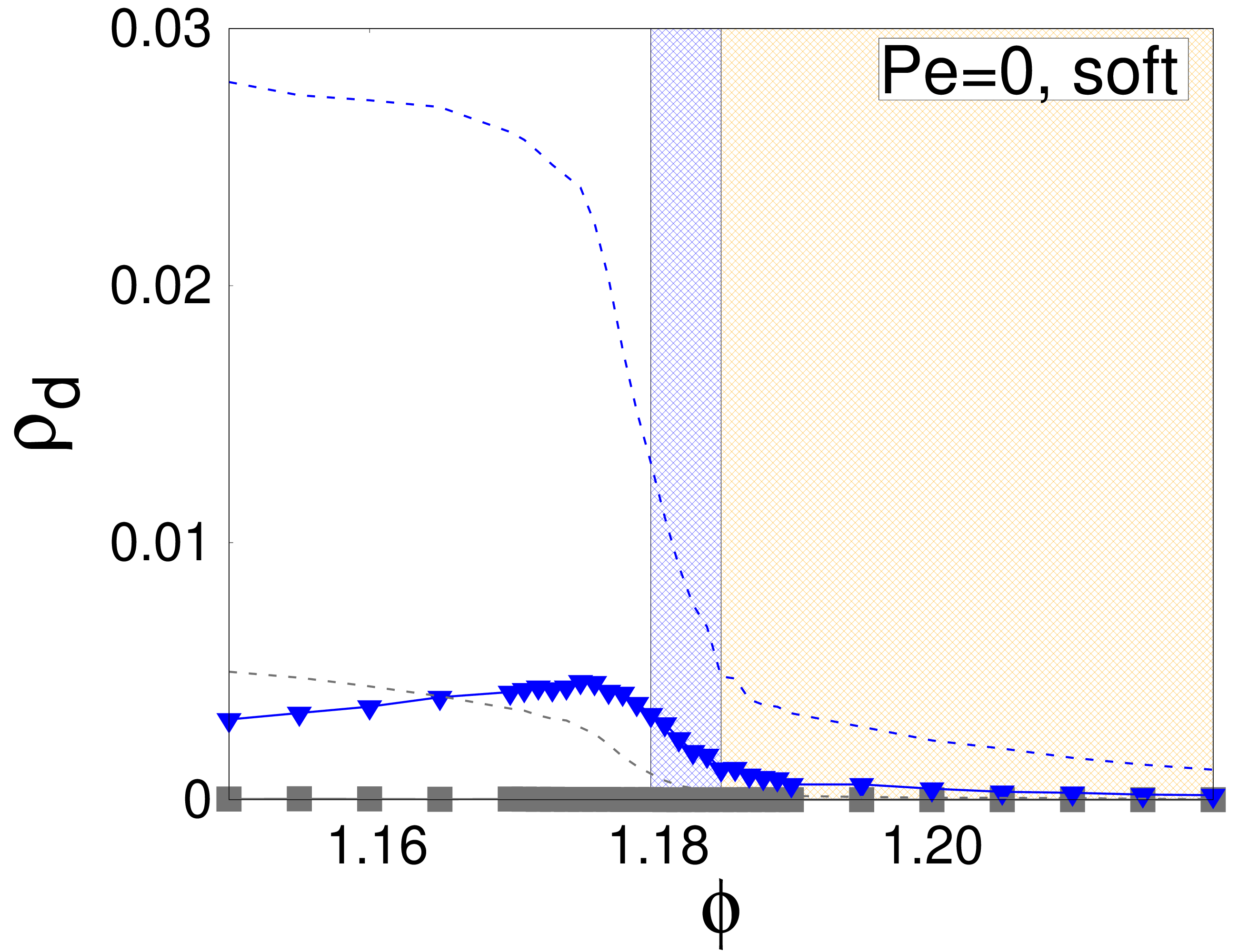}
\caption{Effect of coarse-graining {with $d_s=2$} 
on the {point-like} defect counting in a passive system of soft disks 
The dotted  {curves} show data with no coarse-graining applied.
}
\label{fig:coarse-graining}
\end{figure}

Finally, in Fig.~\ref{fig:coarse-graining} we display the effects of coarse-graining on the isolated defect counting  in the 
soft disk system at Pe = 0. Blue and gray data-points are for dislocations and disclinations, respectively, and 
the dotted curves represent the raw data with no coarse-graining. The figure proves that the excess presence of dislocations in the 
solid phase is strongly diminished as soon as we apply a coarse-graining over a rather short scale, $d_s=2$. 

\vspace{0.25cm}

\begin{figure}[h!]
\vspace{0.25cm}
\centering
\includegraphics[width=8.5cm]{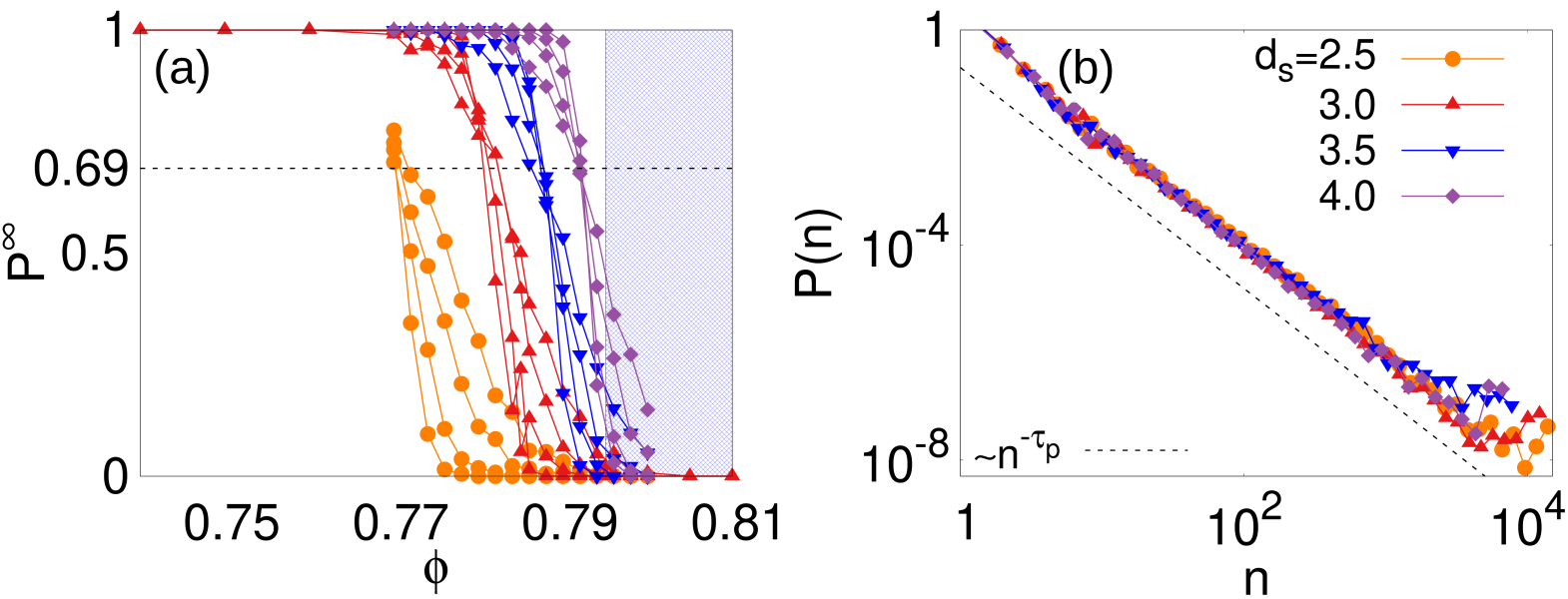}
\caption{Active hard disk system at Pe = 10.
(a) The cluster size probability at $\phi_P$ for different coarse-graining lengths.
(b) The percolation probability of defect clusters $P^\infty$ as a function of $\phi$ for different coarse-graining lengths.
The horizontal dashed line is the percolation threshold for critical percolation in the 2D square lattice with PBC, and it is shown as a reference. 
}
\label{fig:density-defects-ds}
\end{figure}

In Fig.~\ref{fig:density-defects-ds} we study the coarse-graining length influence on the percolation 
properties. In panel (a) we see that the finite size scaling of the percolation probability $P^\infty$ 
depends on $d_s$ and the crossing point, $\phi_P$, moves towards $\phi_l$ for increasing $d_s$. Still, 
the critical properties of the defect clusters are independent of $d_s$ in the sense shown in panel 
(b): the cluster size probability $P(n)$ keeps the same algebraic decay when measured at the 
corresponding $\phi_P$.

\section{Percolating defect clusters at Pe~=~50}
\label{app:df_Pe50}

\begin{figure}
\vspace{0.75cm}
\includegraphics[width=\columnwidth]{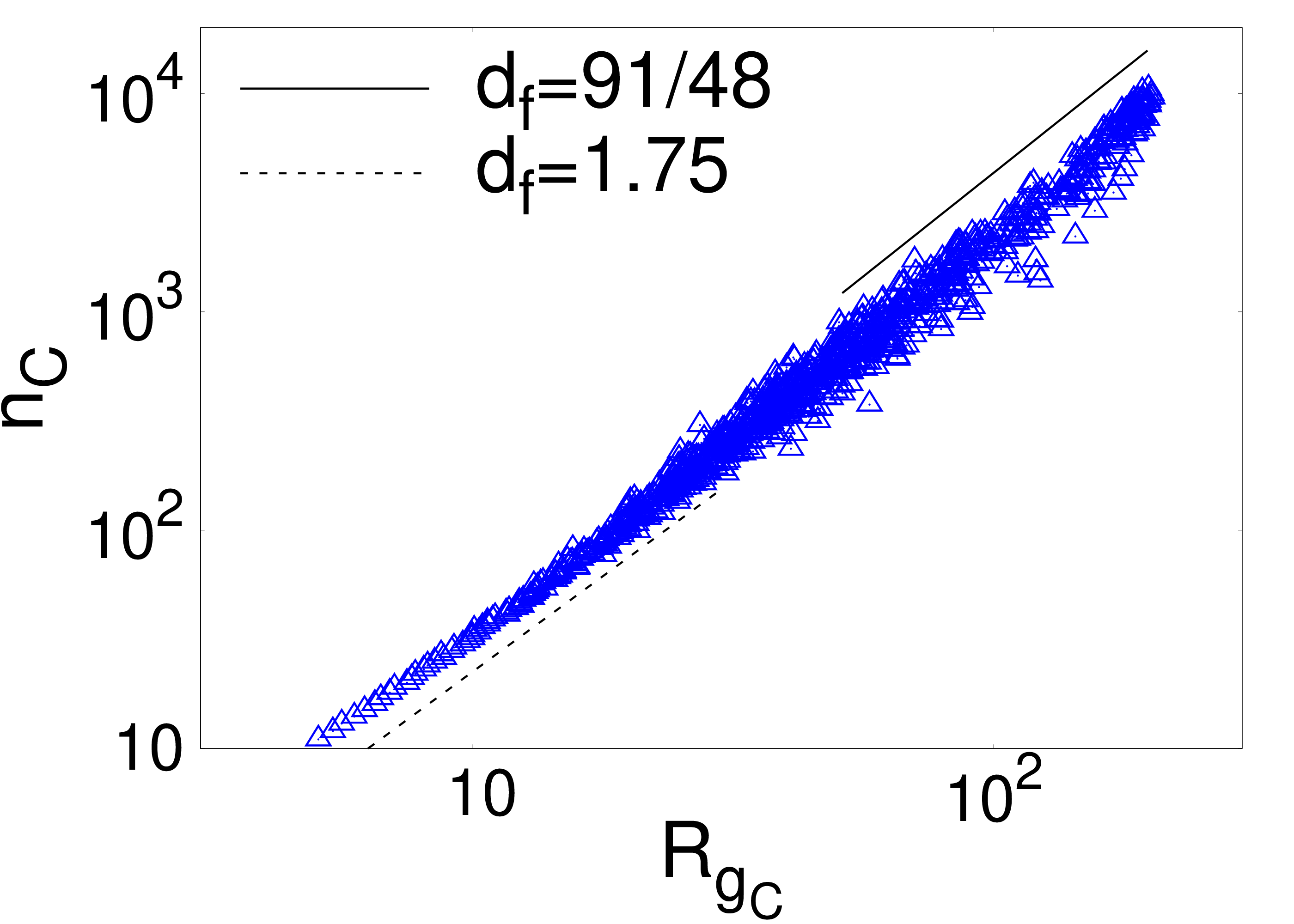}
\caption{Scatter plot  $n_\mathcal{C}$ against the radius of gyration ${R_{\rm{\it g}}}_\mathcal{C}$ of the defect clusters 
for Pe~=~50 and $\phi=\phi_P=0.815$. The continuous and dashed lines are fits of large ($n_{\mathcal C}>100$) 
and small ($n_{\mathcal C}<100$) respectively, {and the extracted fractal dimensions are reported in the key.}
}
\label{fig:two_dfs_Pe50}
\end{figure}

As illustrated in Sec.~\ref{sec:clusters}, we observe percolation of defects at Pe~=~50 in the 
dense liquid, close to the high-density branch of MIPS.  MIPS inhibits the full statistical 
development of percolating clusters, and also allows the existence of defect clusters of mixed nature. 
In this Appendix we provide further details on the morphology of the defect clusters at Pe~=~50.

The scatter plot in Fig.~\ref{fig:two_dfs_Pe50} shows the dependence of the cluster size $n_C$ on the 
radius of gyration ${R_{g}}_C$ at $\phi=\phi_P$, away from MIPS. 
We clearly observe that two different kinds of clusters coexist, characterized by a different fractal dimension.
On the one hand, large clusters of size $n_C\gtrsim100$ have fractal dimension $d_{\rm f}\simeq91/48$, 
related to the critical decay exponent of the 
cluster size distribution shown in Fig.~\ref{fig:percolation_pe50}, as expected in the context of the percolation transition. 
On the other hand, grain-boundary structures of fractal dimension $d_{\rm f}\simeq1.75$ prevail at  smaller scales, 
capturing the presence of particle aggregation in the dense liquid (see Table~\ref{tab:MIPS}). This behavior 
shows how MIPS is partially preempting percolation 
at this Pe,  providing a clear picture of the complete disappearance of percolation at larger Pe numbers.


\bibliographystyle{apsrev4-1}
\bibliography{defects}

\end{document}